\documentclass[a4paper]{article}
\pdfoutput=1
\usepackage{graphicx} % Required for inserting images
\usepackage{jcappub}
\usepackage{url}
\usepackage{graphicx}
\usepackage{subcaption,siunitx,booktabs}
\usepackage{amsmath,amssymb,amstext,amssymb,amsfonts,amsthm}
\usepackage{fontawesome} %for GitHub icons
\usepackage{xspace}
\usepackage{hyperref}
\usepackage{appendix}
\usepackage{physics}
\usepackage{soul}
\usepackage{bm}
\usepackage{comment}
\usepackage{color}
\usepackage[normalem]{ulem}
\usepackage{subcaption}
\usepackage{todonotes}
\usepackage[version=3]{mhchem} % for isotopes/chemistry
\usepackage{aas_macros}

\renewcommand{\dd}{\mathrm{d}}
\newcommand{\vect}[1]{\boldsymbol{\mathbf{#1}}}
\definecolor{deepgreen}{rgb}{0.2,0.8,0.2}

\definecolor{deepblue}{rgb}{0.2,0.2,0.8}

\definecolor{deepred}{rgb}{0.8,0.2,0.2}

\definecolor{linkcolor}{rgb}{0.7752941176470588, 0.22078431372549023, 0.2262745098039215}
\newcommand{\nbicon}{{\color{linkcolor}\faFileCodeO}\xspace}
\newcommand{\nblinkeem}[1]{\href{https://github.com/calvinikchen/expanding-ejecta/tree/main/#1}{\nbicon}}
\newcommand{\githubmaster}{\href{https://github.com/calvinikchen/expanding-ejecta/}{\faGithub}\xspace}
\newcommand\githubicon[1]{\href{#1}{\faGithub}\xspace}

\title{Expanding Ejecta Method:\\
I.~Mapping Supernova Morphology with Intensity Interferometry}

\author[a]{I-Kai Chen,}
 \emailAdd{ic2127@nyu.edu}
 \affiliation[a]{Center for Cosmology and Particle Physics, Department of Physics, New York University, New York, NY 10003, USA}
\author[a]{David Dunsky,}
 \emailAdd{ddunsky@nyu.edu}
 \author[a,b]{Ken Van Tilburg,}
 \emailAdd{kenvt@nyu.edu, kvantilburg@flatironinstitute.org}
 \affiliation[b]{Center for Computational Astrophysics, Flatiron Institute, New York, NY 10010, USA}
\author[c]{Junwu Huang,}
 \emailAdd{jhuang@perimeterinstitute.ca}
 \affiliation[c]{Perimeter Institute for Theoretical Physics, 31 Caroline St.~N., Waterloo, Ontario N2L 2Y5, Canada}
\author[d]{and Robert V.~Wagoner}
\emailAdd{wagoner@stanford.edu}
\affiliation[d]{Department of Physics and KIPAC, Stanford University, Stanford, CA 94305, USA}
 
\abstract{
We explore the potential of optical intensity interferometry to extract angularly resolved information from supernova explosions, introducing the \emph{expanding ejecta method} (EEM) as a robust alternative to the classical expanding photosphere method (EPM). Foreseeing future improvements to intensity interferometers of large light collection area ($25\pi\,\rm{m}^2$ per telescope) equipped with spectral multiplexing ($10^4$ spectral resolution) and fast photodetectors ($10\,\mathrm{ps}$ timing resolution, $50\%$ overall efficiency), we demonstrate that high signal-to-noise measurements of the visibility modulus are achievable for Type IIP (Type Ia) supernovae out to $3~(12)\,\mathrm{Mpc}$. By focusing on generic line emission and absorption in ballistic ejecta, the EEM can relax assumptions about spherical symmetry, blackbody radiation, and extinction. The EEM enables angular diameter distances to be determined with $\sim2\%$ precision for supernovae of apparent magnitude $m = 12$ from a 60-hour observation by an intensity interferometer with those same instrumental specifications. 
We argue that the EEM is significantly more robust to modeling uncertainties and systematic effects than (variants of) the EPM. In a companion paper~\cite{eem-2}, we show how the EEM can be used to provide geometric anchors for cosmic distance ladder calibration, or to construct a wholly independent Hubble diagram based on angular diameter distances.}

\date{\today}

\begin{document}

\maketitle
\flushbottom

\section{Introduction}
\label{sec:introduction}

% History and significance of supernovae
Supernova (SN) explosions are among the brightest and most intensively studied transient astrophysical phenomena. These events are well understood through observations across the electromagnetic spectrum~\cite{alsabti2017handbook}, from radio to $\gamma$-ray frequencies, and even via neutrino detections~\cite{1987PhRvL..58.1494B, 1988NuPhA.478..189T}. Many types of supernovae (SNe) are exceptionally bright in the optical band, remaining luminous for several months and releasing over $10^{51} \,{\rm erg}$ of energy during their explosions. Observations of SNe date back at least to 185 AD, and modern instruments have captured spectra for approximately 2000 SNe, classifying them into distinct categories based on their spectral and temporal properties. The study of SNe has provided critical insights into stellar astrophysics~\cite{1931ApJ....74...81C, 1957RvMP...29..547B,2002RvMP...74.1015W}, severely constrained weakly-coupled new physics beyond the Standard Model~\cite{1990PhR...198....1R,2024arXiv240113728C, 2025PhyR.1117....1C}, and driven major advancements in cosmology~\cite{1999ApJ...517..565P, 1998AJ....116.1009R, 2016ApJ...826...56R,2001ApJ...553...47F}.

% Classical applications of supernovae: distance measurements
Supernovae of Type Ia in particular are widely used as standard(izable) candles for constructing the Hubble diagram and measuring cosmological parameters, including the cosmic expansion rate $H_0$ and the dark energy density~\cite{1999ApJ...517..565P, 1998AJ....116.1009R}. However, Type Ia SNe require independent calibration to anchor the distance ladder, relying on geometry~\cite{2014ApJ...783..130R, 2018ApJ...861..126R, 2020ApJ...891L...1P} and extensions to larger distances with empirical relations in Cepheid variables~\cite{2001ApJ...553...47F, 2016ApJ...826...56R, 2023A&A...672A..85C, 2023ApJ...951..118B,2020JApA...41...23B, 2022ApJ...940...64Y} and the tip of the red giant branch~\cite{2019ApJ...882...34F, 1997MNRAS.289..406S, 2020ApJ...891...57F, 2023AJ....166....2M}, or with other methods~\cite{1997ApJ...475..399T, 2019Natur.567..200P,1977A&A....54..661T, 2024arXiv240803474L}. 

% EPM
The expanding photosphere method (EPM) is a geometric technique for measuring distances to SNe, particularly of Type II, introduced by ref.~\cite{KirshnerKwan} and further developed by refs.~\cite{1981ApJ...250L..65W,1992ApJ...395..366S,1996ApJ...466..911E, 2005A&A...439..671D}. It relies on the principle that as the SN's photosphere expands, its \emph{angular} and \emph{physical} velocities can be related to its distance. By combining observations of the SN's apparent brightness, color temperature, and expansion velocity (derived from spectral lines), an inference can be made about its distance. The method assumes the photosphere radiates as a diluted blackbody, allowing the angular size to be computed from photometry. Dividing the physical expansion velocity by the angular velocity yields the distance to the SN. The EPM has been instrumental in providing independent measurements of cosmological distances, and by extension, the cosmological expansion history~\cite{1977ApJ...214L...5W}. Modern incarnations of the EPM have been able to measure the Hubble constant to an \emph{internal} statistical precision at the few-percent level. Systematic errors are harder to quantify, owing to uncertain SN atmosphere models~\cite{1992ApJ...395..366S, 1994ApJ...432...42S, 2009ApJ...696.1176J, 2014ApJ...782...98B,2018A&A...611A..25G}.
Its reliability is hindered by several challenges, including assumptions of spherical symmetry, uncertainties in the dilution factors used for blackbody corrections, and the influence of extinction by interstellar dust. These limitations have motivated variations on the EPM~\cite{1995ApJ...441..170B,2004ApJ...616L..91B,2002ApJ...566L..63H,2014AJ....148..107R,2015ApJ...815..121D} that partially mitigate some of these systematic uncertainties.
A recent implementation~\cite{2024arXiv241104968V} finds $H_0 = 74.9 \pm 1.9 \, \mathrm{(stat)} \, \mathrm{km/s/Mpc}$, compatible with other local distance indicators.

% limitations of EPM
A fundamental limitation of the EPM and its existing variants is that parameter inference is performed on the \emph{spatially integrated} flux density. Therefore, deviations from spherical symmetry, flux dilution factors, and dust extinction are not knowable \emph{directly}, and indirect measures struggle to attain the requisite \emph{accuracy} even on one of the best-studied examples, SN1999em~\cite{2009ApJ...696.1176J, 2001ApJ...558..615H,2002PASP..114...35L, 2006A&A...447..691D}. 
%spherical symmetry
The EPM assumes spherical expansion of the SN photosphere, but spectropolarimetric observations indicate significant asymmetries in the photosphere and ejecta~\cite{1982ApJ...263..902S, 2001PASP..113..920L, 2024A&A...684A..16D, 2008ARA&A..46..433W}. These asymmetries introduce biases in spatially integrated observations, limiting the method’s accuracy.
%flux dilution
Flux dilution factors, which correct for electron scattering and line blanketing, depend on model atmospheres~\cite{1996ApJ...466..911E,2005A&A...439..671D,2015MNRAS.453.2189D}. Uncertainties in these models, along with intrinsic variability between SNe, propagate into systematic errors in angular diameter and distance estimates.
%dust extinction
Dust along the line of sight dims and reddens SN light, complicating distance measurements. Standard extinction corrections rely on reddening laws~\cite{1989ApJ...345..245C, 1999PASP..111...63F}, which may be improved upon by taking early spectra before maximum light~\cite{2000ApJ...545..444B}, though systematic biases cannot be completely excluded. 

% spatial info
Angularly resolved observations of SNe address each of these limitations by providing (projected) spatial information that cannot be inferred from integrated measurements without some of the aforementioned modeling and assumptions. Asymmetries in the photosphere and ejecta, previously inferred indirectly through spectropolarimetric or line-profile studies, can be directly measured, enabling a precise characterization of deviations from spherical symmetry and eliminating biases in the determination of line-of-sight distances. Flux dilution factors and extinction corrections are irrelevant since the Baade-Wesselink method~\cite{1926AN....228..359B,1946BAN....10...91W} is \emph{not} used to infer the angular size of the SN --- the geometry and structure of the emitting regions are observed directly.
Resolving SN explosions would thus enable a transformative improvement in the accuracy and reliability of distance measurements, free from many of the limitations of the EPM and its variants.

%intensity interferometry, recent revival
Astronomical intensity interferometry exploits the second-order coherence of the optical light of an astrophysical source, in the form of excess correlation of photon arrival times at distant telescopes, to tease out angular information on that source~\cite{1954PMag...45..663B,1956Natur.177...27B,1956Natur.178.1046H}. An intensity interferometer essentially aims to measure the four-point function of the electromagnetic field, as opposed to the two-point function probed by a classical (amplitude) interferometer. A major drawback of the technique is a low signal-to-noise ratio (SNR) on this higher-point correlation function, requiring (or at least desiring) bright sources, large light collection areas, spectral multiplexing, and ultrafast single photon detection~\cite{1974iiaa.book.....B}. Massive improvements in the latter two technologies since the pioneering intensity interferometers~\cite{1967MNRAS.137..375H, 1967MNRAS.137..393H, 1974MNRAS.167..121H} have fueled a recent revival~\cite{2013APh....43..331D,2020MNRAS.491.1540A,2024MNRAS.529.4387A,2020NatAs...4.1164A,2024ApJ...966...28A,2018ExA....46..531R,2024MNRAS.52712243Z} and should make possible a spatially resolved measurement on celestial objects with visible magnitude of 15 or even fainter. As for any interferometer, the spatial resolution is determined by the wavelength divided by the baseline distance. In the case of an \emph{intensity} interferometer, the baseline can be made arbitrarily long, since the light need not be physically recombined --- photons can be recorded and later correlated offline. A bright extragalactic SN at a distance of $3\,\mathrm{Mpc}$ typically has an angular radius of $10^{-10}\,\mathrm{rad} \approx 20 \,\mathrm{\mu as}$, which lies beyond the resolving power of current optical amplitude interferometers and imaging telescopes, but would be an ideal target for a optical intensity interferometer with a $2\,\mathrm{km}$ baseline.

% in this work
In this work, we forecast the capabilities of a spectrally multiplexed intensity interferometer with two baselines to resolve the 3D morphology of SNe. 
Adopting a simple parametric model of H$\alpha$ line absorption and emission around a Type IIP SN, we demonstrate how correlations across spectral channels and varying projected baselines can be used to reconstruct the phase-space distribution of the ballistically expanding ejecta and to geometrically determine angular diameter distances.
This ``expanding ejecta method'' (EEM) is a robust, absolute distance estimator that is less susceptible to systematic uncertainties than other methods. 
The EEM should be broadly applicable to other spectral lines and SN types with appropriate modifications, and could be refined further through more sophisticated forward modeling to enhance accuracy and applicability.
With next-generation intensity interferometry, a SNR of order unity per spectral channel, or total $\mathrm{SNR} \sim 30$ over the bandwidth spanning the P Cygni line feature, can be obtained in ten nights of observation on a SN of visual magnitude 12, an event roughly of annual occurrence (once in a decade for Type IIP). These measurements will illuminate SN physics and establish novel calibrations and determinations of extragalactic distances.

% companion
In a companion paper~\cite{eem-2}, we demonstrate how the EEM enables direct measurements of angular diameter distances to \emph{populations of SNe} as a function of redshift, offering two complementary applications for cosmology. First, the EEM can be used to calibrate Cepheids or Type Ia SNe, providing a geometric anchor for their distance scale with minimal reliance on external methods. Second, it allows for the construction of an independent Hubble diagram, entirely bypassing the cosmic distance ladder. This dual capability makes the EEM a powerful tool for improving the precision of cosmological measurements, addressing tensions in $H_0$, and testing the consistency of the standard cosmological model.
 
% outline
In section~\ref{sec:general}, we review Type IIP SNe during the photosphere phase, focusing on the limitations of spatially unresolved observables --- such as those underlying the EPM and its variants (section~\ref{sec:unresolved}) --- and the benefits offered by spatially resolved observables (section~\ref{sec:resolved}). We describe the potential of intensity interferometry to provide such information (section~\ref{sec:IIobs}). In section~\ref{sec:parametric}, we present a simple parametric model describing the photosphere (section~\ref{sec:photosphere}), ejecta (section~\ref{sec:ejecta}), and asymmetries (section~\ref{sec:asymmetry}). 
Section~\ref{sec:parameter-estimation} forecasts the parameter estimation performance of intensity interferometry within this model. Finally, section~\ref{sec:discussion} suggests future enhancements to the EEM, including more sophisticated modeling, simulations, and additional observables such as polarization. We conclude in section~\ref{sec:conclusions}.

% conventions
We assume a $\Lambda$CDM cosmology with $\Omega_M = 0.3$, $\Omega_\Lambda = 0.7$, and $H_0 = 70 \,\mathrm{km \, s^{-1} \, Mpc^{-1}}$. The code pipeline for our study is available on GitHub (\githubmaster), with interactive links (\nbicon) below each figure providing access to the code used for figure generation.

\section{General Observables} \label{sec:general}

In section~\ref{sec:unresolved}, we first review the important qualitative features of SN optical emission in the context of the traditional EPM, i.e.~from a spatially \emph{unresolved} spectrum.
We then discuss in section~\ref{sec:resolved} how spatially resolved observables may lend more robust, model-independent information on SN morphology.
Finally, we detail how an intensity interferometer with spectral multiplexing can extract spatial information from the image (section~\ref{sec:IIobs}), in anticipation of its use on the parametric model of section~\ref{sec:parametric} to measure the intrinsic properties and angular diameter distances of SNe in section~\ref{sec:parameter-estimation}.

\subsection{Unresolved supernova observables}\label{sec:unresolved}

Optical observations of SNe typically focus on the photosphere phase, around the time when the explosion produces peak luminosity. We will call the surface where the optical depth reaches $\tau = 2/3$ ``the photosphere'', marking the apparent boundary of (the bulk of) the emission, though the frequency dependence of the spectrum is typically determined at greater depths set by the absorptive rather than scattering opacity~\cite{1981ApJ...250L..65W}. This concept of the photosphere has its limitations in terms of temperature and size assignments: in Type Ia SNe, the ejecta may not be optically thick enough for full thermalization, while for Type II SNe, the photosphere radius can depend strongly on wavelength. Despite these complications, the photosphere can provide a useful framework for characterizing SN emission in terms of effective temperature, brightness, and spectral features.

Type Ia SNe result from thermonuclear explosions of carbon-oxygen white dwarfs that exceed the Chandrasekhar limit, releasing $\mathcal{O}(10^{51})\,\text{erg}$ of energy as carbon and oxygen fuse into \ce{^{56} Ni}. The radioactive decay of \ce{^{56} Ni} to \ce{^{56} Co} and subsequently to \ce{^{56} Fe} powers the optical light, which peaks at $2 \times 10^{43} \, \mathrm{erg/s}$ and remains bright for weeks.
Core-collapse SNe, including Types II, Ib, and Ic, arise from the gravitational collapse of massive stars ($M > 8 M_\odot$)~\cite{2009ARA&A..47...63S}. In these events, the collapsing core forms a compact remnant (neutron star or black hole), while a shock wave unbinds and ejects the envelope and outer layers of the progenitor star. Type II SNe retain their hydrogen-rich envelopes, producing strong hydrogen lines and exhibiting luminosities of $\mathcal{O}(10^{42})\, \mathrm{erg/s}$ that remain elevated for months, powered by shock heating, hydrogen recombination, and radioactive decay. Types Ib and Ic result from progenitors that have shed their outer envelopes due to stellar winds or binary interactions, often refered to as stripped-envelope SNe. Type Ib SNe lack hydrogen but exhibit helium lines, while Type Ic SNe, having lost both hydrogen and helium layers, show no such features in their spectra. Both Ib and Ic SNe derive their optical light from radioactive decay, with peak luminosities comparable to those of Type II SNe. The observed average peak luminosity is around $1.5$ magnitude fainter than that of typical Type Ia. However, both Type Ib and Type Ic have a wide spread in absolute magnitude, with some of them even exceeding that of Type Ia~\cite{richardson2006absolute}.

As the continuum emission from the photosphere traverses the expanding ejecta, it produces P Cygni profiles in spectral lines, characterized by blueshifted absorption and symmetric (both blueshifted and redshifted) emission features. These profiles arise from the interplay of ejecta kinematics and composition, and encode key information about the explosion. Hydrogen lines dominate in Type II SNe, while Type Ib SNe show prominent helium features, and Type Ic SNe display lines from heavier elements such as oxygen, magnesium, and calcium. Type Ia SNe are distinguished by the absence of hydrogen and helium lines and the presence of strong silicon absorption near maximum light, along with features from iron-group elements like cobalt and iron at later phases. Spectral line widths and Doppler shifts provide spatially averaged measures of the ranges in ejecta velocity $v_{\rm ej}$ projected along the line of sight. When paired with measurements of brightness and effective temperature from the continuum emission, these observables form the basis of the Baade-Wesselink method applied to SNe --- the expanding photosphere method (EPM) --- for determining SN distances.

The original EPM was first applied to Type Ia SNe in ref.~\cite{1973MNRAS.161...71B} and to Type II SNe in ref.~\cite{KirshnerKwan}, following a proposition by Searle~\cite{1974ARA&A..12..315O}.
The method measures the angular diameter distance $D_A$ to a SN at redshift $z$ as the ratio between the physical expansion velocity $\dot{R}_\text{ph}$ of the expanding SN photosphere and its angular velocity $\dot{\Theta}_\text{ph}$, after correcting for redshift factors appropriately. At a given moment, the photosphere's angular radius $\Theta_\mathrm{ph}$ is:
\begin{align}\label{eq:EPMphotosphere}
    \Theta_\text{ph}(\lambda) = \frac{R_\text{ph}(\lambda)}{D_A} = (1+z)^2 \sqrt{\frac{(1+z)  f_\text{ph}(\lambda)}{\zeta(\lambda')^2 \pi B_{\lambda'}(T_\text{ph}) 10^{-\frac{2}{5}\left[A(\lambda) + A'(\lambda') \right]}}},
\end{align}
where $\lambda = (1+z) \lambda'$ is the observed wavelength and $\lambda'$ is the SN rest-frame wavelength. The observed flux density is $f_\text{ph}(\lambda)$, while $B_{\lambda'}(T)$ is the Planck spectral radiance at temperature $T_\text{ph}$ in the SN rest frame. The exponents $A(\lambda)$ and $A'(\lambda')$ account for dust extinction in the Milky Way and in the SN environment and host galaxy, respectively. The flux dilution factor $\zeta(\lambda')$ captures deviations from blackbody radiation due to e.g.~electron scattering. Inferring $\Theta_\text{ph}(\lambda)$ requires measurements of $f_\text{ph}(\lambda)$, the redshift $z$, and the photospheric temperature $T_\text{ph}$, typically determined by fitting the continuum spectrum. 
The physical photosphere radius evolves approximately as:
\begin{align}
R_\text{ph}(\lambda) \simeq R_0(\lambda) + \dot{R}_\text{ph}(\lambda) \frac{t-t_0}{1+z},
\end{align}
where $R_0 \sim 10^{13}\,\mathrm{cm}$ is the initial radius of the progenitor star for a Type II SN (smaller for Type I SNe), and $t_0$ is the explosion time. In either case, given the high initial expansion speeds of $\dot{R}_\text{ph} \sim 10^4 \, \mathrm{km/s}$, $R_0$ becomes negligible roughly a day after the explosion, before which the ejecta are still accelerated anyway. The photospheric expanion speed $\dot{R}_\mathrm{ph}$ is inferred indirectly by measuring Doppler-shifted spectral features produced by the ejecta, and then using radiative transfer simulations to map these ejecta velocities back to $\dot{R}_\mathrm{ph}$~\cite{1996ApJ...466..911E,  2005A&A...439..671D}. In addition, by measuring $\Theta_\text{ph}(\lambda)$ at a minimum of two different epochs $t$, one can infer $\dot{\Theta}_\text{ph}(\lambda)$ and thus the angular diameter distance:
\begin{align}
    D_A \simeq \frac{\dot{R}_\text{ph}(\lambda)}{(1+z) \dot\Theta_\text{ph}(\lambda)} \, .
\end{align}
The $(1+z)$ factor is necessary because in our convention, all physical speeds are defined in the SN frame at redshift $z$, while angular speeds are measured in the observer's frame at redshift $z=0$.
In what follows, we review and examine features of SNe during the photospheric phase, with a focus on those that have led to major limitations and systematic uncertainties that hamper the traditional EPM. We then follow it with a short discussion of EPM variants. 

\paragraph{Photospheric temperature:}  
The accurate determination of the photospheric temperature $T_\text{ph}$  relies on the assumption that the photosphere emits as a thermalized blackbody. However, this assumption is often violated, particularly for thermonuclear explosions like Type Ia SNe, where the opacity is dominated by complex atomic transitions rather than continuum processes. Consequently, SN spectra (especially in Type Ia) deviate significantly from that of a single-temperature blackbody even after excising spectral lines, with features shaped by line blanketing and scattering effects. In contrast, core-collapse SNe with massive envelopes exhibit conditions closer to local thermal equilibrium (LTE), making Type II SNe more suitable for such analyses. Even so, detailed spectral modeling is required to mitigate uncertainties in temperature assignment, as the photosphere's properties vary significantly with wavelength and depend on the underlying physical conditions.

\paragraph{Flux dilution factors:}  
Dilution factors $\zeta(\lambda')$ are introduced to account for deviations of the SN’s flux from a pure blackbody spectrum, addressing the effects of electron scattering and line blanketing in the atmosphere. Defined as:  
\begin{align}
\zeta(\lambda')^2 = \frac{L(\lambda')}{\pi B_{\lambda'}(T_\mathrm{ph}) 4\pi R_\text{ph}^2} \, ,
\end{align}  
they scale the observed luminosity $L(\lambda')$ to the blackbody luminosity at the photospheric temperature $T_\mathrm{ph}$. These factors depend on the wavelength band used to measure the SN’s color temperature and are derived from synthetic spectra generated by detailed SN atmosphere models, such as E96~\cite{1996ApJ...466..911E} and D05~\cite{2005A&A...439..671D}. While both suites exhibit the same general trends, they differ systematically: D05, with its more advanced treatment of non-LTE effects and a broader range of conditions, yields dilution factors about 15\% higher than E96, along with a greater dispersion (7\% versus 3\%). Even within the context of a \emph{single} simulation suite, EPM applications with different spectral filter combinations yield mutually inconsistent distances~\cite{2009ApJ...696.1176J}. These discrepancies emphasize the sensitivity of dilution factors to underlying model assumptions, including relativistic effects, opacity treatments, and the physical parameters of the ejecta (composition, ionization state, density), introducing systematic uncertainties into the inferred angular radius and distance. 

\paragraph{Photospheric velocity:}  
The SN ejecta undergo homologous expansion, where the velocity $\vect{v}_\mathrm{ej}$ of each mass element is proportional to its  radius $\vect{r}$, resulting in the outer layers expanding faster than the inner layers and allowing velocity to serve as a co-moving coordinate with the ejecta. As the SN evolves, the photosphere recedes inward in the velocity coordinate, corresponding to decreasing enclosed mass. This inward recession reflects the photosphere sampling progressively deeper and slower-moving layers rather than a deceleration of the ejecta.\footnote{The moniker ``expanding photosphere method'' is thus misleading. While the \emph{ejecta} emanate outward from the explosion site at high and nearly constant velocity for some period, the photospheric surface itself typically expands more slowly at early times, and could be quasi-static or even recede inwards in physical space at later stages of the photosphere phase.} Observationally, it manifests as $\dot{R}_\mathrm{ph} \lesssim |\vect{v}_\mathrm{ej}|$ and the gradual narrowing of P Cygni profiles over time. For Type IIP SNe, the evolution of photospheric velocity is notably uniform~\cite{2006ApJ...645..841N}. Expansion velocities are typically measured in de-redshifted spectra as the offset between the absorption minimum of P Cygni profiles and the rest wavelength of the corresponding spectral line, but different lines yield different velocity measurements because the effective photosphere for each element appears to be located at a unique velocity coordinate~\cite{2012MNRAS.419.2783T}. Lighter elements, such as hydrogen, dominate the outer layers of the progenitor and exhibit higher velocities, while heavier elements, such as iron, are concentrated in the inner layers and exhibit lower velocities. This variation in observed velocities introduces complexities for distance measurements, necessitating careful calibration of the chosen spectral lines against theoretical models to ensure consistent and reliable estimates~\cite{gutierrez2017type, bartel2007sn}.

\paragraph{Dust extinction:}
Extinction due to interstellar dust poses another major challenge for the EPM. While extinction in the Milky Way foreground can be reasonably well-corrected using existing maps~\cite{green2015three, schlegel1998maps, burstein1982reddenings, abergel2014planck}, the extinction within the host galaxy and the immediate environment of the progenitor star is often poorly constrained~\cite{duarte2023sample}. This is particularly problematic for core-collapse SNe, which frequently occur in dusty star-forming regions~\cite{quintana2025census}. Host-galaxy extinction $A'(\lambda')$ introduces wavelength-dependent uncertainties that can bias the derived flux and temperature, while circumstellar dust around the progenitor can create additional extinction effects~\cite{li2022dust}. Uncertainties in these extinction corrections directly affect the derived angular radius and distance estimates, which can be up to $20\%$~\cite{jones2009distance}. 

\paragraph{Spherical symmetry:}
The EPM assumes that the SN explosion is spherically symmetric, meaning the photosphere expands uniformly in all directions. However, spectropolarimetric observations have revealed significant asymmetries in many SNe~\cite{1991ApJ...375..264J,1996ApJ...459..307H, leonard2006non, vasylyev2024spectropolarimetry, patat2012vlt}, particularly core-collapse SNe, where deviations from spherical symmetry arise due to the complex dynamics of the explosion mechanism, such as jet formation or anisotropic energy deposition. Additionally, the ejecta are not smooth but exhibit clumping and other irregularities that can alter both the emitted and scattered radiation. These asymmetries introduce angular dependencies into the observed flux. For photospheric radius and expansion velocity inference based on a single event, these uncertainties can yield biased distance measurements. These uncertainties could be averaged down with a large sample of SNe,
taking advantage of the randomized orientation of the SN photosphere. However, a SN's apparent magnitude depends on its absolute magnitude as well as its orientation, which can result in systematic orientation biases coming from selection effects. We discuss some of these selection effects and biases in our companion paper~\cite{eem-2}.

\paragraph{Alternatives to the EPM:}  
To address the limitations of the EPM, several alternative techniques have been developed, each tailored to mitigate specific challenges. The spectral-fitting expanding atmosphere method (SEAM)~\cite{2004ApJ...616L..91B} replaces the simplistic blackbody assumption with detailed radiative transfer models, enabling more accurate distance estimates by fitting observed spectra directly to synthetic spectra generated from numerical simulations. The standard candle method (SCM)~\cite{2002ApJ...566L..63H} leverages empirical correlations between plateau luminosity and expansion velocity in Type IIP SNe, bypassing the need for precise temperature or dilution factor measurements while introducing its own reliance on calibration against well-studied nearby events. The photometric phase method (PPM)~\cite{2014AJ....148..107R} and photometric color method (PCM)~\cite{2015ApJ...815..121D} exploit the relatively uniform photometric properties of Type IIP SNe to determine distances without requiring direct velocity measurements. While some of these approaches offer improvements in precision and reduced reliance on theoretical models, they also introduce new dependencies, such as accurate spectral or photometric calibration, detailed knowledge of SN progenitor properties, and robust modeling of extinction effects. Nevertheless, they may serve as useful cross-checks, improving the reliability of cosmic distance measurements.

Heuristically, one can view the EEM as an extension of SEAM, supplemented by spatially resolved information delivered by intensity interferometry, as we will explore in the next subsection. While multiple spectral features can be used in principle, we demonstrate that the EEM is already robustly applicable to a single spectral line such as $\mathrm{H\alpha}$.

\begin{figure}
    \centering
    \includegraphics[width = 1\textwidth]{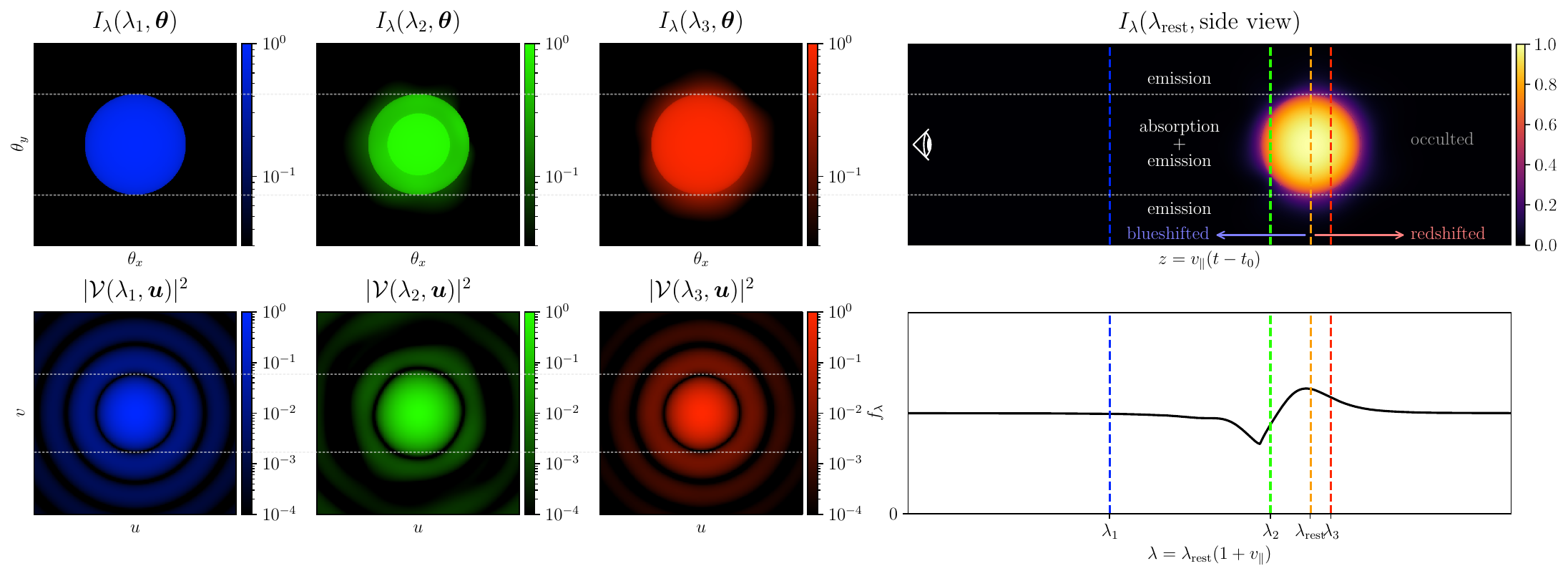}
    \caption{Diagram of a spherical SN explosion with limb darkening and spatially varying but homologously expanding ejecta containing a single spectral line.
    \textbf{Top right:} Spectral radiance $I_\lambda$ at the rest wavelength $\lambda_\mathrm{rest}$ of the spectral line (orange) viewed from the ``side'' with observer to the left. Homologous ejecta expansion implies that line-of-sight (LOS) slices along $z = v_\parallel (t - t_0)$ correspond to slices of equal LOS velocity $v_\parallel$ and thus observed wavelength $\lambda = (1+v_\parallel) \lambda_\mathrm{rest}$ of the spectral line: blueshifted to the left ($\lambda_1$, $\lambda_2$), redshifted to the right ($\lambda_3$). 
    \textbf{Top left:} Spectral radiance $I_\lambda(\lambda, \vect{\theta})$ at wavelengths $\lambda = \lbrace \lambda_1, \lambda_2, \lambda_3\rbrace$ as a function of angle $\vect{\theta} = (\theta_x, \theta_y)$ from the observer's vantage point. The left diagram is so far blueshifted to $\lambda_1$ that only the continuum emission from the photosphere is seen, with little absorption/emission from thin ejecta at these large LOS velocities. The middle diagram (green, $\lambda_2$) shows a partially unobscured photosphere, with significant absorption (and some emission) in an annulus with $|\vect{\theta}| < \Theta_\mathrm{ph}$, and only emission for $|\vect{\theta}| > \Theta_\mathrm{ph}$. The right diagram ($\lambda_3$, red) only shows redshifted emission as the absorbing parts of the ejecta are occulted by the photosphere. \textbf{Bottom left:} Squared visibilities $|\mathcal{V}(\lambda,\vect{u})|^2$ corresponding to the three spectral radiance maps directly above; see section~\ref{sec:IIobs} for definitions of $\mathcal{V}$ and $\vect{u}= (u,v)$. The morphology of the photosphere is captured at $\lambda_1$, while those of the ejecta are exhibited by the differences at $\lambda_2$ and $\lambda_3$. 
    \textbf{Bottom right:} Integrated flux density $f_\lambda$ (linear scale), showing the resulting P Cygni profile as a function of observed wavelength. \nblinkeem{plots/eem_2d_diagrams.ipynb}}
    \label{fig:abs}
\end{figure}

\subsection{Resolved supernova observables} \label{sec:resolved}

Spatially resolving a SN requires instruments with incredible resolution. 
The angular radius $\Theta_\mathrm{ph}$ of a SN photosphere is
\begin{align}
    \Theta_\mathrm{ph} = \frac{R_\mathrm{ph}}{D_A} \approx \underbrace{1.1 \times 10^{-10} \, \mathrm{rad}}_{22 \, \mathrm{\mu as}} \left(\frac{R_\mathrm{ph}}{10^{15} \, \mathrm{cm}}\right) \left(\frac{3\,\mathrm{Mpc}}{D_A} \right), \label{eq:Theta_ph}
\end{align}
where we assumed a typical physical photosphere radius of $R_\mathrm{ph} = 10^{15} \, \mathrm{cm}$ for the numerical estimate, while anticipating dozens of extra-galactic SNe with a line-of-sight (angular diameter) distance of $D_A = 3 \,\mathrm{Mpc}$ to be observed in the next decades.
Even these ``nearby'' events are beyond the angular resolution of individual optical telescopes or amplitude interferometers in the radio or infrared/optical bands~\cite{1642263,EventHorizonTelescope:2019dse,EventHorizonTelescope:2019uob,2005ApJ...628..453T}. However, with optical intensity interferometry consisting of optical telescopes spatially separated by baselines $d$, an angular resolution of
\begin{equation}
    \sigma_{\theta_\mathrm{res}} \simeq \frac{\lambda}{d} \simeq \underbrace{3.28 \times 10^{-10} \, \mathrm{rad}}_{67.7\,\mathrm{\mu as}} \left(\frac{\lambda}{656\,\mathrm{nm}} \right) \left(\frac{2\,\mathrm{km}}{d} \right) \label{eq:resolution}
\end{equation}
can in principle be reached or surpassed with even larger baselines, since the optical light need not be physically recombined as offline correlation is possible. We shall see in section~\ref{sec:IIobs} that the optimal operating resolution for inferring the angular size of the photosphere is such that $\lambda/d \approx 3 \Theta_\mathrm{ph}$ --- so that $d=2\,\mathrm{km}$ is optimal on the H$\alpha$ line for a SN with angular radius $\Theta_\mathrm{ph}$ as in eq.~\eqref{eq:Theta_ph}. The main limitation of intensity interferometry for this purpose is obtaining sufficient signal-to-noise ratio on the interference fringes; this issue is being ameliorated with spectral multiplexing and ultrafast, high-efficiency photon detectors. We leave a review on the capabilities of optical intensity interferometers to the later subsection~\ref{sec:IIobs}, and will first discuss how the EPM can be improved to what we call the expanding ejecta method (EEM).

The spectro-spatial information of the SN luminosity is encapsulated by the spectral radiance, the observed energy per unit time per unit area per unit solid angle per unit wavelength, defined as:
\begin{equation}
    I_{\lambda}(\lambda, \vect{\theta})  \equiv \frac{\dd E}{\dd t  \, \dd A \, \dd^2\vect{\theta} \, \dd \lambda  } (\lambda,\vect{\theta}) \, ,\label{eq:I_lambda_observed}
\end{equation}
where $\vect{\theta}$ marks the angular position on the sky. 
The observed $I_\lambda$ is related to the spectral radiance $I'_{\lambda'}$ in the SN rest frame at redshift $z$ as: 
\begin{align}
    I_{\lambda}(\lambda,\vect{\theta}) = \frac{1}{(1+z)^5} I'_{\lambda'}(\lambda',\vect{\theta},s=0) = \frac{1}{(1+z)^5} 
 \frac{\dd E'}{\dd t'  \, \dd A' \, \dd^2\vect{\theta} \, \dd \lambda'}(\lambda', \vect{\theta},s=0) \, ,
\end{align}
where $s$ is the line-of-sight distance, and ``primed'' quantities are defined in the SN rest frame, e.g.~$\lambda = \lambda'(1+z)$, etc.
In what follows, we will ignore redshift factors for notational clarity and the distinction between primed and unprimed quantities (since these are typically small and easily restored), and simply refer to  $I_\lambda(\lambda,\vect{\theta},s)$, with the understanding that we are interested in the $s \to 0$ limit.
To compute the spectral radiance in full generality, one needs to solve the equation of radiative transfer~\cite{leblanc2010introduction,Rybicki1996}:
\begin{equation}
  \frac{\dd I_{\lambda}}{\dd s} = - \alpha_{\lambda} I_{\lambda} + j_{\lambda} \, , \label{eq:RTE}
\end{equation}
where $\alpha_{\lambda}$ and $j_{\lambda}$ are the absorption and emission coefficients that depend on both wavelength and 3D position. SNe and stellar objects alike generally consist of an optically thick inner region surrounded by an optically thin atmosphere. The optically thick region acts as a source for (pseudo-)continuum emission, whereas sharp and broad spectral features emerge as photons traverse through the optically thin atmosphere~\cite{mihalas1978stellar}.

In the traditional EPM, the continuum emission is assumed to follow a blackbody spectrum, i.e. $I_{\lambda} = B_\lambda(T_{\rm ph})$ at the SN photosphere, which has a radius $R_{\rm ph}$ and temperature $ T_{\rm ph}$ that evolve over time. As discussed in section~\ref{sec:unresolved}, these assumptions introduce uncertainties and biases that limit the method’s precision. A spatially resolved observation of a SN, however, provides significantly more information about the morphology of the optically thick region and the surrounding ejecta. In section~\ref{sec:parametric}, we employ a simple parametric model to extract key SN properties --- such as angular radius, asphericity, orientation, ejecta density, and spectral index --- at moderate signal-to-noise ratio (SNR). Here, we focus on qualitatively demonstrating how various SN properties can, \emph{in principle}, be inferred from high-SNR spatially resolved measurements without relying on a specific model, thereby motivating future studies with more sophisticated approaches than those presented in section~\ref{sec:parametric}.

For this purpose, we describe an idealized observation of a SN with a complex geometry, progressively incorporating the relevant physical effects that modify the spatially resolved observables of the (pseudo-)continuum and spectral lines. While actual observations will ultimately be performed in Fourier space, as discussed in section~\ref{sec:IIobs}, the following descriptions are presented in real (angular) space to aid in developing intuition.

\paragraph{Photosphere without surrounding ejecta:} 
The simplest case is a photosphere \emph{without} surrounding ejecta material. This corresponds to a scenario where photon absorption $\alpha_\lambda$ and emission $j_\lambda$ sharply drop to zero beyond a thin photospheric boundary. 

\begin{figure}
    \centering
    \includegraphics[width=0.48\linewidth]{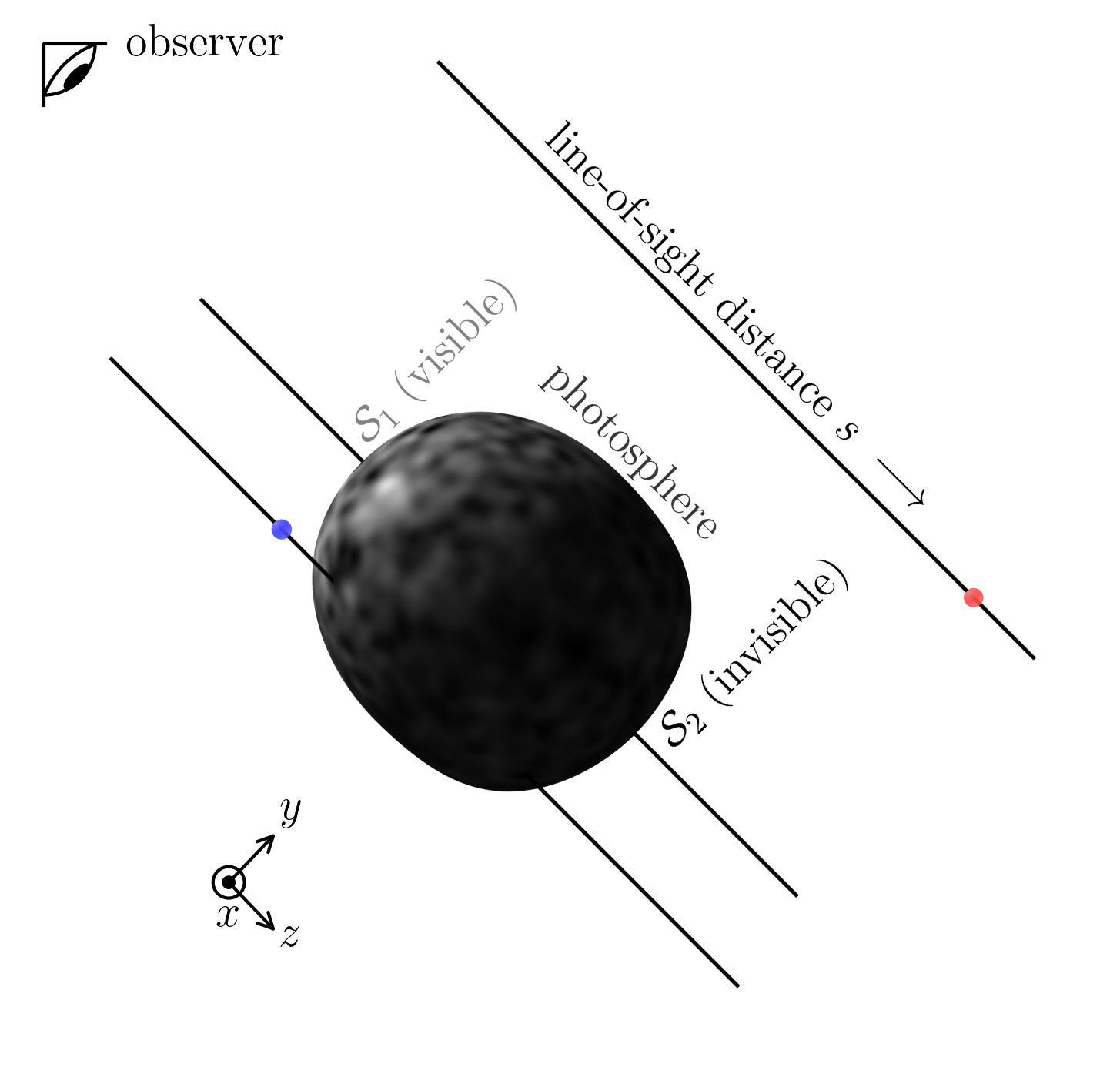}
    \includegraphics[width=0.48\linewidth]
    {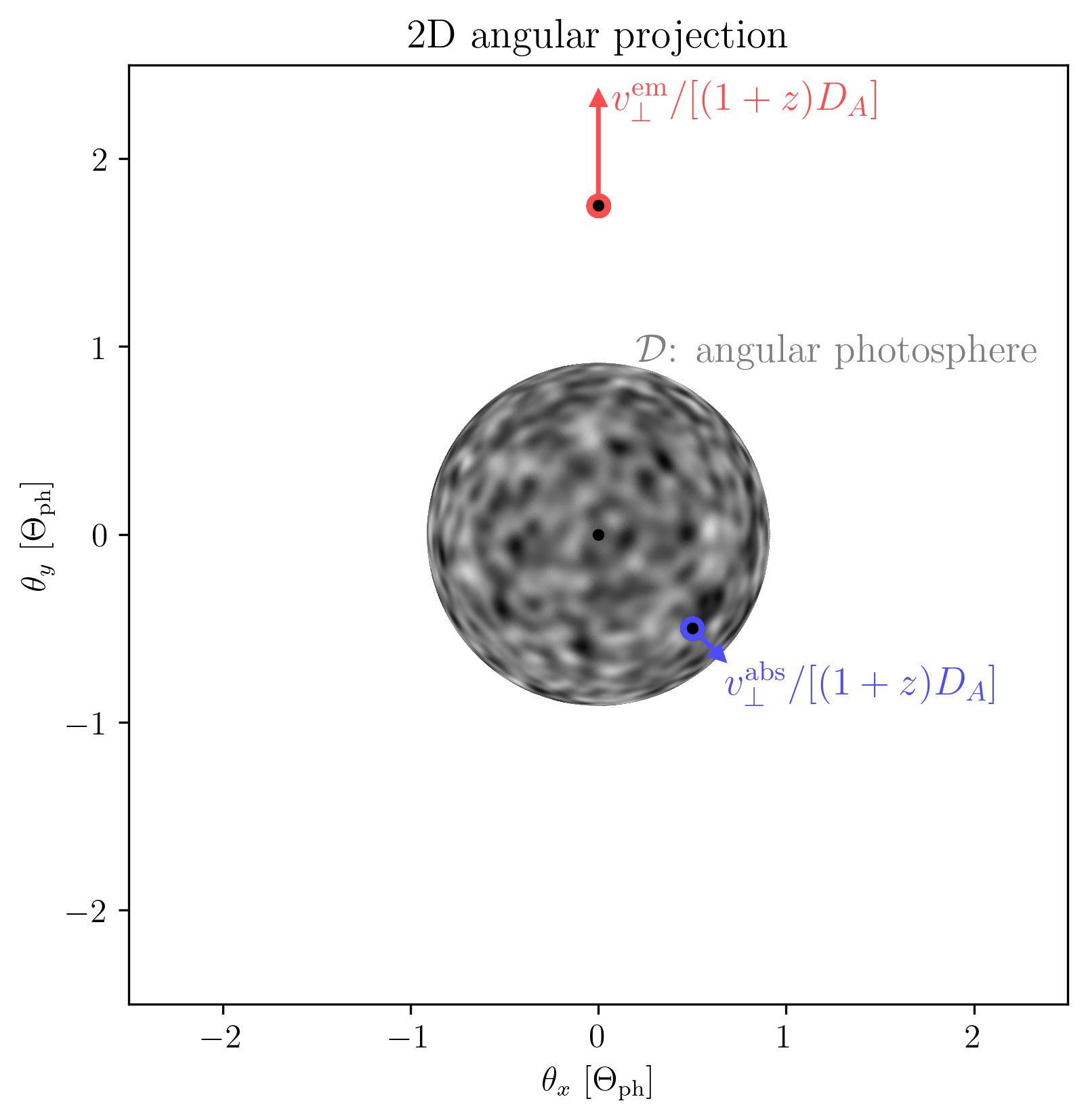}
    \includegraphics[width=0.48\linewidth]{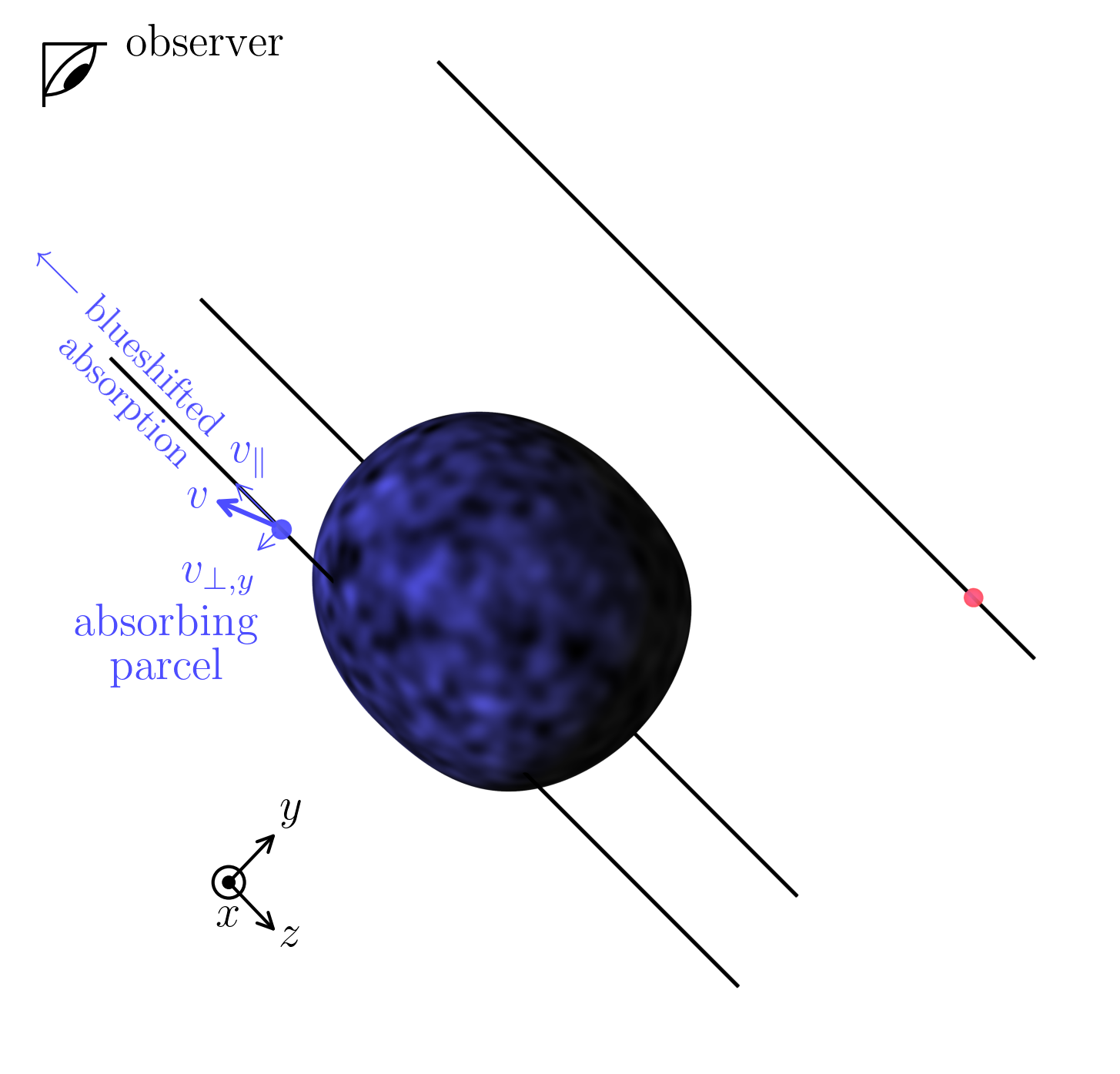}
    \includegraphics[width=0.48\linewidth]{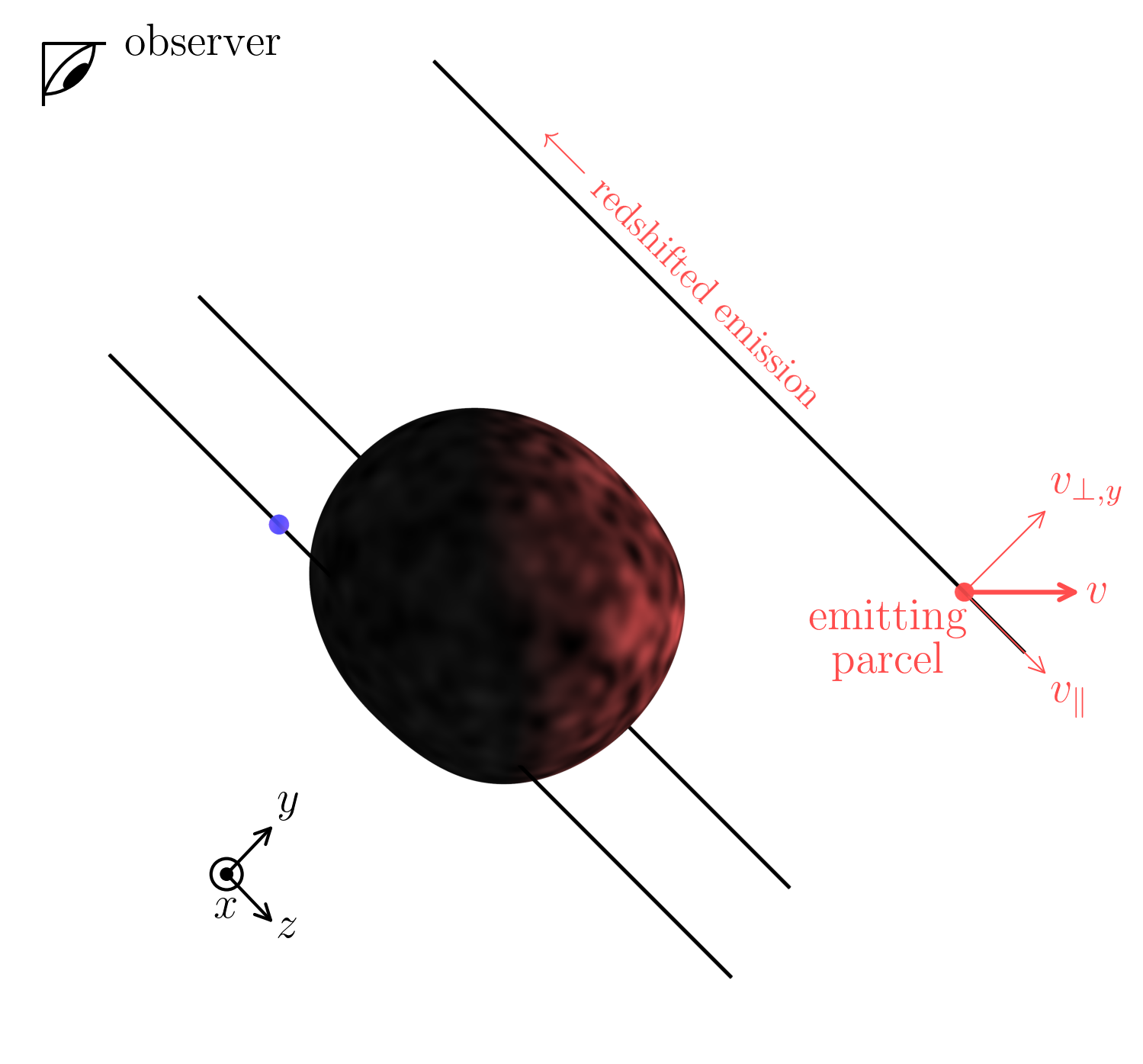}
    \caption{
    \textbf{Top left:} Diagram of general 3D SN geometry. The (convex) photosphere emits continuum emission (grayscale proportional to brightness temperature), and is made up of two hemispherical regions: the surface $\mathcal{S}_1$ visible to the observer (located at $x=y=0$ and $z\to -\infty$), and the occulted surface $\mathcal{S}_2$.
    \textbf{Bottom left:} Part of the photospheric emission (from the blue-lighted portion of the photosphere) is absorbed (some of it also re-emitted) by a parcel of ejecta with negative line-of-sight velocity $v_\parallel^\mathrm{abs}$, reducing the flux density near observed wavelength $\lambda = \lambda_\mathrm{rest} (1+v_\parallel^\mathrm{abs})$, blueshifted from the rest wavelength $\lambda_\mathrm{rest}$.
    \textbf{Bottom right:} A parcel receding from the observer is partially illuminated by the photosphere's ``backside'' $\mathcal{S}_2$ (red-lighted portion of the photosphere) and re-emits radiation through resonant fluorescence, redshifted from the spectral line's wavelength.
    \textbf{Top right:} 2D angular projection of the emission, including the angular region $\mathcal{D}$ ($\mathcal{S}_1$ in projection). The observable properties of the ejected parcels are their angular velocities $\vect{v}_\perp / [(1+z)D_A]$ via intensity interferometry, and their physical line-of-sight velocities $v_\parallel$ via the spectrum. Full 3D shape information is indirectly encoded by the \emph{strength} of the ejecta's emission and absorption of the radiation from the photosphere. \nblinkeem{plots/eem_3d_diagrams.ipynb}
    }    
    \label{fig:photosphere_only}
\end{figure}

We define the photosphere as the set of intersection points $S(\vect{\theta}) = \lbrace s_i(\vect{\theta}) \rbrace$ where each line of sight $\vect{\theta}$ crosses the photospheric surface, ordered in increasing distance from the observer, i.e.~$s_1(\vect{\theta}) < s_2(\vect{\theta}) < \dots$, as illustrated in figure~\ref{fig:photosphere_only}. The set $S(\vect{\theta}) = \emptyset$ is empty if the line of sight misses the photosphere completely, consists of a single point ($|S(\vect{\theta})| = 1$) when grazing the limb at exactly one distance $s_1(\vect{\theta})$, and contains two or more points ($|S(\vect{\theta})| \geq 2$) for lines of sight that pass through the photosphere. If the photosphere is convex, then at most two intersection points exist ($|S(\vect{\theta})| \leq 2$). 
The \emph{visible} part of the photosphere is the 2D projection onto the observer's sky, defined as the region 
\begin{align}
\mathcal{D} \equiv \big \lbrace \vect{\theta} \big| |S(\vect{\theta})| \geq 1 \big\rbrace \, . \label{eq:D_photosphere}
\end{align}
Over this angular region, the ``exposed'' photospheric surface is given by 
\begin{align}
\mathcal{S}_1 = \lbrace \vect{\theta}, s_1(\vect{\theta}) \rbrace \, ,\label{eq:S_1_photosphere}
\end{align}
which remains continuous for a convex photosphere, and is represented by the ``illuminated'' area in the top left panel of figure~\ref{fig:photosphere_only}.

An intensity interferometer can \emph{in principle} map out the spectral radiance $I_\lambda\big[\lambda, \vect{\theta},s_1(\vect{\theta})\big]$ at each point on the exposed surface $\mathcal{S}_1$, and do so at every observation epoch. The locus of the exposed photosphere $\mathcal{S}_1$ itself may depend strongly on wavelength $\lambda$ near spectral lines, e.g.~in a Type Ia SN, though we omit explicit $\lambda$ subscripts for simplicity. However, no information can be gained about the line-of-sight distance $s_1(\vect{\theta})$ at any point on $\mathcal{S}_1$—nor about the ``invisible'' boundaries $s_2(\vect{\theta}), s_3(\vect{\theta}), \dots$—without additional assumptions or external constraints.
If the photosphere emits as a blackbody with temperature $T(\vect{\theta}) = T\big[\vect{\theta},s_1(\vect{\theta})\big]$ on $\mathcal{S}_1$, then the spectral radiance is further restricted to
\begin{align}
I_\lambda\big[\lambda, \vect{\theta},s_1(\vect{\theta})\big] = B_\lambda\left[T(\vect{\theta})\right],
\end{align}
where $B_\lambda[T]$ is the blackbody spectral radiance function. In the special case of a uniform photosphere temperature $T(\vect{\theta}) = T_\mathrm{ph}$, the only measurable quantities are the angular extent of $\mathcal{D}$ and the temperature $T_\mathrm{ph}$.
In all cases above, only transverse asymmetry in the image plane can be measured. No information about asymmetry along the line of sight can be extracted.

\paragraph{Atmospheric effects and the role of spectral lines:}  
In realistic SN environments, the transition from the optically thick photosphere to the surrounding ejecta is governed by a gradual variation in the absorption and emission coefficients, $\alpha_\lambda$ and $j_\lambda$. Consequently, the spectral radiance $I_\lambda(\lambda,\vect{\theta})$ is never purely thermal --- even if the underlying photospheric emission approximates a blackbody --- since local radiative transfer continuously alters the emitted light. In the case of pure Thomson scattering, where radiative transfer is wavelength independent and $\alpha_\lambda$ and $j_\lambda$ are directly related~\cite{Rybicki1996}, well-known effects such as limb darkening and relativistic beaming arise, leading to a brighter central region compared to the edges, as illustrated in figure~\ref{fig:abs}. However, when spectral lines play a significant role, these coefficients vary strongly with wavelength, and the situation becomes considerably more complex.

In systems like SNe~Ia, where the ejecta are composed predominantly of heavy elements and subject to large velocity gradients, the impact of spectral line opacity is particularly dramatic~\cite{2000ApJ...530..757P}. The opacity can be extremely high in spectral regions dominated by strong lines, which can lead to an apparent photospheric radius larger than that inferred from the continuum alone. Radiation absorbed in these lines is very likely to undergo repeated fluorescence cascades, thereby redistributing energy downward in frequency. This efficient downward transport affects the radiative diffusion time and gives rise to an emergent spectrum that may appear thermal without being directly tied to the local gas temperature.

The complex interplay of these effects complicates the notion of a well-defined continuum. Instead of a clear-cut separation between continuum and line contributions, the observed $I_\lambda(\lambda,\vect{\theta})$ should be viewed as a baseline continuum that is strongly modulated by spectral line opacities. Detailed radiative transfer studies such as those of ref.~\cite{2000ApJ...530..757P} are needed to produce high-quality forward models of the emission profile, especially for Type Ia SNe. These insights underscore the challenges and opportunities in augmenting SN spectra with spatially resolved observables, to extract the maximum amount of information out of measuring $I_\lambda(\lambda,\vect{\theta})$ over a broad wavelength range and simultaneously over the entire image plane.

\paragraph{Resonant absorption:}
In the original expanding photosphere method (EPM), characteristic spectral lines are used to infer ejecta velocities from the blueshifted absorption component of the P‑Cygni profile. With angularly resolved observations from intensity interferometry, however, one can extract far more detailed information about the spatial and velocity distribution of the scattering species. The blueshifted absorption modifies the continuum spectral radiance $I_\lambda^{\rm cont}$, the spectral radiance of the photospheric continuum emission as seen without any absorbing ejecta material, according to 
\begin{equation}
    I_\lambda^{\rm abs}(\lambda,\vect{\theta}) = \exp\left[-\tau\big(\lambda,\lambda^i_{\rm rest},\vect{\theta}\big)\right] I_\lambda^{\rm cont}(\lambda,\vect{\theta}),
    \label{eq:absorption}
\end{equation}
where the optical depth $\tau(\lambda,\lambda^i_{\rm rest},\vect{\theta}) \simeq \int \dd s \, \alpha_\lambda(s)$ depends on the density and velocity distribution of the ejecta. For narrow lines, the observed wavelength $\lambda$ and the rest wavelength  $\lambda_\mathrm{rest}^i$ of line $i$ are related via
\begin{equation}\label{eq:wavelength_relation}
    \lambda = \lambda^i_\mathrm{rest} \left( 1 + v^i_\mathrm{ej,\parallel} \right) \, ,
\end{equation}
with $v^i_{\rm ej,\parallel}$ denoting the line-of-sight component of the ejecta velocity. For the remainder of this work, we will mainly consider a \emph{single} spectral line with rest wavelength $\lambda_\mathrm{rest}$ and line-of-sight component of the ejecta velocity $v_{\rm ej,\parallel}$, dropping the $i$ superscripts.

Assuming homologous expansion,  
\begin{equation}
    z = v_{\rm ej,\parallel} (t-t_0) \, , \label{eq:homologous_LOS}
\end{equation}
where $z$ is the displacement along the line of sight and $t-t_0$ is the time since explosion, observing at a particular $\lambda$ effectively selects ejecta with a specific line-of-sight velocity. When combined with the transverse spatial resolution provided by interferometry, this yields a five-dimensional mapping: two spatial dimensions, two angular velocity components, and the line-of-sight velocity of the absorbing species in front of the photosphere. Full 6D phase determination is not generally possible without ``collapsing'' the relative LOS distance and velocity per eq.~\eqref{eq:homologous_LOS}, but crucially this homologous expansion assumption can be tested in the two transverse dimensions for consistency.

This approach is feasible as long as the continuum $I_\lambda^{\rm cont}(\lambda,\vect{\theta})$ is smooth in the vicinity of the sharp spectral feature, and could be applied even to later stages of evolution or to Type Ia SNe.

\paragraph{Resonant re-emission:}

The second component of the P‑Cygni profile is formed by resonant re-emission. In this case, both blueshifted and redshifted emission arise from material outside the projected photosphere. For instance, blueshifted emission originates from ejecta material on the near side of the ejecta (bottom left panel in figure~\ref{fig:photosphere_only}), while redshifted emission comes from the far side (bottom right panel in figure~\ref{fig:photosphere_only}). In absence of radiation re-absorption, the spectral radiance of emission can be expressed as
\begin{equation}
    I_\lambda^\mathrm{em}(\lambda,\vect{\theta}) = \int_0^{s_\mathrm{max}(\vect{\theta})} \dd s \, j_{\lambda}(s) \, ,
\end{equation}
where $s_\mathrm{max} = s_1(\vect{\theta})$ for $\vect{\theta} \in \mathcal{D}$ and $s_\mathrm{max} = +\infty$ otherwise, cfr.~discussion around eqs.~\eqref{eq:D_photosphere} and~\eqref{eq:S_1_photosphere}.
No closed-form solutions exist when both emission and absorption are present. A measurement of the emission spectral radiance
provides independent access to the five-dimensional distribution of the element $i$ around the photosphere. Under the assumption of homologous expansion and when considering an isolated spectral line, the emitted photon is unlikely to be re-absorbed. In more complex scenarios --- where multiple spectral lines overlap or the assumption of homologous expansion breaks down --- a more detailed radiative transfer treatment will be necessary to determine the impact of potential re-absorption on the inferred ejecta morphology.

\paragraph{Combined profile and implications for EEM:}

The sum of resonant absorption and re-emission produces the full P‑Cygni profile,  
\begin{equation}
\label{eq:em_abs}
    I_\lambda(\lambda,\vect{\theta}) = I_\lambda^\mathrm{em}(\lambda,\vect{\theta}) + I_\lambda^{\rm abs}(\lambda,\vect{\theta}) \, ,
\end{equation}
which encodes rich spatial and spectral information about the ejecta. In an idealized, spherically symmetric SN in local thermodynamic equilibrium (LTE), one would expect the emission peak to be centered at zero velocity and the absorption trough to occur near the photospheric velocity, with equal total strengths. However, observations—especially in hydrogen-rich SNe—often reveal a blueshifted emission component, partly due to notable recombination emission. Spatially resolved measurements via intensity interferometry enable us not only to isolate these deviations but also to reconstruct the full five-dimensional distribution $n_i\big(\{\vect{\theta}\},\{\vect{\dot\theta},v_{\rm ej,\parallel}\}\big)$ of the ejecta material. By directly utilizing this detailed mapping, we can bypass the simplifying assumptions of spherical symmetry, pure blackbody emission, and a fixed relation between ejecta and photospheric velocities inherent to the original EPM. This comprehensive information forms the basis of the expanding ejecta method (EEM), which aims to measure distances and infer explosion dynamics directly from the spectro-spatial structure of the ejecta.

\subsection{Intensity interferometer observables}\label{sec:IIobs}

Suppose one observes, in each spectral channel centered on wavelength $\overline{\lambda}$, an intensity of the form:
\begin{align}
    I({\overline{\lambda}},\vect{\theta}) = \epsilon \int \dd \lambda \, \exp\left[\frac{-(\lambda-\overline{\lambda})^2}{2 \sigma_\lambda^2}\right] I_\lambda(\lambda,\vect{\theta}) \simeq \epsilon \sqrt{2\pi} \sigma_\lambda \, I_\lambda(\overline{\lambda},\vect{\theta}) \, ,
\end{align}
where $\epsilon$ is the overall efficiency, accounting for imperfect photodetection and losses in the optical path. 
The Gaussian factor models the effect of a high-resolution spectrograph with spectral resolution $\mathcal{R} = \overline{\lambda}/\sigma_\lambda \gg 1$, selecting only photons within some narrow band $\sigma_\lambda$ in each pixel (channel). The spectral filter is assumed to be narrower than the wavelength variation of the intrinsic spectral radiance $I_\lambda$.

The total expected number of photons observed in each spectral channel is:
\begin{align}
    N(\overline{\lambda}) 
    = A t_\mathrm{obs} \frac{\overline{\lambda}}{2\pi} \int \dd^2 \vect{\theta} \, I(\overline{\lambda},\vect{\theta})  
    = \epsilon A t_\mathrm{obs} \frac{\overline{\lambda}^2}{\sqrt{2\pi}\mathcal{R}} \int \dd^2 \vect{\theta} \,  I_\lambda(\overline{\lambda},\vect{\theta})  \, , \label{eq:photon_number}
\end{align}
where $A$ is the aperture area, and $t_\mathrm{obs}$ is the observation time.

The excess fractional intensity correlation in the spectral channel centered at $\overline{\lambda}$ for a pair of telescopes 1 and 2, separated by a baseline $\vect{d}$ and with relative timing resolution $\sigma_t$, is given by:
\begin{align}
    C(\overline{\lambda},\vect{d}) \equiv \left. \frac{\langle I^{(1)} I^{(2)} \rangle}{\langle I^{(1)}\rangle \langle I^{(2)} \rangle} \right|_{\overline{\lambda}} - 1 \simeq \left(1 + \frac{8\pi^2 \sigma_t^2}{\overline{\lambda}^2 \mathcal{R}^2} \right)^{-1/2} \left| \mathcal{V}\left(\overline{\lambda},\frac{2\pi}{\overline{\lambda}} \vect{d} \right) \right|^2 \, , \label{eq:C}
\end{align}
proportional to the square modulus of the visibility $\mathcal{V}$~\cite{VanTilburg:2023tkl,Galanis:2023gef}:
\begin{align}
    \mathcal{V}\left(\overline{\lambda}, \vect{u} \right) \equiv \frac{\int \dd^2 \vect{\theta} \, I_\lambda\left(\overline{\lambda},\vect{\theta}\right) e^{-i \vect{u} \cdot \vect{\theta}}}{\int \dd^2 \vect{\theta} \, I_\lambda\left(\overline{\lambda},\vect{\theta}\right)} \, . \label{eq:visibility_def}
\end{align}
We see from eq.~\eqref{eq:C} that an intensity interferometer with baseline $\vect{d}$ measures the magnitude of the spectral radiance Fourier transform at angular wavenumber $\vect{u} = (2\pi/{\overline{\lambda}}) \vect{d}_\perp = \hat{\vect{d}_\perp} \, 2\pi/\sigma_{\theta_\mathrm{res}} $, cfr.~eq.~\eqref{eq:resolution}, in each spectral channel centered on $\overline{\lambda}$. The vector $\vect{d}_\perp$ is the perpendicular component of the baseline $\vect{d}$ relative to the LOS direction (i.e.~$\vect{d}_\perp \cdot \vect{\theta} = 0$), with $\hat{\vect{d}}_\perp$ its unit-vector counterpart.
The approximation in eq.~\eqref{eq:C} is valid in the limit that the angular extent of the image is not large relative to the angular dynamic range, which will generally be true for the cases of interest, since the image will be contained in the range $|\vect{\theta}| \lesssim \sigma_{\theta_\mathrm{res}} \mathcal{R}$.
An optimal estimator for $C(\overline{\lambda},\vect{d})$ has a variance~\cite{guerin2025stellar}:
\begin{align}
    \sigma_{C}^2 \simeq \frac{t_\mathrm{obs}}{\sqrt{128\pi}\sigma_t} \frac{1}{N^{(1)}(\overline{\lambda})N^{(2)}(\overline{\lambda})} \, ,
    \label{eq:uncertainty}
\end{align}
with $N^{(j)}(\overline{\lambda})$ the number of photons detected in each telescope $j$ as in eq.~\eqref{eq:photon_number}. Eq.~\eqref{eq:uncertainty} holds in the limit of $C \ll 1$, which will typically hold due to the small prefactor of eq.~\eqref{eq:C}.

It is difficult to reconstruct a fiducial image from the intensity correlation in eq.~\eqref{eq:C} because the phases of the visibilities are not measured in intensity interferometry and because the coverage in the $\vect{u} = (u,v) = ({2\pi}/{\overline{\lambda}})\vect{d}_\perp$ plane is sparse; see refs.~\cite{holmes2010two,holmes2013cramer,2012MNRAS.419..172N,2012MNRAS.424.1006N,dolne2014cramer} for work in this direction. However, given a \emph{parametric model} for the spectral radiance $I_\lambda(\overline{\lambda},\vect{\theta})$ and sufficient SNR, the underlying parameters of the model can be estimated precisely. The goodness-of-fit statistics of the parametric model and the model-independent image reconstruction techniques can serve as useful cross-checks.

The simplest nontrivial example is that of a disk with uniform (constant) spectral radiance $I_{\lambda,0}$ of angular radius $\Theta_\mathrm{ph}$, for which 
\begin{align}
    I_\lambda^\mathrm{UD}(\overline{\lambda},\vect{\theta})
    = \begin{cases}
        I_{\lambda,0} & |\vect{\theta}| < \Theta_\mathrm{ph} \\
        0 & \text{otherwise}
    \end{cases}
    \qquad 
    C^\mathrm{UD} \simeq \frac{\overline{\lambda} \mathcal{R}}{2\sqrt{2}\pi \sigma_t} \left|\frac{2 J_1(x)}{x} \right|^2; \quad x \equiv \frac{2\pi}{\overline{\lambda}} d \Theta_\mathrm{ph} \equiv 2\pi \frac{\Theta_\mathrm{ph}}{\sigma_{\theta_\mathrm{res}}} \, , \label{eq:UD}
\end{align}
expressed in terms of the Bessel function $J_1$ of the first kind, the angular resolution $\sigma_{\theta_\mathrm{res}}$, and $x$, the degree to which the disk is ``over-resolved''.
In this simple one-parameter model, the angular radius parameter $\Theta_\mathrm{ph}$ can be estimated at a precision (per spectral channel) of:
\begin{align}
    \sigma_{\Theta_\mathrm{ph}} = \sigma_{C} \left|\frac{\partial C}{\partial \Theta_\mathrm{ph}} \right|^{-1} 
    = \frac{\sigma_{\theta_\mathrm{res}}}{\mathrm{SNR}_{C}(0)} \left| \frac{x^2}{8 J_1(x) J_2(x)} \right|,
\end{align}
where we have defined $\mathrm{SNR}_{C}(0) \equiv \overline{\lambda} \mathcal{R} / \left[{2\sqrt{2}\pi \sigma_t} \sigma_C\right]$ as the signal-to-noise ratio of the correlation at vanishing baseline $d$.
The fractional precision to which $\Theta_\mathrm{ph}$ can be estimated is then finally:
\begin{align}
    \frac{\sigma_{\Theta_\mathrm{ph}}}{\Theta_\mathrm{ph}} = \frac{1}{\mathrm{SNR}_{C}(0)} \left| \frac{x}{8 J_1(x) J_2(x)}\right| \underset{(x \approx 2.02)}{\approx} \frac{1.23}{\mathrm{SNR}_{C}(0)} \, .
\end{align}
The second equality is at the minimum of the LHS, when one barely resolves the disk at $x \sim 2$, which can be achieved by adjusting/choosing the baseline distance appropriately. At this optimum value, the fractional precision on $\Theta_\mathrm{ph}$ is just the inverse SNR on the zero-baseline intensity correlation (times $1.23$). Over many spectral channels at different $\overline{\lambda}$, the precision is improved by combining the respective uncertainties in inverse quadrature
\begin{align}
    \sigma_{\Theta_\mathrm{ph}} = \left[\sum_{\overline{\lambda}} \sigma_{\Theta_\mathrm{ph},\overline{\lambda}}^{-2} \right]^{-1/2},
\end{align}
which roughly improves the precision as $1/\sqrt{\mathcal{R}}$, i.e.~the inverse square root of the number of independent spectral channels.

Deviations from azimuthal symmetry can be readily detected as well. Suppose eq.~\eqref{eq:UD} were modified to 
\begin{align}
I_\lambda(\overline{\lambda},\vect{\theta})
    = \begin{cases}
        I_{\lambda,0} & |\vect{\theta}| < \Theta_{\mathrm{ph},0} \left[1 + \sum_{m=1}^\infty \eta_m \cos[m(\varphi - \varphi_0) \right] \\
        0 & \text{otherwise} 
    \end{cases},
    \qquad \vect{\theta} \equiv |\vect{\theta}|( \cos \varphi, \sin \varphi) \, ,
\end{align}
with small $0 < \eta_m \ll 1$ deformations of the photosphere boundary with azimuthal phase $\varphi_0$ for each multipole $m = 1, 2, \dots$. For such a deformed photospheric boundary, we find the complex visibility
\begin{align}\label{eq:complexvisibility}
    \mathcal{V}(\overline{\lambda},\vect{u}) \simeq \frac{2 J_1(|\vect{u}| \Theta_\mathrm{ph})}{|\vect{u}| \Theta_\mathrm{ph}} - \sum_{m=1} \eta_m 2 i^{m} J_m(|\vect{u}| \Theta_\mathrm{ph}) \cos\left[m(\varphi_m - \varphi_u) \right] \, , \qquad \vect{u} \equiv |\vect{u}|( \cos \varphi_u,  \sin \varphi_u) \, ,
\end{align}
to leading order in the $\eta_m$, generalizing the zeroth-order result from eq.~\eqref{eq:UD}. There is no first-order signal from the odd-$m$ multipole components $\eta_m$ --- they drop out in the visibility modulus $|\mathcal{V}|$. The even-$m$ components $\eta_m$ can be inferred from their angular dependence $\propto \cos [m(\varphi_m - \varphi_u)]$ as the baseline $\vect{d}$ (and thus $\vect{u} = 2\pi \vect{d}_\perp / \overline{\lambda}$) sweeps the angle $\varphi_u$, with the optimal baseline magnitude still of order $d_\perp \sim \overline{\lambda} / \Theta_\mathrm{ph}$ as before.

In section~\ref{sec:parametric}, we introduce a more complicated parametric model of a SN explosion with ejecta, going beyond these simple projected photospheres of uniform spectral radiance. We will use that more sophisticated model to perform parameter estimation on mock data from intensity interferometric observations in section~\ref{sec:parameter-estimation}.

\section{Parametric Model} \label{sec:parametric}
We present a parametric model to describe how a SN's  photosphere and expanding ejecta produce P Cygni line profiles and how these features appear in observations. 
We model the SN's structure as follows:
\begin{enumerate}
    \item A photosphere with radius $R_\text{ph}$, radiating at effective temperature $T_\text{ph}$ around the spectral line, with deviations from ideal blackbody emission parametrized by an overall normalization $\mathcal{N}$;
    \item Surrounding ejecta with a velocity field $\vect{v}(\vect{r})$;
    \item Line-producing atoms in the ejecta with number densities $n_l(\vect{r})$ and $n_u(\vect{r})$ for lower and upper states, respectively.
\end{enumerate}
Table~\ref{tab:physical_parameter} summarizes the physical parameters used in the model. We adopt simplifying assumptions to isolate the parameters that help break degeneracies between the SN’s intrinsic properties and its angular diameter distance.
The remainder of this section focuses on the influence of expanding ejecta on a single spectral line (e.g.~H$\alpha$), assuming the continuum is modeled and measured independently at line-free wavelengths (see blue panel of figure~\ref{fig:abs}).

 We begin with a spherically symmetric model, connecting our parameterization to the physical descriptions of the photosphere and ejecta in sections~\ref{sec:photosphere} and~\ref{sec:ejecta}. This model illustrates how a SN's image and intensity correlation vary with wavelength near a spectral feature, capturing the effects of blueshifted absorption and both blueshifted and redshifted emission.
Next, in section~\ref{sec:asymmetry}, we extend the model to include a single prolate or oblate axis, forming an ellipsoidal geometry. This allows us to demonstrate how spectral and intensity correlation data can constrain the SN's shape. For both the spherical and ellipsoidal cases, we present model-generated images, spectra, and intensity correlations.

We emphasize that the model presented here is overly simplified, and is introduced to demonstrate the power of intensity correlation measurements. The framework is adaptable and can incorporate more complex treatments of P Cygni profiles, including emission-dominated features and SN Ia line structures.

\begin{figure}
    \centering

\tikzset {_dyjpfu69k/.code = {\pgfsetadditionalshadetransform{ \pgftransformshift{\pgfpoint{0 bp } { 0 bp }  }  \pgftransformscale{1 }  }}}
\pgfdeclareradialshading{_x478jl5ln}{\pgfpoint{0bp}{0bp}}{rgb(0bp)=(0.96,0.65,0.14);
rgb(16.339285714285715bp)=(0.96,0.65,0.14);
rgb(25bp)=(1,1,1);
rgb(400bp)=(1,1,1)}
\tikzset{every picture/.style={line width=0.75pt}} %set default line width to 0.75pt        

\begin{tikzpicture}[x=0.75pt,y=0.75pt,yscale=-.9,xscale=.9]

%Shape: Ellipse [id:dp1612241443957816] 
\draw  [draw opacity=0][shading=_x478jl5ln,_dyjpfu69k][dash pattern={on 4.5pt off 4.5pt}] (33.83,151.4) .. controls (33.93,80.48) and (91.51,23.07) .. (162.43,23.17) .. controls (233.35,23.27) and (290.77,80.84) .. (290.67,151.76) .. controls (290.57,222.69) and (232.99,280.1) .. (162.07,280) .. controls (91.15,279.9) and (33.73,222.33) .. (33.83,151.4) -- cycle ;
%Shape: Ellipse [id:dp7604049916003839] 
\draw  [fill={rgb, 255:red, 248; green, 231; blue, 28 }  ,fill opacity=1 ] (88.25,151.58) .. controls (88.25,110.72) and (121.38,77.59) .. (162.25,77.59) .. controls (203.12,77.59) and (236.25,110.72) .. (236.25,151.58) .. controls (236.25,192.45) and (203.12,225.58) .. (162.25,225.58) .. controls (121.38,225.58) and (88.25,192.45) .. (88.25,151.58) -- cycle ;
%Straight Lines [id:da5575015450361145] 
\draw    (162.25,151.58) -- (211.67,96.67) ;
%Straight Lines [id:da9474347314894263] 
\draw [line width=1.5]    (61,71.33) -- (11.16,21.5) ;
\draw [shift={(8.33,18.67)}, rotate = 45] [fill={rgb, 255:red, 0; green, 0; blue, 0 }  ][line width=0.08]  [draw opacity=0] (13.4,-6.43) -- (0,0) -- (13.4,6.44) -- (8.9,0) -- cycle    ;
%Straight Lines [id:da8019208867444223] 
\draw [line width=1.5]    (262.33,70.67) -- (308.17,24.83) ;
\draw [shift={(311,22)}, rotate = 135] [fill={rgb, 255:red, 0; green, 0; blue, 0 }  ][line width=0.08]  [draw opacity=0] (13.4,-6.43) -- (0,0) -- (13.4,6.44) -- (8.9,0) -- cycle    ;
%Straight Lines [id:da31181564021296104] 
\draw [line width=1.5]    (13.83,297.84) -- (70,241.67) ;
\draw [shift={(11,300.67)}, rotate = 315] [fill={rgb, 255:red, 0; green, 0; blue, 0 }  ][line width=0.08]  [draw opacity=0] (13.4,-6.43) -- (0,0) -- (13.4,6.44) -- (8.9,0) -- cycle    ;
%Straight Lines [id:da9242979256299755] 
\draw [line width=1.5]    (296.84,298.5) -- (247.67,249.33) ;
\draw [shift={(299.67,301.33)}, rotate = 225] [fill={rgb, 255:red, 0; green, 0; blue, 0 }  ][line width=0.08]  [draw opacity=0] (13.4,-6.43) -- (0,0) -- (13.4,6.44) -- (8.9,0) -- cycle    ;
%Straight Lines [id:da6485961236860431] 
\draw [color={rgb, 255:red, 208; green, 2; blue, 27 }  ,draw opacity=1 ][line width=2.25]    (139,221.67) .. controls (138.87,224.02) and (137.63,225.13) .. (135.28,225) .. controls (132.93,224.87) and (131.68,225.99) .. (131.55,228.34) .. controls (131.42,230.69) and (130.18,231.81) .. (127.83,231.68) .. controls (125.48,231.55) and (124.23,232.66) .. (124.1,235.01) .. controls (123.97,237.36) and (122.73,238.48) .. (120.38,238.35) .. controls (118.03,238.22) and (116.78,239.33) .. (116.65,241.68) .. controls (116.52,244.03) and (115.28,245.15) .. (112.93,245.02) .. controls (110.58,244.89) and (109.34,246.01) .. (109.21,248.36) -- (107,250.33) -- (107,250.33) ;
%Straight Lines [id:da12807490742550587] 
\draw [color={rgb, 255:red, 208; green, 2; blue, 27 }  ,draw opacity=1 ][line width=2.25]    (609,150) .. controls (607.69,151.96) and (606.06,152.29) .. (604.1,150.98) .. controls (602.14,149.67) and (600.5,150) .. (599.19,151.96) .. controls (597.88,153.92) and (596.25,154.25) .. (594.29,152.94) .. controls (592.33,151.63) and (590.7,151.96) .. (589.39,153.92) .. controls (588.08,155.88) and (586.44,156.21) .. (584.48,154.9) .. controls (582.52,153.59) and (580.89,153.92) .. (579.58,155.88) .. controls (578.27,157.84) and (576.64,158.17) .. (574.68,156.86) .. controls (572.72,155.55) and (571.09,155.88) .. (569.78,157.84) .. controls (568.47,159.8) and (566.83,160.13) .. (564.87,158.82) .. controls (562.91,157.51) and (561.28,157.84) .. (559.97,159.8) .. controls (558.66,161.76) and (557.03,162.09) .. (555.07,160.78) .. controls (553.11,159.47) and (551.47,159.8) .. (550.16,161.76) .. controls (548.85,163.72) and (547.22,164.05) .. (545.26,162.74) .. controls (543.3,161.43) and (541.67,161.76) .. (540.36,163.72) .. controls (539.05,165.68) and (537.41,166.01) .. (535.45,164.7) .. controls (533.49,163.39) and (531.86,163.72) .. (530.55,165.68) .. controls (529.24,167.64) and (527.61,167.97) .. (525.65,166.66) .. controls (523.69,165.35) and (522.06,165.68) .. (520.75,167.64) .. controls (519.44,169.6) and (517.8,169.93) .. (515.84,168.62) .. controls (513.88,167.31) and (512.25,167.64) .. (510.94,169.6) .. controls (509.63,171.56) and (508,171.89) .. (506.04,170.58) .. controls (504.08,169.27) and (502.44,169.6) .. (501.13,171.56) .. controls (499.82,173.52) and (498.19,173.85) .. (496.23,172.54) .. controls (494.27,171.23) and (492.64,171.56) .. (491.33,173.52) .. controls (490.02,175.48) and (488.38,175.81) .. (486.42,174.5) .. controls (484.46,173.19) and (482.83,173.52) .. (481.52,175.48) .. controls (480.21,177.44) and (478.58,177.77) .. (476.62,176.46) .. controls (474.66,175.15) and (473.03,175.48) .. (471.72,177.44) .. controls (470.41,179.4) and (468.77,179.73) .. (466.81,178.42) .. controls (464.85,177.11) and (463.22,177.44) .. (461.91,179.4) .. controls (460.6,181.36) and (458.97,181.69) .. (457.01,180.38) .. controls (455.05,179.07) and (453.41,179.4) .. (452.1,181.36) .. controls (450.79,183.32) and (449.16,183.65) .. (447.2,182.34) .. controls (445.24,181.03) and (443.61,181.36) .. (442.3,183.32) .. controls (440.99,185.28) and (439.35,185.61) .. (437.39,184.3) .. controls (435.43,182.99) and (433.8,183.32) .. (432.49,185.28) .. controls (431.18,187.24) and (429.55,187.57) .. (427.59,186.26) .. controls (425.63,184.95) and (423.99,185.28) .. (422.68,187.24) .. controls (421.37,189.2) and (419.74,189.53) .. (417.78,188.22) .. controls (415.82,186.91) and (414.19,187.24) .. (412.88,189.2) .. controls (411.57,191.16) and (409.94,191.49) .. (407.98,190.18) .. controls (406.02,188.87) and (404.38,189.2) .. (403.07,191.16) .. controls (401.76,193.12) and (400.13,193.45) .. (398.17,192.14) .. controls (396.21,190.83) and (394.58,191.16) .. (393.27,193.12) .. controls (391.96,195.08) and (390.32,195.41) .. (388.36,194.1) .. controls (386.4,192.79) and (384.77,193.12) .. (383.46,195.08) .. controls (382.15,197.04) and (380.52,197.37) .. (378.56,196.06) .. controls (376.6,194.75) and (374.96,195.08) .. (373.65,197.04) .. controls (372.34,199) and (370.71,199.33) .. (368.75,198.02) .. controls (366.79,196.71) and (365.16,197.04) .. (363.85,199) .. controls (362.54,200.96) and (360.91,201.29) .. (358.95,199.98) .. controls (356.99,198.67) and (355.35,199) .. (354.04,200.96) .. controls (352.73,202.92) and (351.1,203.25) .. (349.14,201.94) .. controls (347.18,200.63) and (345.55,200.96) .. (344.24,202.92) .. controls (342.93,204.88) and (341.29,205.21) .. (339.33,203.9) .. controls (337.37,202.59) and (335.74,202.92) .. (334.43,204.88) .. controls (333.12,206.84) and (331.49,207.17) .. (329.53,205.86) .. controls (327.57,204.55) and (325.93,204.88) .. (324.62,206.84) .. controls (323.31,208.8) and (321.68,209.13) .. (319.72,207.82) .. controls (317.76,206.51) and (316.13,206.84) .. (314.82,208.8) .. controls (313.51,210.76) and (311.88,211.09) .. (309.92,209.78) .. controls (307.96,208.47) and (306.32,208.8) .. (305.01,210.76) .. controls (303.7,212.72) and (302.07,213.05) .. (300.11,211.74) .. controls (298.15,210.43) and (296.52,210.76) .. (295.21,212.72) .. controls (293.9,214.68) and (292.26,215.01) .. (290.3,213.7) .. controls (288.34,212.39) and (286.71,212.72) .. (285.4,214.68) .. controls (284.09,216.64) and (282.46,216.97) .. (280.5,215.66) .. controls (278.54,214.35) and (276.9,214.68) .. (275.59,216.64) .. controls (274.28,218.6) and (272.65,218.93) .. (270.69,217.62) .. controls (268.73,216.31) and (267.1,216.64) .. (265.79,218.6) .. controls (264.48,220.56) and (262.84,220.89) .. (260.88,219.58) .. controls (258.92,218.27) and (257.29,218.6) .. (255.98,220.56) .. controls (254.67,222.52) and (253.04,222.85) .. (251.08,221.54) .. controls (249.12,220.23) and (247.49,220.56) .. (246.18,222.52) .. controls (244.87,224.48) and (243.23,224.81) .. (241.27,223.5) .. controls (239.31,222.19) and (237.68,222.52) .. (236.37,224.48) .. controls (235.06,226.44) and (233.43,226.77) .. (231.47,225.46) .. controls (229.51,224.15) and (227.87,224.48) .. (226.56,226.44) .. controls (225.25,228.4) and (223.62,228.73) .. (221.66,227.42) .. controls (219.7,226.11) and (218.07,226.44) .. (216.76,228.4) .. controls (215.45,230.36) and (213.81,230.69) .. (211.85,229.38) .. controls (209.89,228.07) and (208.26,228.4) .. (206.95,230.36) .. controls (205.64,232.32) and (204.01,232.65) .. (202.05,231.34) .. controls (200.09,230.03) and (198.46,230.36) .. (197.15,232.32) .. controls (195.84,234.28) and (194.2,234.61) .. (192.24,233.3) .. controls (190.28,231.99) and (188.65,232.32) .. (187.34,234.28) .. controls (186.03,236.24) and (184.4,236.57) .. (182.44,235.26) .. controls (180.48,233.95) and (178.84,234.28) .. (177.53,236.24) .. controls (176.22,238.2) and (174.59,238.53) .. (172.63,237.22) .. controls (170.67,235.91) and (169.04,236.24) .. (167.73,238.2) .. controls (166.42,240.16) and (164.78,240.49) .. (162.82,239.18) .. controls (160.86,237.87) and (159.23,238.2) .. (157.92,240.16) .. controls (156.61,242.12) and (154.98,242.45) .. (153.02,241.14) .. controls (151.06,239.83) and (149.43,240.16) .. (148.12,242.12) .. controls (146.81,244.08) and (145.17,244.41) .. (143.21,243.1) .. controls (141.25,241.79) and (139.62,242.12) .. (138.31,244.08) .. controls (137,246.04) and (135.37,246.37) .. (133.41,245.06) .. controls (131.45,243.75) and (129.81,244.08) .. (128.5,246.04) .. controls (127.19,248) and (125.56,248.33) .. (123.6,247.02) .. controls (121.64,245.71) and (120.01,246.04) .. (118.7,248) .. controls (117.39,249.96) and (115.75,250.29) .. (113.79,248.98) .. controls (111.83,247.67) and (110.2,248) .. (108.89,249.96) -- (107,250.33) -- (107,250.33) ;
%Straight Lines [id:da804272661450665] 
\draw [color={rgb, 255:red, 0; green, 0; blue, 255 }  ,draw opacity=1 ][line width=2.25]    (217.67,49.67) .. controls (217.44,52.01) and (216.15,53.06) .. (213.8,52.83) .. controls (211.45,52.6) and (210.16,53.65) .. (209.93,56) .. controls (209.7,58.35) and (208.41,59.4) .. (206.06,59.17) .. controls (203.71,58.93) and (202.42,59.98) .. (202.19,62.33) .. controls (201.96,64.68) and (200.67,65.73) .. (198.32,65.5) .. controls (195.97,65.26) and (194.68,66.31) .. (194.45,68.66) .. controls (194.22,71.01) and (192.93,72.06) .. (190.58,71.83) .. controls (188.23,71.6) and (186.94,72.65) .. (186.71,75) .. controls (186.48,77.35) and (185.19,78.4) .. (182.84,78.16) -- (181,79.67) -- (181,79.67) ;
%Straight Lines [id:da1824546871121273] 
\draw [color={rgb, 255:red, 0; green, 0; blue, 255 }  ,draw opacity=1 ][line width=2.25]    (217.67,49.67) .. controls (219.7,48.47) and (221.31,48.88) .. (222.51,50.91) .. controls (223.71,52.94) and (225.32,53.35) .. (227.35,52.15) .. controls (229.38,50.95) and (230.99,51.36) .. (232.2,53.39) .. controls (233.4,55.42) and (235.01,55.83) .. (237.04,54.63) .. controls (239.07,53.43) and (240.69,53.85) .. (241.88,55.88) .. controls (243.09,57.91) and (244.7,58.32) .. (246.73,57.12) .. controls (248.76,55.92) and (250.37,56.33) .. (251.57,58.36) .. controls (252.77,60.39) and (254.38,60.8) .. (256.41,59.6) .. controls (258.44,58.4) and (260.05,58.81) .. (261.26,60.84) .. controls (262.46,62.87) and (264.07,63.28) .. (266.1,62.08) .. controls (268.13,60.88) and (269.75,61.3) .. (270.94,63.33) .. controls (272.15,65.36) and (273.76,65.77) .. (275.79,64.57) .. controls (277.82,63.37) and (279.43,63.78) .. (280.63,65.81) .. controls (281.83,67.84) and (283.44,68.25) .. (285.47,67.05) .. controls (287.5,65.85) and (289.11,66.26) .. (290.32,68.29) .. controls (291.51,70.32) and (293.13,70.74) .. (295.16,69.54) .. controls (297.19,68.34) and (298.8,68.75) .. (300,70.78) .. controls (301.21,72.81) and (302.82,73.22) .. (304.85,72.02) .. controls (306.88,70.82) and (308.49,71.23) .. (309.69,73.26) .. controls (310.89,75.29) and (312.5,75.7) .. (314.53,74.5) .. controls (316.56,73.3) and (318.17,73.71) .. (319.38,75.74) .. controls (320.57,77.77) and (322.19,78.19) .. (324.22,76.99) .. controls (326.25,75.79) and (327.86,76.2) .. (329.06,78.23) .. controls (330.27,80.26) and (331.88,80.67) .. (333.91,79.47) .. controls (335.94,78.27) and (337.55,78.68) .. (338.75,80.71) .. controls (339.95,82.74) and (341.56,83.15) .. (343.59,81.95) .. controls (345.62,80.75) and (347.23,81.16) .. (348.44,83.19) .. controls (349.63,85.22) and (351.25,85.64) .. (353.28,84.44) .. controls (355.31,83.24) and (356.92,83.65) .. (358.12,85.68) .. controls (359.33,87.71) and (360.94,88.12) .. (362.97,86.92) .. controls (365,85.72) and (366.61,86.13) .. (367.81,88.16) .. controls (369.01,90.19) and (370.62,90.6) .. (372.65,89.4) .. controls (374.68,88.2) and (376.3,88.62) .. (377.5,90.65) .. controls (378.7,92.68) and (380.31,93.09) .. (382.34,91.89) .. controls (384.37,90.69) and (385.98,91.1) .. (387.18,93.13) .. controls (388.39,95.16) and (390,95.57) .. (392.03,94.37) .. controls (394.06,93.17) and (395.67,93.58) .. (396.87,95.61) .. controls (398.07,97.64) and (399.68,98.05) .. (401.71,96.85) .. controls (403.74,95.65) and (405.36,96.07) .. (406.56,98.1) .. controls (407.76,100.13) and (409.37,100.54) .. (411.4,99.34) .. controls (413.43,98.14) and (415.04,98.55) .. (416.24,100.58) .. controls (417.45,102.61) and (419.06,103.02) .. (421.09,101.82) .. controls (423.12,100.62) and (424.73,101.03) .. (425.93,103.06) .. controls (427.13,105.09) and (428.74,105.5) .. (430.77,104.3) .. controls (432.8,103.1) and (434.42,103.52) .. (435.62,105.55) .. controls (436.82,107.58) and (438.43,107.99) .. (440.46,106.79) .. controls (442.49,105.59) and (444.1,106) .. (445.3,108.03) .. controls (446.51,110.06) and (448.12,110.47) .. (450.15,109.27) .. controls (452.18,108.07) and (453.79,108.48) .. (454.99,110.51) .. controls (456.18,112.54) and (457.8,112.96) .. (459.83,111.76) .. controls (461.86,110.56) and (463.47,110.97) .. (464.68,113) .. controls (465.88,115.03) and (467.49,115.44) .. (469.52,114.24) .. controls (471.55,113.04) and (473.16,113.45) .. (474.36,115.48) .. controls (475.57,117.51) and (477.18,117.92) .. (479.21,116.72) .. controls (481.24,115.52) and (482.85,115.93) .. (484.05,117.96) .. controls (485.24,119.99) and (486.86,120.41) .. (488.89,119.21) .. controls (490.92,118.01) and (492.53,118.42) .. (493.74,120.45) .. controls (494.94,122.48) and (496.55,122.89) .. (498.58,121.69) .. controls (500.61,120.49) and (502.22,120.9) .. (503.42,122.93) .. controls (504.63,124.96) and (506.24,125.37) .. (508.27,124.17) .. controls (510.3,122.97) and (511.92,123.39) .. (513.11,125.42) .. controls (514.31,127.45) and (515.92,127.86) .. (517.95,126.66) .. controls (519.98,125.46) and (521.59,125.87) .. (522.8,127.9) .. controls (524,129.93) and (525.61,130.34) .. (527.64,129.14) .. controls (529.67,127.94) and (531.28,128.35) .. (532.48,130.38) .. controls (533.69,132.41) and (535.3,132.82) .. (537.33,131.62) .. controls (539.36,130.42) and (540.98,130.84) .. (542.17,132.87) .. controls (543.37,134.9) and (544.98,135.31) .. (547.01,134.11) .. controls (549.04,132.91) and (550.65,133.32) .. (551.86,135.35) .. controls (553.06,137.38) and (554.67,137.79) .. (556.7,136.59) .. controls (558.73,135.39) and (560.34,135.8) .. (561.54,137.83) .. controls (562.75,139.86) and (564.36,140.27) .. (566.39,139.07) .. controls (568.42,137.87) and (570.04,138.29) .. (571.23,140.32) .. controls (572.43,142.35) and (574.04,142.76) .. (576.07,141.56) .. controls (578.1,140.36) and (579.71,140.77) .. (580.92,142.8) .. controls (582.12,144.83) and (583.73,145.24) .. (585.76,144.04) .. controls (587.79,142.84) and (589.4,143.25) .. (590.6,145.28) .. controls (591.8,147.31) and (593.42,147.73) .. (595.45,146.53) .. controls (597.48,145.33) and (599.09,145.74) .. (600.29,147.77) .. controls (601.49,149.8) and (603.1,150.21) .. (605.13,149.01) -- (609,150) -- (609,150) ;
%Straight Lines [id:da5051072077749463] 
\draw [color={rgb, 255:red, 189; green, 16; blue, 224 }  ,draw opacity=1 ][line width=2.25]    (236.25,151.58) .. controls (237.91,149.91) and (239.58,149.9) .. (241.25,151.56) .. controls (242.92,153.22) and (244.59,153.21) .. (246.25,151.54) .. controls (247.91,149.87) and (249.58,149.86) .. (251.25,151.52) .. controls (252.92,153.18) and (254.59,153.17) .. (256.25,151.5) .. controls (257.91,149.83) and (259.58,149.82) .. (261.25,151.48) .. controls (262.92,153.14) and (264.59,153.13) .. (266.25,151.46) .. controls (267.91,149.79) and (269.58,149.78) .. (271.25,151.43) .. controls (272.92,153.09) and (274.59,153.08) .. (276.25,151.41) .. controls (277.91,149.74) and (279.58,149.73) .. (281.25,151.39) .. controls (282.92,153.05) and (284.59,153.04) .. (286.25,151.37) .. controls (287.91,149.7) and (289.58,149.69) .. (291.25,151.35) .. controls (292.92,153.01) and (294.59,153) .. (296.25,151.33) .. controls (297.91,149.66) and (299.58,149.65) .. (301.25,151.31) .. controls (302.92,152.97) and (304.59,152.96) .. (306.25,151.29) .. controls (307.91,149.62) and (309.58,149.61) .. (311.25,151.26) .. controls (312.92,152.92) and (314.59,152.91) .. (316.25,151.24) .. controls (317.91,149.57) and (319.58,149.56) .. (321.25,151.22) .. controls (322.92,152.88) and (324.59,152.87) .. (326.25,151.2) .. controls (327.91,149.53) and (329.58,149.52) .. (331.25,151.18) .. controls (332.92,152.84) and (334.59,152.83) .. (336.25,151.16) .. controls (337.91,149.49) and (339.58,149.48) .. (341.25,151.14) .. controls (342.92,152.8) and (344.59,152.79) .. (346.25,151.12) .. controls (347.91,149.45) and (349.58,149.44) .. (351.25,151.1) .. controls (352.92,152.75) and (354.59,152.74) .. (356.25,151.07) .. controls (357.91,149.4) and (359.58,149.39) .. (361.25,151.05) .. controls (362.92,152.71) and (364.59,152.7) .. (366.25,151.03) .. controls (367.91,149.36) and (369.58,149.35) .. (371.25,151.01) .. controls (372.92,152.67) and (374.59,152.66) .. (376.25,150.99) .. controls (377.91,149.32) and (379.58,149.31) .. (381.25,150.97) .. controls (382.92,152.63) and (384.59,152.62) .. (386.25,150.95) .. controls (387.91,149.28) and (389.58,149.27) .. (391.25,150.93) .. controls (392.92,152.58) and (394.59,152.57) .. (396.25,150.9) .. controls (397.91,149.23) and (399.58,149.22) .. (401.25,150.88) .. controls (402.92,152.54) and (404.59,152.53) .. (406.25,150.86) .. controls (407.91,149.19) and (409.58,149.18) .. (411.25,150.84) .. controls (412.92,152.5) and (414.59,152.49) .. (416.25,150.82) .. controls (417.91,149.15) and (419.58,149.14) .. (421.25,150.8) .. controls (422.92,152.46) and (424.59,152.45) .. (426.25,150.78) .. controls (427.91,149.11) and (429.58,149.1) .. (431.25,150.76) .. controls (432.92,152.41) and (434.59,152.4) .. (436.25,150.73) .. controls (437.91,149.06) and (439.58,149.05) .. (441.25,150.71) .. controls (442.92,152.37) and (444.59,152.36) .. (446.25,150.69) .. controls (447.91,149.02) and (449.58,149.01) .. (451.25,150.67) .. controls (452.92,152.33) and (454.59,152.32) .. (456.25,150.65) .. controls (457.91,148.98) and (459.58,148.97) .. (461.25,150.63) .. controls (462.92,152.29) and (464.59,152.28) .. (466.25,150.61) .. controls (467.91,148.94) and (469.58,148.93) .. (471.25,150.59) .. controls (472.92,152.24) and (474.59,152.23) .. (476.25,150.56) .. controls (477.91,148.89) and (479.58,148.88) .. (481.25,150.54) .. controls (482.92,152.2) and (484.59,152.19) .. (486.25,150.52) .. controls (487.91,148.85) and (489.58,148.84) .. (491.25,150.5) .. controls (492.92,152.16) and (494.59,152.15) .. (496.25,150.48) .. controls (497.91,148.81) and (499.58,148.8) .. (501.25,150.46) .. controls (502.92,152.12) and (504.59,152.11) .. (506.25,150.44) .. controls (507.91,148.77) and (509.58,148.76) .. (511.25,150.42) .. controls (512.92,152.07) and (514.59,152.06) .. (516.25,150.39) .. controls (517.91,148.72) and (519.58,148.71) .. (521.25,150.37) .. controls (522.92,152.03) and (524.59,152.02) .. (526.25,150.35) .. controls (527.91,148.68) and (529.58,148.67) .. (531.25,150.33) .. controls (532.92,151.99) and (534.59,151.98) .. (536.25,150.31) .. controls (537.91,148.64) and (539.58,148.63) .. (541.25,150.29) .. controls (542.92,151.95) and (544.59,151.94) .. (546.25,150.27) .. controls (547.91,148.6) and (549.58,148.59) .. (551.25,150.25) .. controls (552.92,151.9) and (554.59,151.89) .. (556.25,150.22) .. controls (557.91,148.55) and (559.58,148.54) .. (561.25,150.2) .. controls (562.92,151.86) and (564.59,151.85) .. (566.25,150.18) .. controls (567.9,148.51) and (569.57,148.5) .. (571.24,150.16) .. controls (572.91,151.82) and (574.58,151.81) .. (576.24,150.14) .. controls (577.9,148.47) and (579.57,148.46) .. (581.24,150.12) .. controls (582.91,151.78) and (584.58,151.77) .. (586.24,150.1) .. controls (587.9,148.43) and (589.57,148.42) .. (591.24,150.08) .. controls (592.91,151.73) and (594.58,151.72) .. (596.24,150.05) .. controls (597.9,148.38) and (599.57,148.37) .. (601.24,150.03) .. controls (602.91,151.69) and (604.58,151.68) .. (606.24,150.01) -- (609,150) -- (609,150) ;
%Shape: Circle [id:dp30932357841579616] 
\draw  [fill={rgb, 255:red, 189; green, 16; blue, 224 }  ,fill opacity=1 ] (251.33,150.17) .. controls (251.33,144.55) and (255.89,140) .. (261.5,140) .. controls (267.11,140) and (271.67,144.55) .. (271.67,150.17) .. controls (271.67,155.78) and (267.11,160.33) .. (261.5,160.33) .. controls (255.89,160.33) and (251.33,155.78) .. (251.33,150.17) -- cycle ;
%Shape: Arc [id:dp9137858302149506] 
\draw  [draw opacity=0][line width=1.5]  (640.7,137.05) .. controls (644.46,142.07) and (646.68,148.3) .. (646.67,155.06) .. controls (646.64,171.24) and (633.8,184.4) .. (617.77,184.98) -- (616.67,155) -- cycle ; \draw  [line width=1.5]  (640.7,137.05) .. controls (644.46,142.07) and (646.68,148.3) .. (646.67,155.06) .. controls (646.64,171.24) and (633.8,184.4) .. (617.77,184.98) ;  
%Shape: Triangle [id:dp9942947146675327] 
\draw  [fill={rgb, 255:red, 155; green, 155; blue, 155 }  ,fill opacity=1 ][line width=1.5]  (644.17,166.67) -- (653.67,206.33) -- (634.67,206.33) -- cycle ;
%Shape: Triangle [id:dp11546338112531562] 
\draw  [line width=1.5]  (610.58,150.57) -- (645.5,163.33) -- (639.4,174.05) -- cycle ;

%Straight Lines [id:da21816573183961085] 
\draw    (160.33,315) -- (644.33,312) ;
\draw [shift={(644.33,312)}, rotate = 179.64] [color={rgb, 255:red, 0; green, 0; blue, 0 }  ][line width=0.75]    (0,5.59) -- (0,-5.59)   ;
\draw [shift={(160.33,315)}, rotate = 179.64] [color={rgb, 255:red, 0; green, 0; blue, 0 }  ][line width=0.75]    (0,5.59) -- (0,-5.59)   ;
%Straight Lines [id:da23941858590866527] 
\draw [color={rgb, 255:red, 208; green, 2; blue, 27 }  ,draw opacity=1 ][line width=1.5]  [dash pattern={on 5.63pt off 4.5pt}]  (106.8,19) -- (106.8,301.67) ;
%Straight Lines [id:da08732945869301556] 
\draw [color={rgb, 255:red, 0; green, 0; blue, 255 }  ,draw opacity=1 ][line width=1.5]  [dash pattern={on 5.63pt off 4.5pt}]  (217.8,19) -- (217.8,301.67) ;
%Straight Lines [id:da028238248991550696] 
\draw [line width=1.5]    (561,280.5) -- (515,280.5) ;
\draw [shift={(511,280.5)}, rotate = 360] [fill={rgb, 255:red, 0; green, 0; blue, 0 }  ][line width=0.08]  [draw opacity=0] (13.4,-6.43) -- (0,0) -- (13.4,6.44) -- (8.9,0) -- cycle    ;
%Straight Lines [id:da1923690308649766] 
\draw [line width=1.5]    (561,280.5) -- (561,234.5) ;
\draw [shift={(561,230.5)}, rotate = 90] [fill={rgb, 255:red, 0; green, 0; blue, 0 }  ][line width=0.08]  [draw opacity=0] (13.4,-6.43) -- (0,0) -- (13.4,6.44) -- (8.9,0) -- cycle    ;

% Text Node
\draw (119.33,172.33) node [anchor=north west][inner sep=0.75pt]  [font=\normalsize] [align=left] {\begin{minipage}[lt]{60.57pt}\setlength\topsep{0pt}
\begin{center}
Supernova
\end{center}
Photosphere
\end{minipage}};
% Text Node
\draw (178.5,107.5) node   [align=left] {\begin{minipage}[lt]{20.63pt}\setlength\topsep{0pt}
$\displaystyle R_{\text{ph}}$
\end{minipage}};
% Text Node
\draw (51,87.67) node [anchor=north west][inner sep=0.75pt]   [align=left] {\begin{minipage}[lt]{31.07pt}\setlength\topsep{0pt}
\begin{center}
Ejecta
\end{center}

\end{minipage}};
% Text Node
\draw (141.17,156.5) node   [align=left] {\begin{minipage}[lt]{20.63pt}\setlength\topsep{0pt}
$\displaystyle T_{\text{ph}}$
\end{minipage}};
% Text Node
\draw (268.5,25.83) node   [align=left] {\begin{minipage}[lt]{20.63pt}\setlength\topsep{0pt}
$\displaystyle \vect{v}( \vect{r})$
\end{minipage}};
% Text Node
\draw (64.5,169.5) node   [align=left] {\begin{minipage}[lt]{20.63pt}\setlength\topsep{0pt}
$\displaystyle  \begin{array}{{>{\displaystyle}l}}
n_{l}( \vect{r})\\
n_{u}( \vect{r})
\end{array}$
\end{minipage}};
% Text Node
\draw (312.33,211.33) node [anchor=north west][inner sep=0.75pt]  [color={rgb, 255:red, 208; green, 2; blue, 27 }  ,opacity=1 ] [align=left] {\begin{minipage}[lt]{93.99pt}\setlength\topsep{0pt}
\begin{center}
redshifted emission
\end{center}

\end{minipage}};
% Text Node
\draw (312.33,127.67) node [anchor=north west][inner sep=0.75pt]  [color={rgb, 255:red, 189; green, 16; blue, 224 }  ,opacity=1 ] [align=left] {\begin{minipage}[lt]{105.92pt}\setlength\topsep{0pt}
\begin{center}
blueshifted absorption
\end{center}

\end{minipage}};
% Text Node
\draw (312.33,52.67) node [anchor=north west][inner sep=0.75pt]  [color={rgb, 255:red, 0; green, 0; blue, 255 }  ,opacity=1 ] [align=left] {\begin{minipage}[lt]{98.54pt}\setlength\topsep{0pt}
\begin{center}
blueshifted emission
\end{center}

\end{minipage}};
% Text Node
\draw (401.83,327.83) node   [align=left] {\begin{minipage}[lt]{20.63pt}\setlength\topsep{0pt}
$\displaystyle D_{A}$
\end{minipage}};
% Text Node
\draw (557,210) node [anchor=north west][inner sep=0.75pt]   [align=left] {$\displaystyle x$};
% Text Node
\draw (495,272) node [anchor=north west][inner sep=0.75pt]   [align=left] {$\displaystyle z$};

\end{tikzpicture}

    \caption{An illustration of the parameters of a supernova and its ejecta, highlighting the photon path for the blueshifted absorption and the redshifted/blueshifted emission for P Cygni profile. The vertical redshifted/blue dashed lines are the wavelength bins shown in the red/blue vertical lines in figures~\ref{fig:emission},~\ref{fig:sphere}, and~\ref{fig:asymmetry}. The physical parameters listed in table~\ref{tab:physical_parameter} are also indicated in the illustration.} 
    \label{fig:handdrawn}
\end{figure}
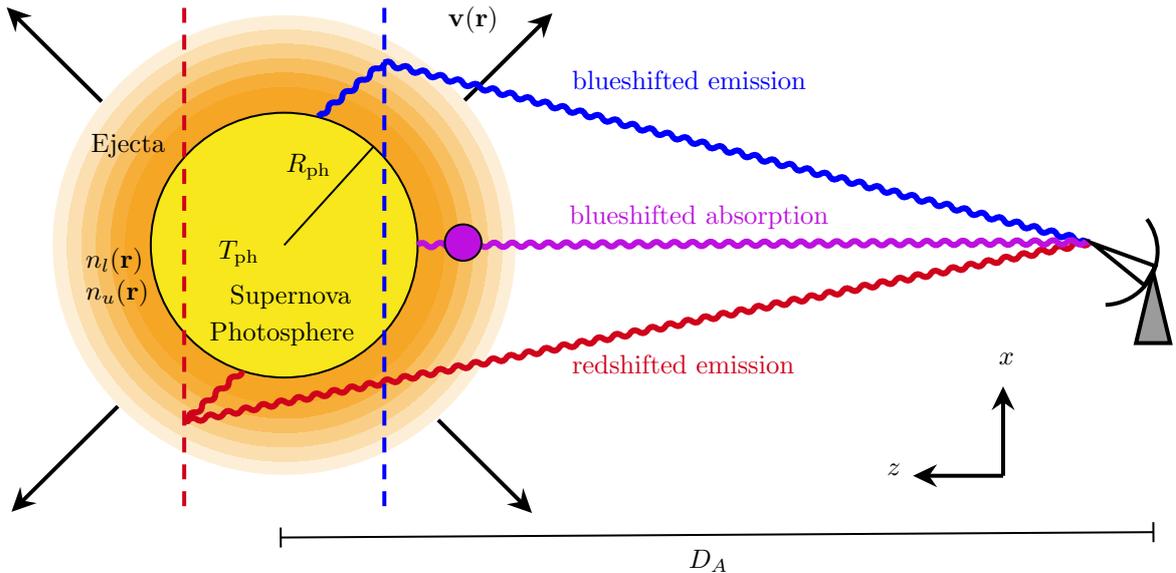

\begin{table}[]
\centering
\begin{tabular}{@{}cl@{}}
\toprule
\multicolumn{1}{l}{\textbf{General Parameters}}  & \textbf{Description}                                                    \\ \midrule
\multicolumn{1}{l}{\textbf{Photosphere}} &                                                                         \\
$R_\text{ph}$                                     & surface of photosphere                                                  \\
$T_\text{ph}$                                   & effective temperature of photosphere                                     \\
$D_A$                                     & angular diameter distance to supernova                                  \\ \midrule
\multicolumn{1}{l}{\textbf{Ejecta}}      &                                                                         \\
$\vect{v}(\vect{r})$                                     & velocity field of the ejecta                                            \\
$n_l(\vect{r}),~n_u(\vect{r})$                         & number density of lower and upper population of line-producing atom \\ \bottomrule
\end{tabular}
\caption{List of the physical parameters of a supernova and its ejecta.}
\label{tab:physical_parameter}
\end{table}

\subsection{Spherically symmetric model}\label{sec:spherical}
We begin with a spherically symmetric model of the SN photosphere and its surrounding ejecta. Figure~\ref{fig:handdrawn} illustrates the assumed structure.

\subsubsection{Photosphere}
\label{sec:photosphere}
The photosphere is modeled as an opaque spherical surface radiating as a blackbody at temperature $T_\text{ph}$. In a coordinate system centered on the explosion, it satisfies:
\begin{equation}
\label{eq:photosphere}
x^2 + y^2 + z^2 = R^2_\text{ph} \, ,
\end{equation}
where $R_\text{ph}$ is the physical radius of the photosphere. Its angular radius, as observed from Earth, is:
\begin{equation}
{\Theta_\text{ph}} = \frac{R_\text{ph}}{D_A} \, .
\end{equation}
The photosphere appears as a uniform disk with spectral radiance given by Planck’s law:
\begin{equation}\label{eq:Icont}
I^{\text{cont}}_\lambda(\lambda, \vect{\theta}) = \mathcal{N} B_\lambda(T_\mathrm{ph}) \, \Theta_\text{H}\left(\frac{R_\text{ph}}{D_A} - |\vect{\theta}|\right),
\end{equation}
 where $\Theta_\text{H}$ is the Heaviside step function. Limb darkening is neglected for simplicity, though it could be included in more detailed models (see ref.~\cite{acharyya2024angular}).
 To account for deviations from ideal blackbody emission e.g.~due to dust extinction or non-LTE effects), we introduce --- and will marginalize over --- the normalization factor $
 \mathcal{N}$. These photosphere-related parameters $\lbrace R_\text{ph}, T_\text{ph}, \mathcal{N}\rbrace$ are listed in table~\ref{tab:model}. Note that $v_\text{ph}$ corresponds to the velocity of the ejecta at the photosphere at radius $R_\text{ph}$ under the assumption of homologous expansion (see section~\ref{sec:resolved}), not the velocity $\dot{R}_\text{ph}$ of the photosphere itself.
\begin{table}[]
\centering
\begin{tabular}{@{}cl@{}}
\toprule
\multicolumn{1}{l}{\textbf{Model Parameters}}  & \textbf{Description}                                                    \\ \midrule
\multicolumn{1}{l}{\textbf{Geometric}} &                                                                         \\
$v_\text{ph}$                                     & velocity of the ejecta at photosphere                                                  \\
$D_A$                                     & angular diameter distance to supernova                                  \\
$\eta$ & degree of asymmetry \\$\theta,\phi$ &two Euler angle of the ellipsoidal photosphere\\
\midrule
\multicolumn{1}{l}{\textbf{Optical}}      &                                                                         \\
$\tau_\text{ph}$                                     & line optical depth at the photosphere                                            \\
$n$ & spectral index of the optical depth power law\\
\midrule
\multicolumn{1}{l}{\textbf{Spectral}} &                                          \\
$T_\text{ph}$ & effective temperature of the spectrum\\
$\mathcal{N}$ & overall normalization of the spectrum to account for unknown attenuation
\\ \bottomrule
\end{tabular}
\caption{List of the model parameters used in our parametric model to describe a supernova. The parameters $\lbrace \eta, \theta, \phi \rbrace$ specify the deformation and orientation of the spherical model to an asymmetric explosion; all others go into the spherical model.}
\label{tab:model}
\end{table}

\subsubsection{Ejecta}
\label{sec:ejecta}
The ejecta consist of line-producing atoms surrounding the SN, characterized by a velocity field $\vect{v}(\vect{r})$ and number densities of lower and upper energy levels, $n_l(\vect{r})$ and $n_u(\vect{r})$, respectively. We assume homologous expansion: ejecta velocity is proportional to distance from the explosion center, so position and velocity are related by
\begin{equation*}
    \vect{v}(\vect{r}) = \frac{\vect{r}}{t - t_0}\, ,
\end{equation*}
where $t$ is the observation time and $t_0$ is the explosion time.

To model the optical properties of the ejecta, we solve the radiative transfer equation for a two-level system to obtain the line optical depth:
\begin{equation}
\begin{split}
    \tau(\lambda, \vect{\theta}) &= \int_{0}^{s_\mathrm{max}(\vect{\theta)}} \dd s \, \alpha_\lambda(s) 
    = \int_{0}^{s_\mathrm{max}(\vect{\theta)}} \dd s \,\frac{B_{lu}h\nu_0}{4\pi}n_l(s, \vect{\theta})\left(1-\frac{n_u(s, \vect{\theta})}{n_l(s, \vect{\theta})}\frac{g_l}{g_u}\right)\phi\left[\lambda - \lambda_\text{rest}(s, \vect{\theta})\right] \, ,
\end{split}
\end{equation}
where $s$ is the line-of-sight coordinate, $B_{lu}$ is the Einstein coefficient, $g_l, g_u$ are statistical weights, and $\phi(\lambda)$ is the normalized line profile function which describes the shape of the emission/absorption line with the normalization $\int\phi(\lambda)\dd\lambda = 1$.
Assuming a narrow line profile and homologous expansion, each wavelength corresponds to a unique position along the line of sight, so the integral simplifies under the Sobolev approximation~\cite{sobolev1960moving}:
\begin{equation}
    \tau(\lambda, \vect{\theta}) = \frac{B_{lu}h\nu_0}{4\pi}n_l(\lambda,\vect{\theta})\left(1-\frac{n_u(\lambda,\vect{\theta})}{n_l(\lambda,\vect{\theta})}\frac{g_l}{g_u}\right).
\end{equation}

We assume the number densities follow power laws with radius (or equivalently, with velocity under homologous expansion):
\begin{equation}
    \begin{split}
        \frac{n_l(\vect{r})}{n_l(R_\text{ph})}&= \left(\frac{|\vect{r}|}{R_\text{ph}}\right)^{-\alpha}= \left(\frac{|\vect{v}|}{|\vect{v_\text{ph}}|}\right)^{-\alpha},\\
        \left(1-\frac{n_u(\vect{r})}{n_l(\vect{r})}\frac{g_l}{g_u}\right)/\left(1-\frac{n_u(R_\text{ph})}{n_l(R_\text{ph})}\frac{g_l}{g_u}\right)&= \left(\frac{|\vect{r}|}{R_\text{ph}}\right)^{-\beta}= \left(\frac{|\vect{v}|}{|\vect{v_\text{ph}}|}\right)^{-\beta},
    \end{split}
    \label{eq:alpha_beta}
\end{equation}
where  $v_\text{ph}$ is the \emph{velocity of the ejecta at the photosphere} (\emph{not} the velocity of the photospheric surface itself) and $\alpha, \beta$ are the respective power-law indices.
Combining these gives the optical depth as a power law in velocity:
\begin{equation}\label{eq:tau_model}
    \tau(\vect{v})= \tau_\text{ph}\left(\frac{|\vect{v_\text{ph}}|}{|\vect{v}|}\right)^{n},
\end{equation}
where $\tau_\text{ph}$ is the line optical depth at the photosphere and  $n = \alpha +\beta$. This profile determines the influence of line-producing ejecta on the observed image.

\begin{figure}
    \centering
    \includegraphics[width = \textwidth]{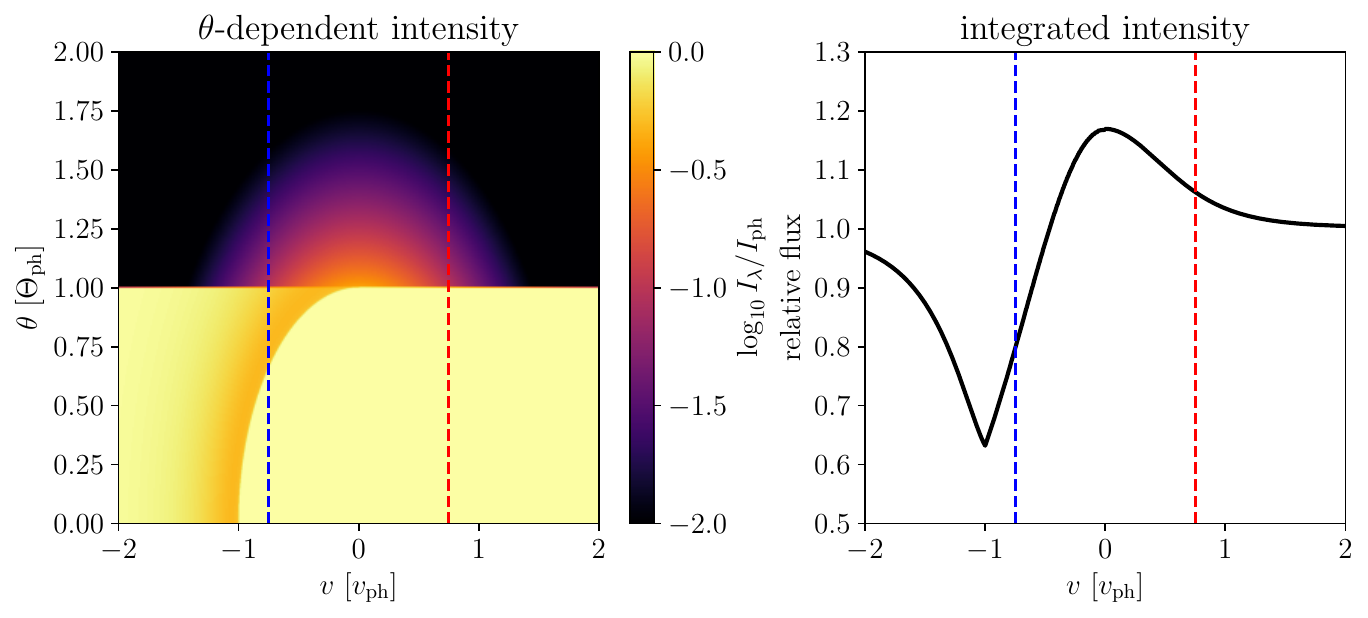}
    \caption{P Cygni line profile for a spherically symmetric supernova, using $\tau_\text{ph} = 2$ and $n = 5$. The vertical red (blue) dashed line marks velocities of $0.75~v_\text{ph}~(-0.75~v_\text{ph})$, corresponding to the spectral channels shown in figures~\ref{fig:sphere} and~\ref{fig:ellipsoid}. \textbf{Left:} Spectral radiance as a function of velocity (wavelength) and angular position $\vect{\theta}$ across the SN image. The central region with $|\vect{\theta}| < \Theta_\text{ph}$ represents the photosphere. The dark-orange arc illustrates the P Cygni absorption feature, which appears at $v_\text{ph}$ at the center of the photosphere and shifts toward zero velocity at the limb, tracing the line-of-sight velocity of ejecta at the photosphere’s surface. The absorption extends to higher (blueshifted) velocities for ejecta further out. Emission appears as a halo outside the photosphere, centered at the rest wavelength, originating from redshifted and blueshifted ejecta beyond the photospheric disk. \textbf{Right:} Spatially integrated P Cygni spectrum, obtained by integrating the left panel over all angles $\vect{\theta}$. The spectrum shows a classic P Cygni profile: a blueshifted absorption dip peaking at $v_\text{ph}$ and a symmetric emission bump centered at the rest wavelength. The plotted wavelength range corresponds to velocities of $\pm0.05~c$, or $1.5\times10^4~\text{km/s}$. \nblinkeem{plots/P_cygni_spectrum.ipynb}}
    \label{fig:emission}
\end{figure}

\subsubsection{Spectrum and intensity correlations}
The optical depth profile derived above allows us to compute both the observed spectrum and intensity correlations. In this section, we describe the contributions from absorption and emission, as illustrated in figure~\ref{fig:emission}.

\paragraph{Absorption}
The suppression from absorption by the ejecta is described in eq.~\eqref{eq:absorption}. In our model, absorption begins at the rest wavelength of the line at the edge of the photosphere ($|\vect{\theta}| = \Theta_\mathrm{ph}$) and effectively blueshifts going inward toward the center ($|\vect{\theta}| = 0$). This creates the characteristic P Cygni absorption feature, visible as the darker arc in the left panel of figure~\ref{fig:emission}.
At the center of the photosphere, the absorption line corresponds to the velocity $v_\text{ph}$; towards the limb, it approaches the rest wavelength due to decreasing line-of-sight velocity. Ejecta at greater distances from the explosion center (but towards the observer) produce absorption at higher blueshifted velocities, broadening the feature.

\begin{figure}
    \centering
    \includegraphics[width = \textwidth]{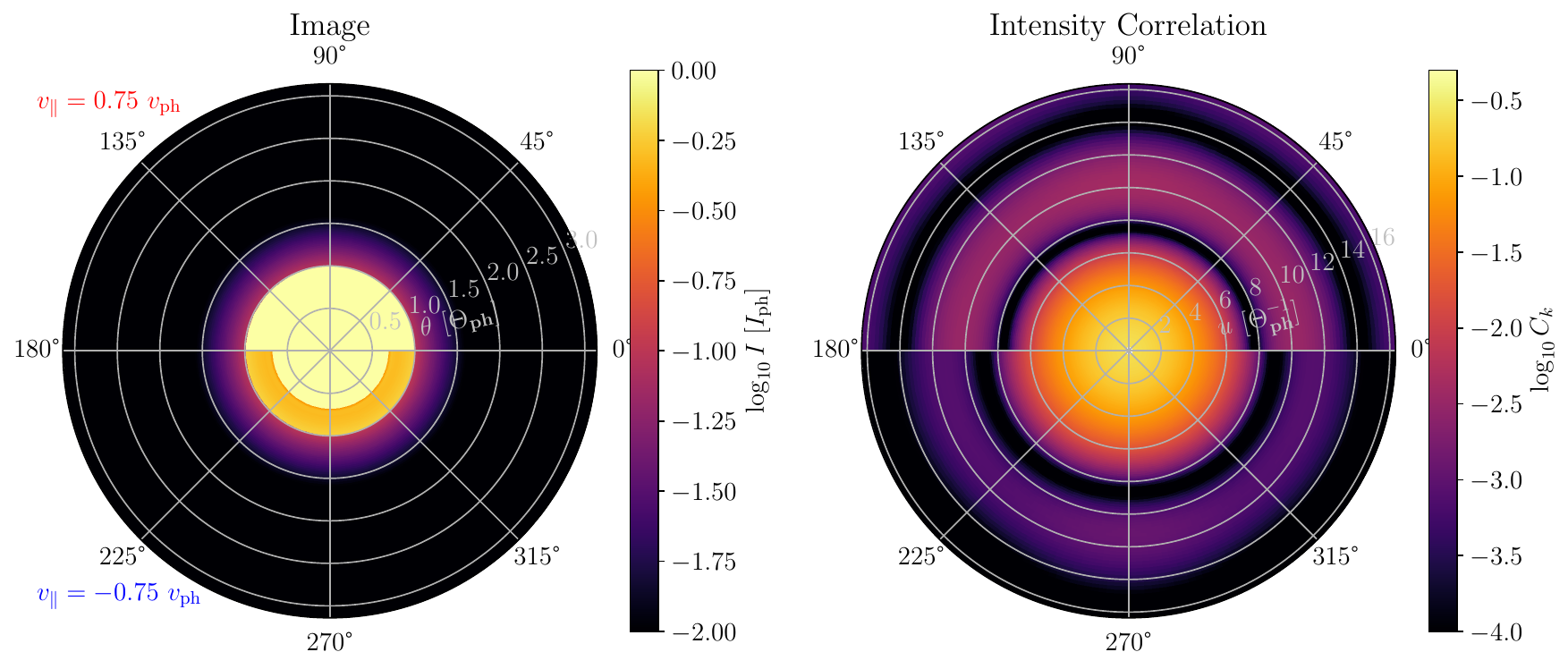}
    \caption{\textbf{Left:} The image of a spherical SN. The radial coordinate is in units of the angular photosphere radius $\Theta_\mathrm{ph}$. The upper half of the image is taken at a redshifted frequency with Doppler velocity $0.75\,v_\text{ph}$ (the red dashed line in figure~\ref{fig:emission}), where no absorption occurs. The lower half of the image is taken at a blueshifted frequency at $-0.75\,v_\text{ph}$ (the blue dashed line in figure~\ref{fig:emission}), where absorption occurs in an annulus on the outside of the projected photosphere. Both  upper and lower halves of the panel show emission features around the photosphere. \textbf{Right:} The intensity correlation maps of the images on the left. 
    The lower half of the intensity correlation map exhibits a larger main lobe, because the bright part of the photosphere is smaller in the corresponding image on the left due to absorption.
    The radial coordinate is the angular wavenumber magnitude $|\vect{u}|$ in units of $\Theta_\mathrm{ph}^{-1}$. A dense intensity interferometry array could in principle sample this square modulus of the visibility function in its entirety and thus reconstruct the real-space images on the left at high fidelity.
    \nblinkeem{demo.ipynb}}
    \label{fig:sphere}
\end{figure}

The lower panel of figure~\ref{fig:sphere} shows a 2D image at a blueshifted spectral channel, clearly revealing the edge-absorbed feature described above.

\begin{figure}
    \centering
    \includegraphics[width=\linewidth]{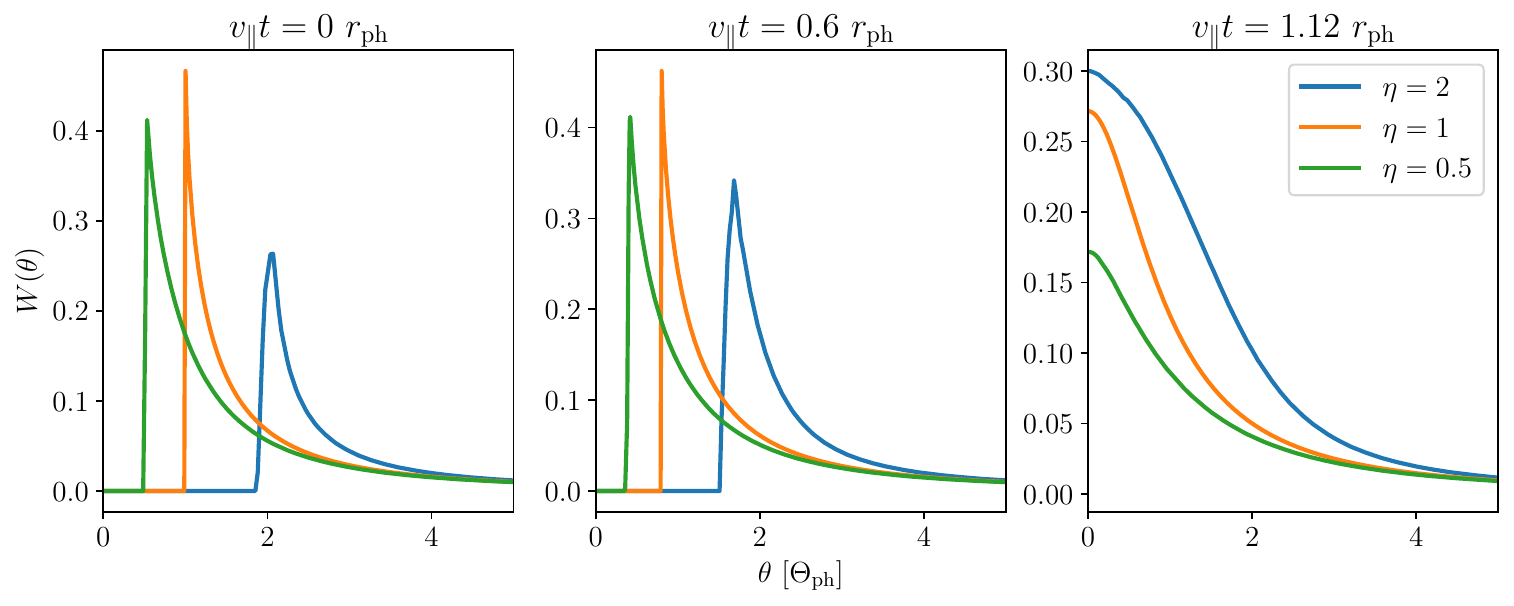}
    \caption{The geometric dilution factor for an oblate, spherical, and prolate spheroid as a function of $r_\perp = \theta D_A$ with three different $z = v_\parallel \, t$, with the short/long axis perpendicular to the line of sight. The geometric dilution factor vanishes inside the photosphere and monotonically decreases as a function of $r_\perp$  outside the photosphere. \nblinkeem{plots/geo_w.ipynb}}
    \label{fig:geo_w}
\end{figure}

\paragraph{Emission}
For a patch of ejecta material located at $\vect{r}(\lambda, \vect{\theta})$, assuming LTE, the emitted spectral radiance is given by:
\begin{equation}\label{eq:em_model}
    I^\text{em}_\lambda(\lambda,  \vect{\theta}) = S_\lambda(\lambda,\vect{\theta})\left(1-e^{-\tau(\lambda, \vect{\theta})}\right) \, ,
\end{equation}
where $S_\lambda$ is the line source function. Assuming the photospheric radiance is uniform across the narrow bandwidth of the P Cygni line and isotropic due to blackbody emission, the source function simplifies to:
\begin{equation}
    S_\lambda(\lambda, \vect{\theta}) = \frac{j_\lambda}{\alpha_\lambda} 
    = \frac{1}{4\pi}\int \dd \Omega \, I_\lambda(\lambda, \vect{\theta)}
    = I^\text{cont}_\lambda(\lambda)~W(\lambda, \vect{\theta}) \, .
    \label{eq:emission}
\end{equation}
The factor $W(\lambda, \vect{\theta})$ is the geometric dilution factor, which represents the fraction of the sky subtended by the photosphere as seen from the emitting point. For a spherical photosphere of radius $R_\text{ph}$:
\begin{equation}\label{eq:Wfunc}
    W(\lambda, \vect{\theta}) = \frac{1}{2}\left[1-\sqrt{1 - \frac{R_\text{ph}^2}{|\vect{r}(\lambda, \vect{\theta})|^2}}\right].
\end{equation}
Figure~\ref{fig:geo_w} illustrates how $W$ varies with geometry --- spherical, prolate, and oblate --- and wavelength. Inside the photosphere $(|\vect{\theta}|<\Theta_\text{ph}), W = 0$; beyond the photosphere, it decreases with distance. This effect contributes to the emission ring seen outside the photosphere in both figure~\ref{fig:emission} (left) and the 2D image in figure~\ref{fig:sphere}.
In the spatially integrated spectrum (figure~\ref{fig:emission}, right), we observe a classic P Cygni profile: blueshifted absorption from material in front of the photosphere and emission --- both redshifted and blueshifted --- from ejecta surrounding it.

Figure~\ref{fig:sphere} further shows 2D SN images and intensity correlations in two spectral channels: redshifted (blueshifted) by $0.75 \, v_\text{ph}~(-0.75 \,v_\text{ph})$. In the blueshifted image, absorption appears at the limb of the photosphere; the redshifted image, by contrast, shows pure emission in a ring, as the line of sight to the far-side ejecta outside $\mathcal{D}$ is not blocked.
The intensity correlation reflects these structures. In the redshifted channel, it resembles a disk’s diffraction pattern with the first zero at $u = 1.22\times 2 \pi \Theta_\text{ph}^{-1}$. In the blueshifted channel, absorption shrinks the effective photosphere, widening the intensity correlation. Additionally, the second lobe is weaker due to the absorption annulus.

\subsection{Ellipsoidal model}
\label{sec:asymmetry}
In section~\ref{sec:spherical}, we developed a simple parametric model for a spherically symmetric SN. Here, we generalize the model to describe an ellipsoidal SN photosphere, allowing for deviations from spherical symmetry by introducing one oblate or prolate axis. Specifically, we relax eq.~\eqref{eq:photosphere} to describe the surface of an ellipsoidal photosphere as:
\begin{equation}
    \frac{x'^2}{\eta^2} + y'^2 + z'^2 = R^2_\text{ph} \, ,
\end{equation}
where the primed coordinate $(x', y', z')$ is the coordinate system centered at the center of the SN with the prolate/oblate axis in the $x$-axis. The parameter $\eta$ controls the shape of the ellipsoid: $\eta > 1~(\eta<1)$ corresponds to a prolate (oblate) ellipsoid.

\begin{figure}
    \centering

\tikzset{every picture/.style={line width=0.75pt}} %set default line width to 0.75pt        

\begin{tikzpicture}[x=0.75pt,y=0.75pt,yscale=-1,xscale=1]

%Shape: Square [id:dp9669684187656598] 
\draw  [fill={rgb, 255:red, 164; green, 197; blue, 235 }  ,fill opacity=0.99 ] (325.27,76.17) -- (495.07,76.17) -- (495.07,245.97) -- (325.27,245.97) -- cycle ;
%Shape: Ellipse [id:dp5344986179541649] 
\draw  [fill={rgb, 255:red, 245; green, 166; blue, 35 }  ,fill opacity=1 ][line width=1.5]  (63.35,204.76) .. controls (52.22,192.92) and (64.3,163.47) .. (90.34,138.97) .. controls (116.38,114.47) and (146.52,104.21) .. (157.66,116.04) .. controls (168.79,127.88) and (156.71,157.34) .. (130.67,181.83) .. controls (104.62,206.33) and (74.49,216.59) .. (63.35,204.76) -- cycle ;
%Shape: Arc [id:dp4907978013925036] 
\draw  [draw opacity=0][line width=1.5]  (287.04,146.71) .. controls (290.79,151.73) and (293.01,157.97) .. (293,164.72) .. controls (292.97,180.91) and (280.13,194.07) .. (264.1,194.65) -- (263,164.67) -- cycle ; \draw  [line width=1.5]  (287.04,146.71) .. controls (290.79,151.73) and (293.01,157.97) .. (293,164.72) .. controls (292.97,180.91) and (280.13,194.07) .. (264.1,194.65) ;  
%Shape: Triangle [id:dp6938741201081657] 
\draw  [fill={rgb, 255:red, 155; green, 155; blue, 155 }  ,fill opacity=1 ][line width=1.5]  (290.5,176.33) -- (300,216) -- (281,216) -- cycle ;
%Shape: Triangle [id:dp7219737120148402] 
\draw  [line width=1.5]  (256.91,160.24) -- (291.83,173) -- (285.74,183.72) -- cycle ;

%Straight Lines [id:da07062737540510944] 
\draw  [dash pattern={on 4.5pt off 4.5pt}]  (110.5,160.4) -- (249.67,160.4) ;
%Straight Lines [id:da6679263440216234] 
\draw  [dash pattern={on 4.5pt off 4.5pt}]  (110.5,160.4) -- (110.5,40.67) ;
%Straight Lines [id:da3410172379403327] 
\draw    (110.5,160.4) -- (157.66,116.04) ;
%Straight Lines [id:da8558483240400936] 
\draw  [dash pattern={on 4.5pt off 4.5pt}]  (226.37,47.33) -- (157.66,116.04) ;
%Shape: Arc [id:dp1804470793579258] 
\draw  [draw opacity=0] (111.86,81.44) .. controls (134.62,81.75) and (155.28,89.89) .. (170.67,103.01) -- (110.5,160.4) -- cycle ; \draw   (111.86,81.44) .. controls (134.62,81.75) and (155.28,89.89) .. (170.67,103.01) ;  
%Shape: Ellipse [id:dp744489891512651] 
\draw  [fill={rgb, 255:red, 245; green, 166; blue, 35 }  ,fill opacity=1 ][line width=1.5]  (363.02,205.42) .. controls (351.88,193.59) and (363.97,164.13) .. (390.01,139.64) .. controls (416.05,115.14) and (446.19,104.87) .. (457.32,116.71) .. controls (468.46,128.55) and (456.37,158) .. (430.33,182.5) .. controls (404.29,207) and (374.15,217.26) .. (363.02,205.42) -- cycle ;
%Straight Lines [id:da49003833908831584] 
\draw  [dash pattern={on 4.5pt off 4.5pt}]  (410.17,161.07) -- (495,161.07) ;
%Straight Lines [id:da12817286102343317] 
\draw  [dash pattern={on 4.5pt off 4.5pt}]  (410.17,161.07) -- (410.17,76) ;
%Straight Lines [id:da9644891821447268] 
\draw    (410.17,161.07) -- (457.32,116.71) ;
%Straight Lines [id:da03988455053075768] 
\draw  [dash pattern={on 4.5pt off 4.5pt}]  (497.87,76.17) -- (457.32,116.71) ;
%Shape: Arc [id:dp7188380085147374] 
\draw  [draw opacity=0] (462.69,112.25) .. controls (476.71,124.51) and (485.47,141.63) .. (485.61,160.61) -- (410.17,161.07) -- cycle ; \draw   (462.69,112.25) .. controls (476.71,124.51) and (485.47,141.63) .. (485.61,160.61) ;  
%Straight Lines [id:da9505753837129147] 
\draw [line width=1.5]    (382,234.5) -- (336,234.5) ;
\draw [shift={(386,234.5)}, rotate = 180] [fill={rgb, 255:red, 0; green, 0; blue, 0 }  ][line width=0.08]  [draw opacity=0] (13.4,-6.43) -- (0,0) -- (13.4,6.44) -- (8.9,0) -- cycle    ;
%Straight Lines [id:da34701297915145046] 
\draw [line width=1.5]    (336,234.5) -- (336,188.5) ;
\draw [shift={(336,184.5)}, rotate = 90] [fill={rgb, 255:red, 0; green, 0; blue, 0 }  ][line width=0.08]  [draw opacity=0] (13.4,-6.43) -- (0,0) -- (13.4,6.44) -- (8.9,0) -- cycle    ;
%Straight Lines [id:da23492274223758858] 
\draw [line width=1.5]    (237,234.5) -- (191,234.5) ;
\draw [shift={(187,234.5)}, rotate = 360] [fill={rgb, 255:red, 0; green, 0; blue, 0 }  ][line width=0.08]  [draw opacity=0] (13.4,-6.43) -- (0,0) -- (13.4,6.44) -- (8.9,0) -- cycle    ;
%Straight Lines [id:da38817040423692384] 
\draw [line width=1.5]    (237,234.5) -- (237,188.5) ;
\draw [shift={(237,184.5)}, rotate = 90] [fill={rgb, 255:red, 0; green, 0; blue, 0 }  ][line width=0.08]  [draw opacity=0] (13.4,-6.43) -- (0,0) -- (13.4,6.44) -- (8.9,0) -- cycle    ;

% Text Node
\draw (136,135) node [anchor=north west][inner sep=0.75pt]   [align=left] {$\displaystyle \eta $};
% Text Node
\draw (137.33,64.33) node [anchor=north west][inner sep=0.75pt]   [align=left] {$\displaystyle \theta $};
% Text Node
\draw (463,134.33) node [anchor=north west][inner sep=0.75pt]   [align=left] {$\displaystyle \phi $};
% Text Node
\draw (74,224.3) node [anchor=north west][inner sep=0.75pt]   [align=left] {supernova};
% Text Node
\draw (410.33,224.33) node [anchor=north west][inner sep=0.75pt]   [align=left] {image plane};
% Text Node
\draw (332,164) node [anchor=north west][inner sep=0.75pt]   [align=left] {$\displaystyle y$};
% Text Node
\draw (390,226) node [anchor=north west][inner sep=0.75pt]   [align=left] {$\displaystyle x$};
% Text Node
\draw (233,164) node [anchor=north west][inner sep=0.75pt]   [align=left] {$\displaystyle x$};
% Text Node
\draw (171,226) node [anchor=north west][inner sep=0.75pt]   [align=left] {$\displaystyle z$};

\end{tikzpicture}

    \caption{Illustration of the asymmetry of the supernova. The parameter $\eta$ controls the elongation/oblateness of the one axis of the ellipsoid. $\theta$ and $\phi$ are two Euler angles defined by the illustration.}
    \label{fig:asymmetry}
\end{figure}
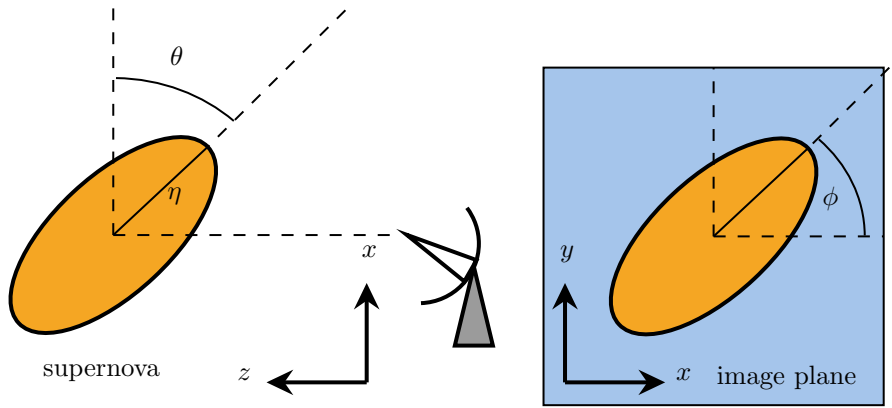

To incorporate the ellipsoidal geometry into our model, we apply the following spatial coordinate transformation to eqs.~\eqref{eq:photosphere}--\eqref{eq:emission}:
\begin{equation}
\begin{pmatrix}
x\\
y\\
z
\end{pmatrix} = \begin{pmatrix}
    \cos\phi  &-\sin\phi & 0\\
    \sin\phi & \cos\phi & 0\\
    0 & 0 & 1
\end{pmatrix}
\begin{pmatrix}
    \cos\theta & 0 & \sin\theta\\
    0 & 1 & 0\\
    -\sin\theta & 0 & \cos\theta
\end{pmatrix}
\begin{pmatrix}
x'/\eta\\
y'\\
z'
\end{pmatrix} \, .
\end{equation}
That is, we rescale the $x$-axis by $\eta$ to match the ellipsoidal shape and rotate the SN with a rotation, modifying the geometry of the ejecta accordingly. This introduces three additional geometric parameters --- $\eta, \theta, \phi$ --- described in table~\ref{tab:model}. The Euler angles $\theta$ and $\phi$, defined in figure~\ref{fig:asymmetry}, determine the orientation of the ellipsoid: $\theta$ is a rotation of the ellipsoid's axis in the $x$--$z$~plane (inclination), while $\phi$ is the azimuthal rotation in the image ($x$--$y$) plane. 
We use Monte Carlo integration to compute the dilution factor $W(\lambda,\vect{\theta})$ for an ellipsoidal geometry, i.e.~the solid angle subtended by the ellipsoidal photosphere from each position $\vect{r}(\lambda,\vect{\theta})$.

\begin{figure}
    \centering
    \includegraphics[width = \textwidth]{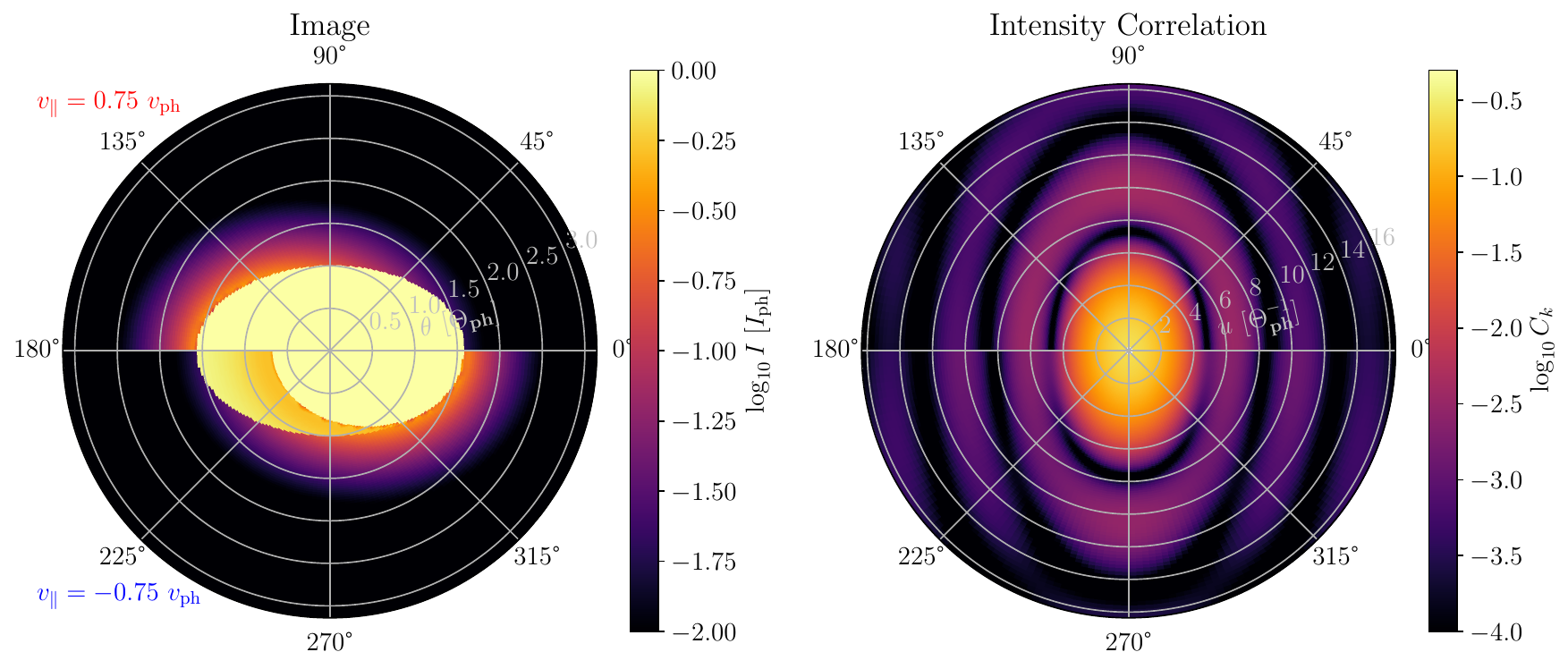}
    \caption{\textbf{Left:} The image of an asymmetric SN with $\eta = 2$ and the long axis rotated $\theta = \pi/4$ (and $\phi = 0$). The radial coordinate is in unit of the angular size of the SN in the \emph{non-special direction}. The upper-half of the image is taken at a redshifted frequency $0.75~v_\text{ph}$ (the red dashed line in figure~\ref{fig:emission}) and we can see that there is no absorption in the photosphere and there is emission from the left-half of the SN ejecta. The lower-half of the image is taken at a blueshifted frequency $-0.75~v_\text{ph}$ (the blue dashed line in figure~\ref{fig:emission}) and we can see that there is absorption in the photosphere and the emission is coming from the right-half of the SN. \textbf{Right:} The intensity correlation map of the image on the left. The radial coordinate is the angular wavenumber of the Fourier transform, in unit of the inverse of the angular size of the SN in the \emph{non-special direction}. This would be the image the intensity interferometer see if we have a whole array of telescopes covering different baselines. We can see that the interference pattern shows the asymmetric between the long and short axis. The lower-half of the intensity correlation map also shows a larger main lobe. This is because the effective photosphere is smaller in the corresponding image on the left from the absorption. \nblinkeem{demo.ipynb}}
    \label{fig:ellipsoid}
\end{figure}

Figure~\ref{fig:ellipsoid} shows the spatially resolved SN image and corresponding intensity correlation for a prolate ellipsoidal model using this extended parametric description. The upper panel shows the image in a redshifted spectral channel; the lower panel shows the blueshifted counterpart. The asymmetry in the redshifted/blueshifted emission regions is a direct consequence of the ellipsoidal geometry.

Compared to the spherical case (figure~\ref{fig:sphere}), the ellipsoidal SN shows elongation along one axis in the image plane, depending on $\eta$. The inclination angle $\theta$ introduces a tilt to the image, shifting the observed halo and absorption patterns. Notably, in the blueshifted channel, both the emission and absorption structures rotate in the opposite direction compared to the redshifted channel, reflecting the orientation of denser ejecta regions. The Euler angle $\phi$ produces an in-plane rotation of the entire image, and this geometric asymmetry is also imprinted on the intensity correlation function observed by an optical intensity interferometer.

The intensity correlation of a ellipsoidal SN has the following properties: the intensity correlation becomes ellipsoidal with deformation occurring perpendicular to the ellipsoid’s major axis. The inclination $\theta$ shifts the image centroid, producing a phase shift in the visibility function that reverses in the redshifted/blueshifted channels. However, since intensity interferometry measures only the magnitude (modulus) of the complex visibility function, the parity of this asymmetry is not observable. The angle $\phi$ manifests as a rotation of the intensity correlation in the $u$--$v$~plane.

Comparison of figures~\ref{fig:sphere} and~\ref{fig:ellipsoid} reveals the differences between spherical and ellipsoidal SNe. Most notably, in the asymmetric case, the width of the intensity correlation varies with baseline angle in the $u$--$v$~plane, unlike the spherical case, where it is constant. However, this variation disappears when the SN is viewed along its symmetry axis (i.e.~when the image is axisymmetric).
We will discuss in detail the degeneracy between the degree of asymmetry $(\eta)$ and inclination $(\theta)$ in interpreting intensity interferometric data in section~\ref{sec:morphology} and figure~\ref{fig:morphology}.

\section{Sensitivity Forecast} 
\label{sec:parameter-estimation}
In this section, we forecast the sensitivity of an optical intensity interferometer for observing a SN IIP during its photospheric phase, using this case as a benchmark scenario. Section~\ref{sec:morphology} focuses on the sensitivity to measuring the SN’s asymmetry using the ellipsoidal model described in section~\ref{sec:asymmetry}. Section~\ref{sec:distance} turns to the uncertainty in determining the angular diameter distance via EEM, and quantifies how various supernova parameters systematically affect the inferred distance.

Throughout this section, we adopt a fiducial SN IIP observed over ten nights (60 hours total), with a photospheric radius of $10^{15}~\text{cm}$ at a distance of $3~\text{Mpc}$. This corresponds to an apparent magnitude in the visible band of $m_V = 12.0$, which is roughly the brightness of the most luminous SNe observed annually (though most are not of Type IIP). In principle, SNe of this brightness can be observed even during twilight conditions~\cite{2008A&A...481..575P}. However, in practice, securing 6 hours of usable observation time per night can be optimistic. Constraints such as instrument calibration, telescope pointing and synchronization, the target’s declination relative to the observatory latitude, and atmospheric air mass all reduce observing efficiency.

For the interferometer, we assume the following benchmark parameters: telescope collecting area $A = 25\pi~\text{m}^2$, spectral resolution $\mathcal{R} = 10^4$, timing resolution $\sigma_t = 10~\text{ps}$, and efficiency $\epsilon = 0.5$. This configuration yields a combined performance parameter of $\sqrt{\mathcal{R}/\sigma_t}A\epsilon = 1.2\times10^3~\text{m}^2~\text{ps}^{1/2}$.

\begin{figure}
    \centering
    \includegraphics[width=0.6\linewidth]{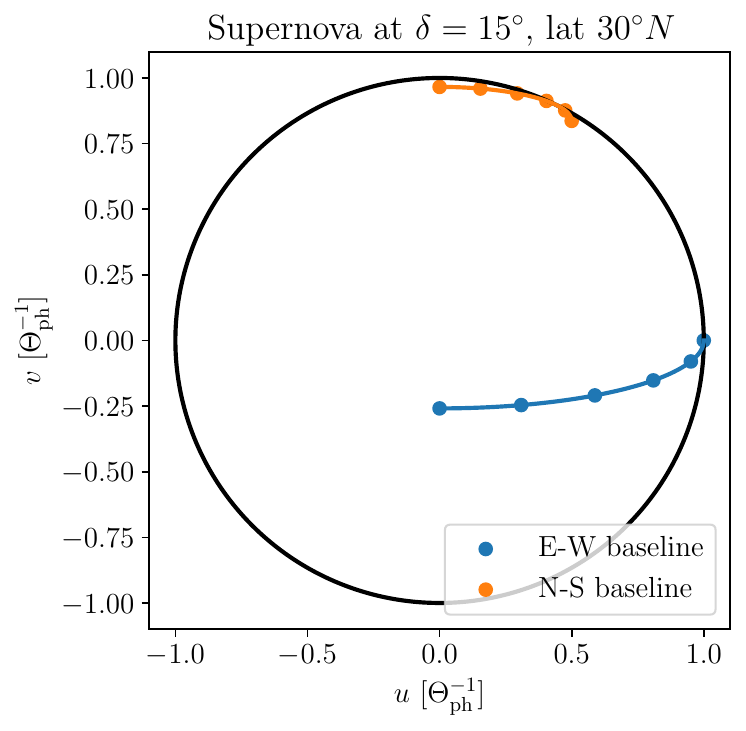}
    \caption{Tracks of two pairs of intensity interferometers with fixed baselines in the East-West and North-South directions, both located at latitude $30^\circ$N, observing a target at declination $15^\circ$ as Earth rotates. The axes $u$ and $v$ are the angular wavenumber coordinates of the Fourier transform defined in eq.~\eqref{eq:visibility_def}. We use this interferometric configuration as a realistic sample to demonstrate the sensitivity. We plot the 6-hour fiducial observation used in this paper. The optimal baseline for this setup for observing a $R_\mathrm{ph} = 10^{15}\, \rm{cm}$ supernova at $3\,\mathrm{Mpc}$ is $d = 2\,{\rm km}$. It is worth noting that with an \emph{array} of intensity interferometer telescopes, we could optimize and obtain denser information in the $u$--$v$-plane. \nblinkeem{plots/baseline.ipynb}} 
    \label{fig:baseline}
\end{figure}

Figure~\ref{fig:baseline} illustrates the interferometer's configuration in the $u$--$v$~plane. The SN is placed at a declination of $15^\circ$ with the telescope array located at a latitude of $30^\circ N$. We employ two interferometer baselines of optimal length $\sim2~\text{km}$, as justified in section~\ref{sec:IIobs}. Under this setup, the intensity correlation in each spectral bin can be measured at $\text{SNR}\sim1$. 

From observations of SN IIP, the ejecta velocity at the photosphere is typically $v_\text{ph}\lesssim0.05~c$ \cite{takats2012measuring}, which confines the P Cygni spectral features to roughly $\sim10^3$ spectral bins at $\mathcal{R} = 10^4$. This allows for a total combined signal-to-noise ratio of $\text{SNR}\sim30$.

We construct likelihood functions using both the observational data and the parametric model described in section~\ref{sec:parametric}. The total likelihood is separated into two components: $(1)$ the likelihood associated with the intensity interferometric measurements, $\mathcal{L}_\text{II}$, and $(2)$ the likelihood derived from the spatially integrated spectrum, $\mathcal{L}_\text{spectrum}$. The latter can be obtained either from the photon counts in each spectral bin in an individual telescope within the intensity interferometer array --- simultaneously obtained with the intensity correlation measurement --- or through supplementary observations conducted by dedicated spectroscopic surveys~\cite{Soumagnac:2024dhr,Anderson:2024uwn}. The combined log-likelihood function is expressed as:
\begin{equation}
    \log\mathcal{L}_\text{total} = \log\mathcal{L}_\text{II} + \log\mathcal{L}_\text{spectrum} \, .
\end{equation}
In this analysis, we assume simple Gaussian likelihoods for both components. To estimate the uncertainties on the inferred parameters, we employ the Fisher information formalism. Specifically, the Fisher information matrix is computed as the Hessian of the negative log-likelihood function, evaluated at the maximum likelihood point corresponding to the true model parameters.

We use eq.~\eqref{eq:uncertainty} to scale the uncertainty in the inferred parameters with the brightness of the supernova and the characteristics of the intensity interferometer. For the spectral likelihood, we assume a fiducial uncertainty of 10\%, following previous studies of P Cygni line profile modeling and observations~\cite{najarro1997spectroscopic, rivet2020intensity}. This accounts for systematic errors arising from both instrumental limitations and uncertainties in the spectral models.

\subsection{Morphology}
\label{sec:morphology}

The intensity interferometer provides a direct measurement of SN morphology by mapping the shape of the intensity correlation across various baseline vectors (and thus the $u$--$v$~plane), like in the simple example of eq.~\eqref{eq:complexvisibility}. As Earth rotates over the course of a night, the projected baseline naturally traces different angles in the $u$--$v$~plane, allowing a single telescope pair to sample multiple directions (as illustrated in figure~\ref{fig:baseline}). By comparing the intensity correlation maps in figures~\ref{fig:sphere} and~\ref{fig:ellipsoid}, one can see how different baseline orientations reveal structural anisotropies in the SN photosphere.

To visualize this more concretely, consider the intensity correlation of a uniform circular disk of angular radius $\Theta_\text{ph}$, which exhibits its first zero at
\begin{equation}
    |\vect{u_0}| = \frac{2\pi d_0}{\lambda} \approx2\pi \frac{1.22}{\Theta_\text{ph}} \, .
\end{equation}
For the spherical SN shown in figure~\ref{fig:sphere}, this zero point is independent of baseline angle due to axisymmetry. However, in the case of an ellipsoidal (prolate) SN with $\eta = 2$ and inclination $\theta = \pi/4$ and $\phi = 0$, as shown in figure~\ref{fig:ellipsoid}, the effective angular radius in the $u$-direction is 
\begin{equation}
    \Theta_\text{ph,eff} = \sqrt{\eta^2\cos^2\theta + \sin^2\theta}\, \Theta_\text{ph} \approx 1.58 \, \Theta_\text{ph} \, .
\end{equation}
Thus, the visibility zero shifts to $|\vect{u}|\approx7.67/1.58\,\Theta_\text{ph}^{-1}\approx 4.85\, \Theta_\text{ph}^{-1}$ in the $u$-direction.

While spectral measurements provide integrated flux information, they lack the spatial resolution to directly measure SN morphology. However, using the parametric model in section~\ref{sec:parametric}, one can infer morphological properties from the line profile. In particular, the absorption component in the P Cygni profile depends solely on the column density of the ejecta and is therefore relatively insensitive to asymmetry. The emission component, however, is affected by geometric dilution, which varies with inclination angle and asymmetry.

For example, in a prolate SN $(\eta>1)$ with the symmetry axis perpendicular to the line of sight $(\theta = 0)$, the projected size of the photosphere from the point of view of a parcel of ejecta along the semi-major axis is reduced, leading to weaker emission. Conversely, for $\theta = \pi/2$, the projected photosphere appears larger from the point of view of ejecta along the minor axes of the ellipsoid, enhancing the emission. A similar methodology has been applied to the kilonova AT2017gfo/GW170817~\cite{Sneppen:2023vkk}, where deviations from spherical symmetry were inferred by comparing transverse and line-of-sight emission, combined with an  inclination fixed from radio jet observations obtained via Hubble Space Telescope precision astrometry~\cite{mooley2022optical,mooley2018superluminal}.

Intensity interferometry removes the need for a separate measurement on the inclination $\theta$ by providing independent constraints on both the degree of asymmetry $\eta$ and the inclination angle $\theta$, in conjunction with the parametric P Cygni line model. Polarization measurements are easily incorporated into intensity interferometry and could provide further information about SN asymmetry~\cite{2008ARA&A..46..433W}; we defer discussion of polarization effects to future work.

\begin{figure}
    \centering
    \includegraphics[width=\linewidth]{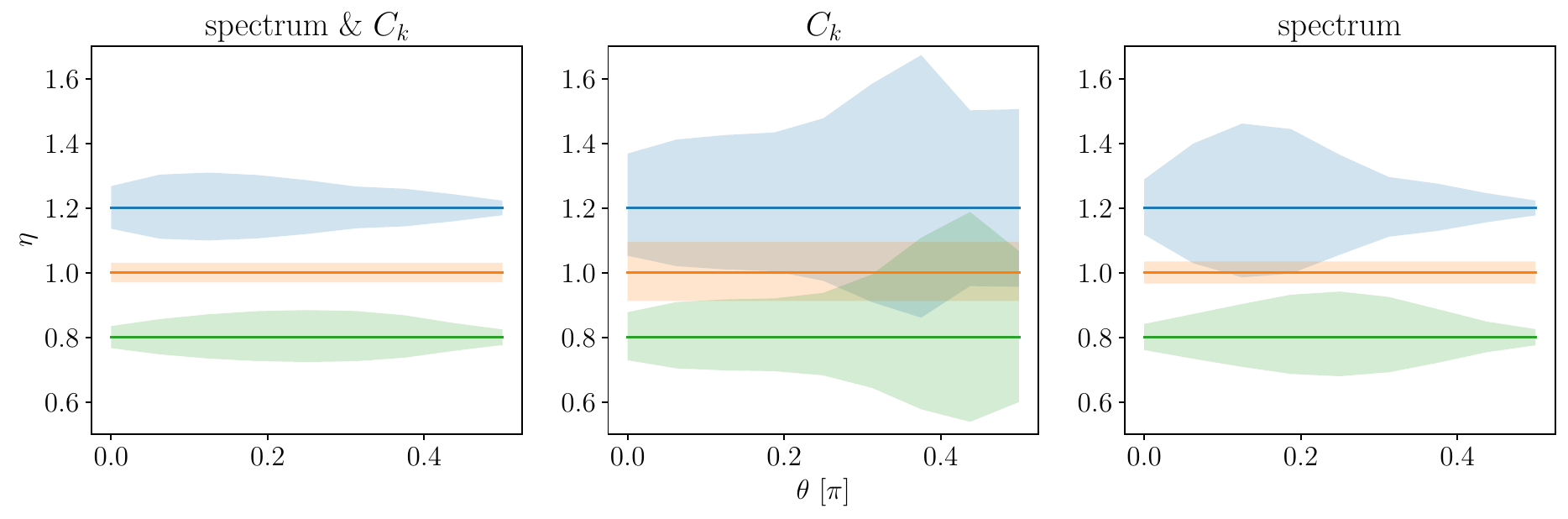}
    \caption{Fisher information analysis of shape $\eta$ inferences for prolate, spherical, and oblate SNe as a function of the Euler angle $\theta$. \textbf{Left:} uncertainties using both the intensity correlations and the integrated spectra. \textbf{Middle:} uncertainties using only the intensity correlations. \textbf{Right:} uncertainties using only the integrated spectrum. As expected, the intensity correlations contain more information when $\theta$ is close to zero than at $\pi/2$. \nblinkeem{plots/eta_theta.ipynb}}
    \label{fig:morphology}
\end{figure}

\begin{figure}
    \centering
    \includegraphics[width=\linewidth]{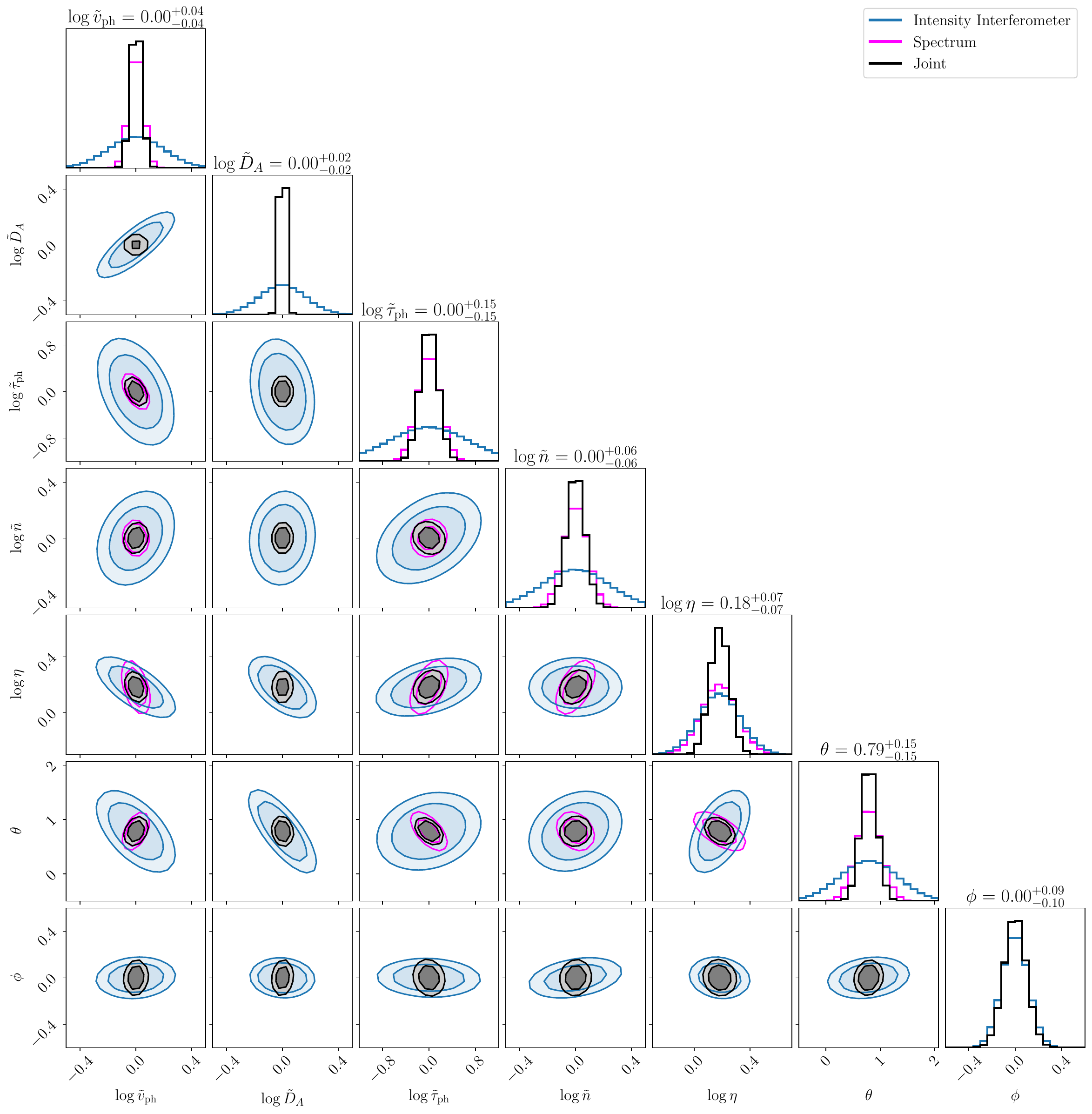}
    \caption{Corner plot of the Fisher information in intensity correlation (blue), the spectrum (magenta), and their combination (black). The tilde parameters in the labels are short-hand notation for $\tilde{x} = x/x_\text{truth}$. The spectrum does \emph{not} contain information about the distance $D_A$ (because we erase information about the overall amplitude by marginalizing over $\mathcal{N}$) and the rotation in the image plane $\phi$. As can be seen in the $v_\text{ph}$--$D_A$ corner plot, intensity interferometry primarily measures the angular size of the SN, leading to an approximate degeneracy of  $v_\text{ph} \propto D_A$.  The spectrum contains more information about the ejecta velocities. The joint information can constrain the distance to a mag 12 SN to $\sim 2\%$ with 10 nights of observation, assuming observatory specifications described in this section. The $\eta$--$\theta$ panel shows that intensity interferometric and spectroscopic measurements have different approximate degeneracy directions and thus complement each other. The combined precision on $\eta$ is $\sim7\%$. The total signal-to-noise is $\text{SNR}_\text{total}\sim30$. The truth parameters are $v_\text{ph} = 6\times10^3 \, \text{km/s}~D_A = 3 \, \text{Mpc},~\tau_\text{ph} = 2,~n = 4,~\eta = 1.2,~\theta = \pi/4,~\text{and}~\phi = 0$. 
    \nblinkeem{plots/corner_fisher.ipynb}}
    \label{fig:corner}
\end{figure}

Figure~\ref{fig:morphology} illustrates how the uncertainty in $\eta$ depends on the inclination $\theta$, for three benchmark values: $\eta = 0.8, 1.0, 1.2$, representing oblate, spherical, and prolate shapes respectively. These values are motivated by both polarization observations~\cite{2008ARA&A..46..433W} and recent hydrodynamical simulations~\cite{Vartanyan:2024bhx}. The total uncertainty is shown as well as contributions from intensity interferometry and spectroscopy alone.
The center panel shows that the uncertainty in $\eta$ from intensity interferometry is minimized $(\sim10\%)$ when $\theta = 0$, and increases up to $\sim30\%$ at $\theta = \pi/2$, where the SN appears axisymmetric. This is expected, as morphological asymmetry is best constrained when the intensity correlation can be sampled across directions with the largest contrast. The right panel shows that the spectroscopic uncertainty is minimized at $\theta=\pi/2$, where differences between emission and absorption are most pronounced. This reflects the geometry-dependent dilution effect discussed earlier and is consistent with findings in ref.~\cite{Sneppen:2023vkk}. As shown in figure~\ref{fig:corner}, there exists a mild degeneracy between $\eta$ and $\theta$, especially near configurations where emission and absorption contributions become comparable. In particular, the largest uncertainty $(\sim30\%)$ occurs when $\theta$ is such that the ratio of emission to absorption becomes insensitive to changes in $\eta$. 
These results underscore the complementary nature of spectroscopic and intensity interferometric observations in constraining SN morphology, a fact especially apparent in the $\eta$--$\theta$ panel in figure~\ref{fig:corner}.


It is important to note that these forecasts depend on how well our simple parametric model describes real SNe. For instance, recombination emission has been observed in many Type II SN spectra~\cite{KirshnerKwan}. As discussed further in appendix~\ref{app:recomb}, we model this effect as independent scaling factors for emission and absorption optical depths. Crucially, this excess emission component is independently measurable from both the spectrum and intensity correlation, and in our model, it is not degenerate with the asymmetry parameter $\eta$.

\subsection{Angular diameter distance}
\label{sec:distance}

This section demonstrates how intensity interferometry can be used to directly measure the angular diameter distance to SNe IIP, addressing several limitations of traditional spectroscopic methods such as EPM. We begin by describing how  intensity interferometry overcomes the systematic uncertainties inherent to the EPM by spatially resolving the photosphere. We then use a parametric model to simulate observations at multiple epochs during the plateau phase of a SN IIP, highlighting how the evolving line velocities provide a direct handle on distance. Finally, we quantify the achievable precision and outline how these measurements impact the construction of the cosmic distance ladder.

As discussed in section~\ref{sec:unresolved}, EPM-based SN distance inference suffers from numerous systematic uncertainties. These include assumptions about the geometry of the photosphere, deviations from blackbody radiation, dust attenuation, and the challenge of mapping the velocity of the P Cygni spectral feature to that of the photosphere. Most of these uncertainties cannot be directly addressed with spectroscopic data alone, as there are numerous degenerate directions (caused by integration over the image plane).

Intensity interferometry overcomes many of these challenges by directly resolving the angular distribution of the spectral radiance across many spectral channels. It can provide a model-independent measurement of the angular size of the SN without relying on blackbody assumptions. It also directly probes morphological asymmetries via the directional width of the intensity correlation across different baselines. Furthermore, it disentangles emission and absorption components within the P Cygni profile by analyzing changes in the intensity correlation across narrow spectral channels, thereby resolving the emission and absorption regions in velocity space.

To demonstrate the utility of intensity interferometry in tracking the evolution of a SN IIP during its plateau phase, we model a SN observed at days 30, 45, and 60 post-explosion. During this plateau phase, the photosphere is assumed to remain at a constant radius, while ejecta continue to flow outward. Using the parametric model introduced in section~\ref{sec:parametric}, we simulate observations at these epochs to probe progressively deeper layers within the SN.

\begin{figure}
    \centering
    \includegraphics[width=0.7\linewidth]{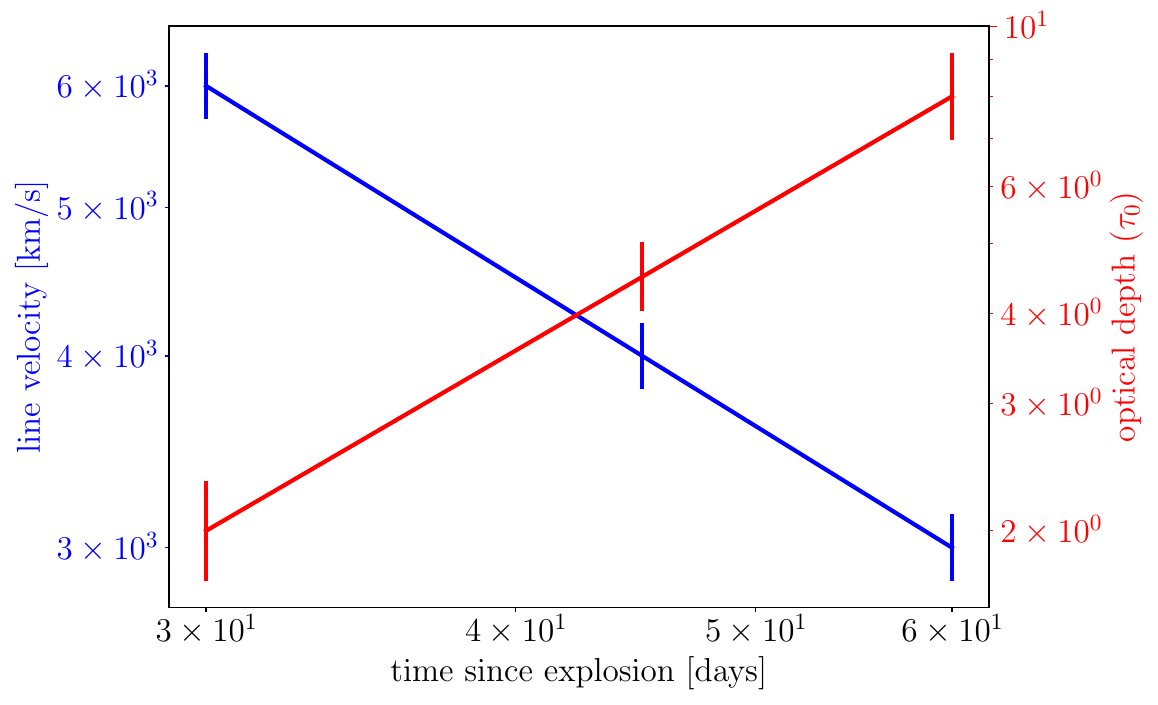}
    \caption{Fisher Information of a SN IIP measured with the intensity interferometer specified in this section at 3 different time points $t = 30, 45, 60~\text{day}$ with parameters $v_\text{ph} = 6000, 4500, 3000~\text{km/s}$ and $\tau_\text{ph} = 2.0, 4.5, 8.0$. We demonstrated that the intensity interferometer can measure the evolution of the SN IIP when it is at the plateau phase as it measures the changes in ejecta velocity at the photosphere and the optical depth even though the photosphere remains constant in space. \nblinkeem{plots/v_tau.ipynb}}
    \label{fig:tauv}
\end{figure}

If we assume that the number of atoms in the lower energy level remains constant during the expansion, then continuity implies that $n_l\propto(t-t_0)^{\alpha - 3}$ (see eq.~\eqref{eq:alpha_beta}). For $\alpha>3$, the atomic density at the photosphere increases over time. Figure~\ref{fig:tauv} shows the relevant parameters used, assuming that the optical depth at the photosphere evolves as $\tau_\text{ph}\propto (t-t_0)^2$, consistent with previous modeling efforts~\cite{castor1979atlas}. Appendix~\ref{app:spectral_index} explores how variations in $n = \alpha + \beta$ impact the measurement precision. While the power-law density profile is a simplification, figure~\ref{fig:tauv} demonstrates that the optical depth profile can, in principle, be extracted from the measured intensity correlation.

It is worth emphasizing that in our model, the photosphere is  held stationary to highlight the fact that EEM measures the velocity of expanding ejecta material --- not the velocity of the photosphere itself. Although the photosphere in a realistic SN will not be exactly static over time, EEM measures the velocity of the ejecta newly emerging from the photosphere. Importantly, this velocity --- shown in figure~\ref{fig:tauv} --- is the same as that inferred from spectral P Cygni profiles. Thus, the EEM approach eliminates the need to assume a relation between the ejecta velocities and photospheric surface velocity, a critical limitation in traditional EPM as discussed in section~\ref{sec:unresolved}.

Figure~\ref{fig:corner} shows corner plots of the Fisher information from intensity interferometry alone, spectroscopy alone, and a joint likelihood analysis combining both. Spectral data alone, marginalized over the normalization $\mathcal{N}$, lack information about the angular diameter distance $D_A$ and image-plane orientation $\phi$. In contrast, intensity interferometry provides an angular size measurement, leading to a degeneracy in the $v_\text{ph}$--$D_A$~plane, with the constraint $v_\text{ph}/D_A = \text{constant}$. Intensity interferometry alone measures this combination of parameters to $\sim9\%$ from figure~\ref{fig:corner}. Spectroscopy, while more precise in measuring $v_\text{ph}$, gains additional power when combined with angular size measurements to break this degeneracy, enabling precise distance determination. 

With 10 nights of observation and the telescope specifications described earlier, we estimate that the distance to a magnitude 12 SN can be measured to $\sim2\%$ precision using the EEM. The $\eta$--$\theta$~panel in figure~\ref{fig:corner} further illustrates how interferometric and spectroscopic methods probe different degeneracy directions. Intensity interferometry measures asymmetry through the projected aspect ratio $\eta_\text{eff}$ --- the degree of asymmetry projected onto the image plane --- $\eta_\text{eff}^2 = \eta^2\cos^2\theta + \sin^2\theta$, which appears as the blue contours in the plot. Spectroscopy, by contrast, constrains the line-of-sight component of the velocity via $v_\parallel^2 = v_\text{ph}^2~(\eta^2\sin^2\theta + \cos^2\theta)$, leading to a distinct, orthogonal degeneracy direction.

These complementary constraints from spectroscopy and intensity interferometry allow for simultaneous inference of SN asymmetry, inclination, velocity, and distance, improving parameter precision significantly. In particular, the joint analysis achieves an uncertainty of $\sim7\%$ on the asymmetry parameter $\eta$ using the interferometric configuration described in this section.

According to the Fisher formalism, the uncertainty in angular diameter distance $\sigma_D$ scales with the measurement uncertainty as given by eq.~\eqref{eq:uncertainty}. Applying this scaling to our case yields:
\begin{equation}
\begin{split}
    \frac{\sigma_D}{D} &\propto \left(\frac{N_0}{N}\right)\left(\frac{t_{\text{obs},0}}{t_\text{obs}}\right)^{-1/2}\\
    & = 2\%\cdot10^{0.4(m_V - 12)}\left(\frac{\text{60 hrs}}{t_\text{obs}}\right)^{1/2}\left(\frac{25\pi~\text{m}^2}{A}\right)\left(\frac{\sigma_t}{10~\text{ps}}\right)^{1/2}\left(\frac{10^4}{\mathcal{R}}\right)^{1/2}\left(\frac{0.5}{\epsilon}\right) \, ,
\end{split}
\end{equation}
where $m_V$ is the apparent visible magnitude of the SN. This equation quantifies the precision of the angular diameter distance measurement and is utilized in our companion paper~\cite{eem-2} to forecast the uncertainty on Hubble constant inferences employing the EEM technique on populations of SNe.

\section{Generalizations and improvements} 
\label{sec:discussion}

We have mainly investigated how optical intensity interferometry can be used to measure the morphology and distance of a SN from the wavelength-dependent intensity correlations ($\propto |\mathcal{V}|^2)$ around spectral lines. Another qualitative feature that has already been utilized to measure SN shape is spectropolarimetry (see ref.~\cite{2008ARA&A..46..433W} for a review). Light scattering with electrons in an aspherically expanding envelope of ejecta material leads to intrinsic linear polarization~\cite{shakhovskoi1976optical,shapiro1982polarization}. Follow-up studies suggest that a SN atmosphere with both scattering and absorption will have a higher degree of linear polarization, and that the lines forming P Cygni scattering profiles have increased polarization at the absorption minimum~\cite{mccall1984supernovae}. These early studies led to an observational campaign that found linear polarization from core-collapse SNe and demonstrated that they generically have substantial asphericity~\cite{Leonard:2000vh,2008ARA&A..46..433W}. 

\begin{figure}
    \centering
% Gradient Info
  
\tikzset {_1v7sl5lv5/.code = {\pgfsetadditionalshadetransform{ \pgftransformshift{\pgfpoint{0 bp } { 0 bp }  }  \pgftransformscale{1 }  }}}
\pgfdeclareradialshading{_m2gxkf92g}{\pgfpoint{0bp}{0bp}}{rgb(0bp)=(0.96,0.65,0.14);
rgb(11.875bp)=(0.96,0.65,0.14);
rgb(25bp)=(1,1,1);
rgb(400bp)=(1,1,1)}

% Gradient Info
  
\tikzset {_0qvbzrkzb/.code = {\pgfsetadditionalshadetransform{ \pgftransformshift{\pgfpoint{0 bp } { 0 bp }  }  \pgftransformscale{1 }  }}}
\pgfdeclareradialshading{_2yx3zwvoy}{\pgfpoint{0bp}{0bp}}{rgb(0bp)=(0.96,0.65,0.14);
rgb(11.875bp)=(0.96,0.65,0.14);
rgb(25bp)=(1,1,1);
rgb(400bp)=(1,1,1)}
\tikzset{every picture/.style={line width=0.75pt}} %set default line width to 0.75pt        

\begin{tikzpicture}[x=0.75pt,y=0.75pt,yscale=-1,xscale=1]

%Shape: Ellipse [id:dp5523394632944416] 
\draw  [draw opacity=0][shading=_m2gxkf92g,_1v7sl5lv5][dash pattern={on 4.5pt off 4.5pt}] (300.7,139.9) .. controls (300.77,92.62) and (350.31,54.37) .. (411.37,54.45) .. controls (472.42,54.54) and (521.87,92.93) .. (521.8,140.21) .. controls (521.73,187.48) and (472.19,225.74) .. (411.13,225.65) .. controls (350.08,225.57) and (300.63,187.18) .. (300.7,139.9) -- cycle ;
%Shape: Ellipse [id:dp9056726775340112] 
\draw  [draw opacity=0][shading=_2yx3zwvoy,_0qvbzrkzb][dash pattern={on 4.5pt off 4.5pt}] (96.58,139.9) .. controls (96.65,92.62) and (135.02,54.35) .. (182.3,54.42) .. controls (229.58,54.49) and (267.85,92.86) .. (267.78,140.14) .. controls (267.71,187.41) and (229.34,225.69) .. (182.06,225.62) .. controls (134.79,225.55) and (96.51,187.18) .. (96.58,139.9) -- cycle ;
%Straight Lines [id:da07190503811079973] 
\draw    (148.3,90.6) -- (216.3,90.6) ;
\draw [shift={(219.3,90.6)}, rotate = 180] [fill={rgb, 255:red, 0; green, 0; blue, 0 }  ][line width=0.08]  [draw opacity=0] (8.93,-4.29) -- (0,0) -- (8.93,4.29) -- cycle    ;
\draw [shift={(145.3,90.6)}, rotate = 0] [fill={rgb, 255:red, 0; green, 0; blue, 0 }  ][line width=0.08]  [draw opacity=0] (8.93,-4.29) -- (0,0) -- (8.93,4.29) -- cycle    ;
%Straight Lines [id:da8671006945692108] 
\draw    (148.3,189.2) -- (183.3,189.2) -- (216.3,189.2) ;
\draw [shift={(219.3,189.2)}, rotate = 180] [fill={rgb, 255:red, 0; green, 0; blue, 0 }  ][line width=0.08]  [draw opacity=0] (8.93,-4.29) -- (0,0) -- (8.93,4.29) -- cycle    ;
\draw [shift={(145.3,189.2)}, rotate = 0] [fill={rgb, 255:red, 0; green, 0; blue, 0 }  ][line width=0.08]  [draw opacity=0] (8.93,-4.29) -- (0,0) -- (8.93,4.29) -- cycle    ;
%Straight Lines [id:da08626517678007928] 
\draw    (133,105.9) -- (133,173.9) ;
\draw [shift={(133,176.9)}, rotate = 270] [fill={rgb, 255:red, 0; green, 0; blue, 0 }  ][line width=0.08]  [draw opacity=0] (8.93,-4.29) -- (0,0) -- (8.93,4.29) -- cycle    ;
\draw [shift={(133,102.9)}, rotate = 90] [fill={rgb, 255:red, 0; green, 0; blue, 0 }  ][line width=0.08]  [draw opacity=0] (8.93,-4.29) -- (0,0) -- (8.93,4.29) -- cycle    ;
%Straight Lines [id:da9427336025322093] 
\draw    (231.6,105.9) -- (231.6,173.9) ;
\draw [shift={(231.6,176.9)}, rotate = 270] [fill={rgb, 255:red, 0; green, 0; blue, 0 }  ][line width=0.08]  [draw opacity=0] (8.93,-4.29) -- (0,0) -- (8.93,4.29) -- cycle    ;
\draw [shift={(231.6,102.9)}, rotate = 90] [fill={rgb, 255:red, 0; green, 0; blue, 0 }  ][line width=0.08]  [draw opacity=0] (8.93,-4.29) -- (0,0) -- (8.93,4.29) -- cycle    ;
%Straight Lines [id:da3246916649323137] 
\draw    (190.42,103.02) -- (219.18,131.78) ;
\draw [shift={(221.3,133.9)}, rotate = 225] [fill={rgb, 255:red, 0; green, 0; blue, 0 }  ][line width=0.08]  [draw opacity=0] (8.93,-4.29) -- (0,0) -- (8.93,4.29) -- cycle    ;
\draw [shift={(188.3,100.9)}, rotate = 45] [fill={rgb, 255:red, 0; green, 0; blue, 0 }  ][line width=0.08]  [draw opacity=0] (8.93,-4.29) -- (0,0) -- (8.93,4.29) -- cycle    ;
%Straight Lines [id:da5452390910103797] 
\draw    (174.18,103.02) -- (145.42,131.78) ;
\draw [shift={(143.3,133.9)}, rotate = 315] [fill={rgb, 255:red, 0; green, 0; blue, 0 }  ][line width=0.08]  [draw opacity=0] (8.93,-4.29) -- (0,0) -- (8.93,4.29) -- cycle    ;
\draw [shift={(176.3,100.9)}, rotate = 135] [fill={rgb, 255:red, 0; green, 0; blue, 0 }  ][line width=0.08]  [draw opacity=0] (8.93,-4.29) -- (0,0) -- (8.93,4.29) -- cycle    ;
%Straight Lines [id:da37953021824763034] 
\draw    (174.18,176.78) -- (145.42,148.02) ;
\draw [shift={(143.3,145.9)}, rotate = 45] [fill={rgb, 255:red, 0; green, 0; blue, 0 }  ][line width=0.08]  [draw opacity=0] (8.93,-4.29) -- (0,0) -- (8.93,4.29) -- cycle    ;
\draw [shift={(176.3,178.9)}, rotate = 225] [fill={rgb, 255:red, 0; green, 0; blue, 0 }  ][line width=0.08]  [draw opacity=0] (8.93,-4.29) -- (0,0) -- (8.93,4.29) -- cycle    ;
%Straight Lines [id:da7362893970424282] 
\draw    (219.18,148.02) -- (190.42,176.78) ;
\draw [shift={(188.3,178.9)}, rotate = 315] [fill={rgb, 255:red, 0; green, 0; blue, 0 }  ][line width=0.08]  [draw opacity=0] (8.93,-4.29) -- (0,0) -- (8.93,4.29) -- cycle    ;
\draw [shift={(221.3,145.9)}, rotate = 135] [fill={rgb, 255:red, 0; green, 0; blue, 0 }  ][line width=0.08]  [draw opacity=0] (8.93,-4.29) -- (0,0) -- (8.93,4.29) -- cycle    ;
%Straight Lines [id:da7830783304538511] 
\draw    (377.32,90.4) -- (445.32,90.4) ;
\draw [shift={(448.32,90.4)}, rotate = 180] [fill={rgb, 255:red, 0; green, 0; blue, 0 }  ][line width=0.08]  [draw opacity=0] (8.93,-4.29) -- (0,0) -- (8.93,4.29) -- cycle    ;
\draw [shift={(374.32,90.4)}, rotate = 0] [fill={rgb, 255:red, 0; green, 0; blue, 0 }  ][line width=0.08]  [draw opacity=0] (8.93,-4.29) -- (0,0) -- (8.93,4.29) -- cycle    ;
%Straight Lines [id:da4215881945170177] 
\draw    (377.32,189) -- (412.32,189) -- (445.32,189) ;
\draw [shift={(448.32,189)}, rotate = 180] [fill={rgb, 255:red, 0; green, 0; blue, 0 }  ][line width=0.08]  [draw opacity=0] (8.93,-4.29) -- (0,0) -- (8.93,4.29) -- cycle    ;
\draw [shift={(374.32,189)}, rotate = 0] [fill={rgb, 255:red, 0; green, 0; blue, 0 }  ][line width=0.08]  [draw opacity=0] (8.93,-4.29) -- (0,0) -- (8.93,4.29) -- cycle    ;
%Straight Lines [id:da7744833705332708] 
\draw    (362.02,105.7) -- (362.02,173.7) ;
\draw [shift={(362.02,176.7)}, rotate = 270] [fill={rgb, 255:red, 0; green, 0; blue, 0 }  ][line width=0.08]  [draw opacity=0] (8.93,-4.29) -- (0,0) -- (8.93,4.29) -- cycle    ;
\draw [shift={(362.02,102.7)}, rotate = 90] [fill={rgb, 255:red, 0; green, 0; blue, 0 }  ][line width=0.08]  [draw opacity=0] (8.93,-4.29) -- (0,0) -- (8.93,4.29) -- cycle    ;
%Straight Lines [id:da35885273965719433] 
\draw    (460.62,105.7) -- (460.62,173.7) ;
\draw [shift={(460.62,176.7)}, rotate = 270] [fill={rgb, 255:red, 0; green, 0; blue, 0 }  ][line width=0.08]  [draw opacity=0] (8.93,-4.29) -- (0,0) -- (8.93,4.29) -- cycle    ;
\draw [shift={(460.62,102.7)}, rotate = 90] [fill={rgb, 255:red, 0; green, 0; blue, 0 }  ][line width=0.08]  [draw opacity=0] (8.93,-4.29) -- (0,0) -- (8.93,4.29) -- cycle    ;
%Straight Lines [id:da8257726798707905] 
\draw    (419.44,102.82) -- (448.2,131.58) ;
\draw [shift={(450.32,133.7)}, rotate = 225] [fill={rgb, 255:red, 0; green, 0; blue, 0 }  ][line width=0.08]  [draw opacity=0] (8.93,-4.29) -- (0,0) -- (8.93,4.29) -- cycle    ;
\draw [shift={(417.32,100.7)}, rotate = 45] [fill={rgb, 255:red, 0; green, 0; blue, 0 }  ][line width=0.08]  [draw opacity=0] (8.93,-4.29) -- (0,0) -- (8.93,4.29) -- cycle    ;
%Straight Lines [id:da0992441856365518] 
\draw    (403.2,102.82) -- (374.44,131.58) ;
\draw [shift={(372.32,133.7)}, rotate = 315] [fill={rgb, 255:red, 0; green, 0; blue, 0 }  ][line width=0.08]  [draw opacity=0] (8.93,-4.29) -- (0,0) -- (8.93,4.29) -- cycle    ;
\draw [shift={(405.32,100.7)}, rotate = 135] [fill={rgb, 255:red, 0; green, 0; blue, 0 }  ][line width=0.08]  [draw opacity=0] (8.93,-4.29) -- (0,0) -- (8.93,4.29) -- cycle    ;
%Straight Lines [id:da9389246973750913] 
\draw    (403.2,176.58) -- (374.44,147.82) ;
\draw [shift={(372.32,145.7)}, rotate = 45] [fill={rgb, 255:red, 0; green, 0; blue, 0 }  ][line width=0.08]  [draw opacity=0] (8.93,-4.29) -- (0,0) -- (8.93,4.29) -- cycle    ;
\draw [shift={(405.32,178.7)}, rotate = 225] [fill={rgb, 255:red, 0; green, 0; blue, 0 }  ][line width=0.08]  [draw opacity=0] (8.93,-4.29) -- (0,0) -- (8.93,4.29) -- cycle    ;
%Straight Lines [id:da7494138884006627] 
\draw    (448.2,147.82) -- (419.44,176.58) ;
\draw [shift={(417.32,178.7)}, rotate = 315] [fill={rgb, 255:red, 0; green, 0; blue, 0 }  ][line width=0.08]  [draw opacity=0] (8.93,-4.29) -- (0,0) -- (8.93,4.29) -- cycle    ;
\draw [shift={(450.32,145.7)}, rotate = 135] [fill={rgb, 255:red, 0; green, 0; blue, 0 }  ][line width=0.08]  [draw opacity=0] (8.93,-4.29) -- (0,0) -- (8.93,4.29) -- cycle    ;
%Straight Lines [id:da8229531130185956] 
\draw    (344.02,100.7) -- (344.02,178.7) ;
\draw [shift={(344.02,181.7)}, rotate = 270] [fill={rgb, 255:red, 0; green, 0; blue, 0 }  ][line width=0.08]  [draw opacity=0] (8.93,-4.29) -- (0,0) -- (8.93,4.29) -- cycle    ;
\draw [shift={(344.02,97.7)}, rotate = 90] [fill={rgb, 255:red, 0; green, 0; blue, 0 }  ][line width=0.08]  [draw opacity=0] (8.93,-4.29) -- (0,0) -- (8.93,4.29) -- cycle    ;
%Straight Lines [id:da4258359287041571] 
\draw    (478.62,100.7) -- (478.62,178.7) ;
\draw [shift={(478.62,181.7)}, rotate = 270] [fill={rgb, 255:red, 0; green, 0; blue, 0 }  ][line width=0.08]  [draw opacity=0] (8.93,-4.29) -- (0,0) -- (8.93,4.29) -- cycle    ;
\draw [shift={(478.62,97.7)}, rotate = 90] [fill={rgb, 255:red, 0; green, 0; blue, 0 }  ][line width=0.08]  [draw opacity=0] (8.93,-4.29) -- (0,0) -- (8.93,4.29) -- cycle    ;

\end{tikzpicture}
\caption{A sketch of the polarization magnitude and direction of an angularly resolved SN. The black arrows indicate the direction and magnitude of light polarization, which arises from interactions between light and the ejecta material around the supernova (indicated by the orange shading).  For an unresolved source, only the net polarization direction and magnitude can be observable, and an ellipsoidal SN (right panel) will appear linearly polarized. For a resolved SN, the E-mode polarization pattern can be directly measured. The amplitude of this E-mode polarization can be used to measure the elongation in the line of sight direction (see more details in the main text).}
    \label{fig:polarization}
\end{figure}
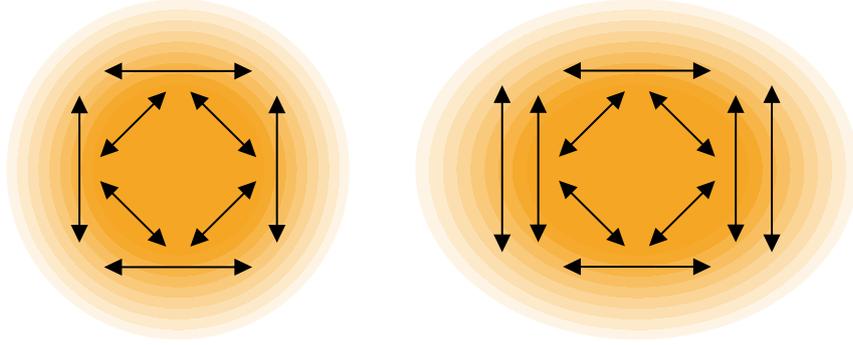

The origin of this net linear polarization of an aspherical SN is shown in figure~\ref{fig:polarization} (adapted from ref.~\cite{Leonard:1999xp}). Whereas the linearly polarized light from different parts of a spherical SN would cancel due to axisymmetry, for an aspherical SN, the cancellation is imperfect, and a net linear polarization remains. 
The linear polarization direction and fraction, and their time evolution since the explosion of the SN, has been used to study the evolution of the shape of a SN, especially SN1987A~\cite{1991ApJ...375..264J,arnett1989supernova}. 

More information can be obtained via intensity interferometry separated into two distinct polarization channels.
The most important additional feature, sketched in figure~\ref{fig:polarization}, is the ``E-mode'' polarization pattern around the SN (using terminology borrowed from literature of cosmic microwave background polarization). 
The amplitude of the E-mode polarization (relative to the total intensity) contains information about the SN elongation along the line of sight.
We expect intensity interferometric polarimetry can augment the measurement of the three geometric parameters in our simple parametric model. The degree of asymmetry $\eta$ and the two Euler angles can be extracted from the fraction and direction of the linear polarization and the fraction of the E-mode polarization.
The intensity correlation can be measured for the two linear polarization directions independently, and subsequently combined to extract the amplitude of the modes corresponding to different polarization patterns. 
These measurements would allow us to measure the other polarization patterns in the multipole expansion,  which can be used to measure properties of the ejecta beyond our simple parametric model.
We leave a detailed study of spectral-multiplexed \emph{polarimetric} intensity interferometry on SNe to future work. 

For a realistic SN, spectral features may overlap~\cite{2017hsn..book..769S}. This provides both a challenge and an opportunity. On one hand, this makes it hard to independently determine the photosphere radius (see figure~\ref{fig:abs}). When the P Cygni profiles of different spectral lines $i$ and $j$ overlap:
\begin{equation}
    \lambda \simeq \lambda^i_{\rm rest} (1+v^i_{\rm ej,\parallel}) \simeq \lambda^j_{\rm rest} (1+v^j_{\rm ej,\parallel}) \, ,
\end{equation}
the intensity correlations will also contain contributions from both spectral lines. These effects need to be taken into account in a parametric model that is used to extract both the morphology of and the distance to a SN from an observation. On the other hand, comparing the intensity correlations measured at different spectral lines can provide valuable information about the velocity distribution of the ejecta material at different times after the explosion, and how these distributions vary for different atomic elements. For example, a question that could be addressed is whether the distribution of certain elements in the ejecta are more spherically symmetric than the photosphere.

In this work, we focused on the EEM's ability to break degeneracies constituting the dominant systematic uncertainties of the original EPM in the simple parametric model of a Type IIP SN in section~\ref{sec:parametric} and~\ref{sec:parameter-estimation}. However, this methodology can also be employed on different types of SNe (see recent work on Type Ia SNe~\cite{Kim:2025ggf}) to extract spatial features of interest. Effects that we hope to include in future studies include limb darkening, relativistic beaming~\cite{1990ApJ...361..367H}, and recombination emission~\cite{KirshnerKwan}. 
We leave a detailed study that incorporates these more sophisticated effects in a Fisher analysis and a MCMC pipeline to future work.  

In this study, we considered an intensity interferometer consisting of a pair of telescopes separated by a baseline that is optimized to resolve the angular size of a SN. Longer baselines can be used to 
resolve smaller-scale features of the image. With an array of telescopes with different baseline lengths and orientations, the coefficients of a multipole expansion of  $\left|\mathcal{V}\left(\overline{\lambda},\vect{u} \right)\right|^2$ can be determined without reference to a parametric model. This would provide valuable information about the SN shape, chemical composition, and velocity distribution of the ejecta material.  
Machine-learning techniques~\cite{IIAI} could also be utilized to optimize the intensity interferometer array for studying SN morphology and distances, and to improve parameter estimation and image reconstruction based on dedicated numerical simulations.

We stress that our parametric model of SN provides a proof-of-concept for using intensity interferometers to resolve SNe and measure their morphology and distance. For more accurate modeling of SNe, radiative transfer simulation codes such as \texttt{TARDIS}~\cite{kerzendorf2014spectral}, \texttt{ARTIS}~\cite{sim2007multidimensional,kromer2009time}, and \texttt{SEDONA}~\cite{kasen2006time} could be used to  repeat the analysis in this work, and to explore the effects of different SN models on the measurement uncertainty projections.

\section{Conclusions} \label{sec:conclusions}

We have introduced the expanding ejecta method (EEM), a novel approach for a geometric determination of the morphology and angular diameter distance of SNe using spectrally-multiplexed intensity interferometry. This technique leverages the expanding line-emitting and line-absorbing ejecta material --- largely independent of the photosphere --- to determine the SN's shape, orientation, and line-of-sight distance. By focusing on the H$\alpha$ spectral feature, we demonstrated how intensity interferometry can resolve both transverse and radial velocity distributions of the ejecta. A SN of apparent magnitude $12$ or brighter (at peak light) is expected to occur roughly annually. Such a Type IIP SN will be $\sim 3 \,{\rm Mpc}$ away from Earth, with an angular radius of $\sim 10^{-10}\,{\rm rad} \approx 20\, {\rm \mu as}$. With baselines of $\sim 2\,{\rm km}$, optical intensity interferometers can spatially resolve such explosions, enabling precise measurements of their evolving angular structure.

To forecast the power of EEM, we developed a simple yet flexible parametric model of a spheroidal SN with homologously expanding ejecta surrounding a static photosphere. The model includes three geometric parameters: the elongation $\eta$ and two Euler angles, $\theta$ and $\phi$ (see table~\ref{tab:model}). Using combined measurements of the SN spectrum and intensity correlations, these parameters can be inferred to $\sim 10\%$ precision for both oblate and prolate morphologies, assuming benchmark specifications for the intensity interferometer of $A = 25\pi\,\mathrm{m}^2$, $\mathcal{R} = 10^4$, $\sigma_t = 10\,\text{ps}$, and $\epsilon = 0.5$, or equivalently, a figure of merit $\sqrt{\mathcal{R}/\sigma_t}A\epsilon = 1.2\times10^3~\text{m}^2~\text{ps}^{1/2}$ over a 60-hour integration. We identified mild degeneracies between $\eta$ and $\theta$ when using either spectral or interferometric data alone, but these are cleanly broken when both datasets are jointly analyzed, as illustrated in figures~\ref{fig:morphology} and~\ref{fig:corner}.

Crucially, accurate inference of SN morphology enables precise angular diameter distance measurements. By tracking the evolving angular size and line-of-sight velocity distribution of the same ejecta material, EEM achieves a fractional uncertainty of $\sim 2\%$ in the distance to a $m = 12$ Type IIP SN (figure~\ref{fig:corner}). This level of precision is already sufficient to support competitive cosmological applications~\cite{eem-2}. With rapid advances in intensity interferometry, we foresee further improvements in precision and extensions to fainter SNe. 

The basic parametric model presented here can be systematically extended to incorporate additional physical and geometric complexities, such as recombination radiation, limb darkening, polarization, spectropolarimetry, and spherical harmonic expansions of the ejecta geometry. These generalizations will be essential for robustly disentangling the physical characteristics of SNe, and for demonstrating that the EEM provides an unbiased, direct measurement of SN distances across a wide range of conditions.

The application of the EEM to moderate-distance SNe ($D_A \lesssim 100\,\mathrm{Mpc}$) opens the door to independent calibrations of the cosmic distance ladder. With more capable intensity interferometers, SNe at larger distances ($D_A \gtrsim 100\,\mathrm{Mpc}$) come into view, enabling geometric distance measurements well within the Hubble flow, and thus a direct inference of the Hubble constant $H_0$. These broader cosmological implications are explored in our companion paper~\cite{eem-2}.

\paragraph*{Note Added:} This method was first presented by I-Kai Chen in the workshop on \href{https://events.perimeterinstitute.ca/event/347/contributions/1397/}{Future Prospects of Intensity Interferometry}. While this work was being completed, a similar approach for Type Ia SNe was independently proposed in ref.~\cite{Kim:2025ggf}, which incorporates more sophisticated emission profiles generated by two spectral synthesis code packages.

\acknowledgments
We thank Masha Baryakhtar, Lars Bildsten, Neal Dalal, Marios Galanis, Jared Goldberg, and Yong-Zhong Qian for helpful discussions.
This material is based upon work supported by the National Science Foundation under Grant No.~PHY-2210551. %KVT grant 
IC is supported by the James Arthur Graduate Associate Fellowship and the NYU GSAS Dissertation Writing Fellowship.
DD is supported by the James Arthur Postdoctoral Fellowship.
JH is grateful for the hospitality of New York University and Center for Computational Astrophysics (CCA) at the Flatiron Institute, where part of this work was carried out.

Research at Perimeter Institute is supported in part by the Government of Canada through the Department of Innovation, Science and Economic Development and by the Province of Ontario through the Ministry of Colleges and Universities. The Center for Computational Astrophysics at the Flatiron Institute is supported by the Simons Foundation. We have made use of the software packages \texttt{corner}~\cite{corner}, \texttt{SciPy}~\cite{scipy}, \texttt{NumPy}~\cite{numpy}, and \texttt{JAX}~\cite{jax2018github}.

\bibliography{eem.bib}
\bibliographystyle{JHEP}

\appendix
\section{Analysis Pipeline}
\begin{figure}
    \centering
    \includegraphics[width=\linewidth]{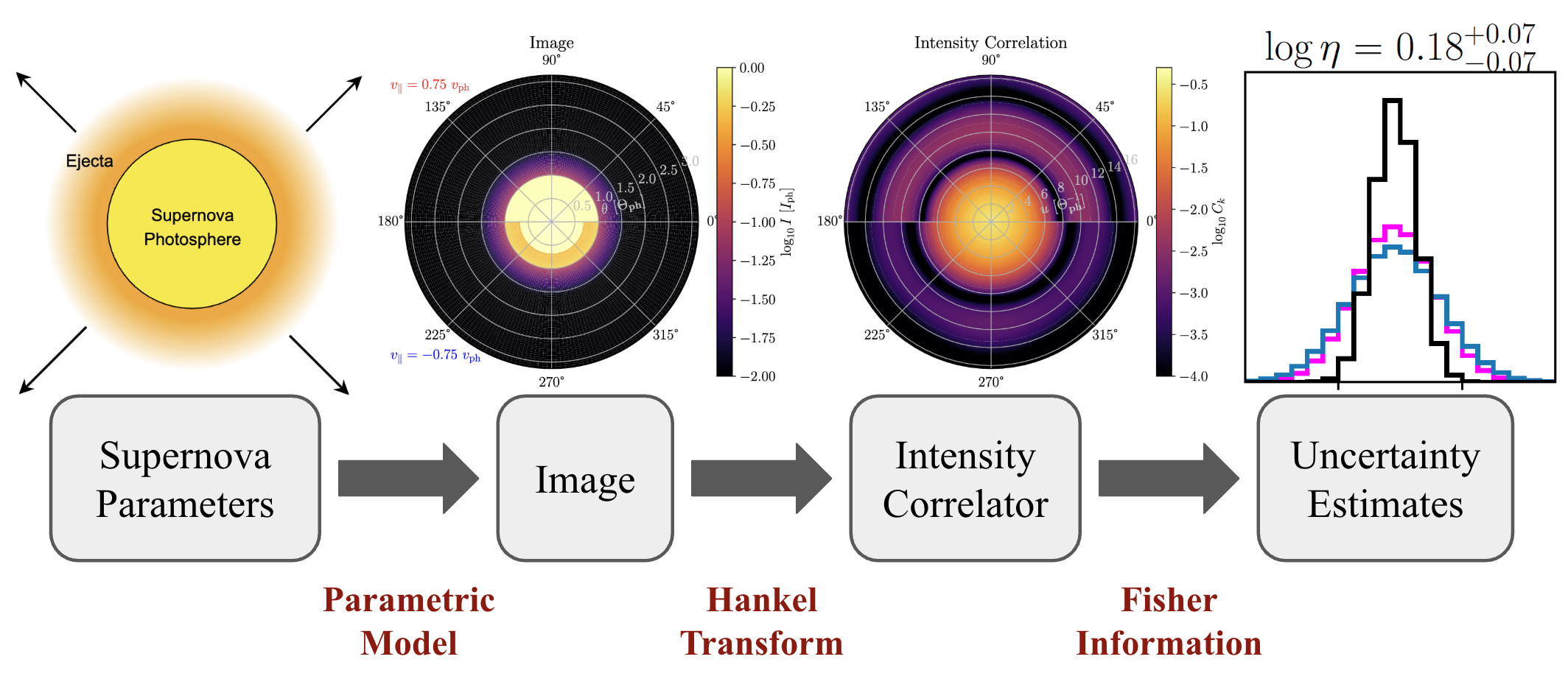}
    \caption{Graphical representation of the analysis pipeline we use in this paper to generate SN images, their corresponding intensity correlation, and uncertainty estimates using Fisher information.}
    \label{fig:pipeline}
\end{figure}

We outline the procedure used to generate SN images around a P Cygni line feature and to produce the corresponding sensitivity projections. A graphical overview of the analysis pipeline is shown in figure~\ref{fig:pipeline}.
\begin{enumerate}
    \item \textbf{Specify SN parameters:} Set key model parameters such as the ejecta velocity at the photosphere $v_\text{ph}$, the angular diameter distance $D_A$, and other relevant quantities.
    \item \textbf{Calculate line-of-sight coordinates:} For each spectral channel at wavelength $\lambda$ (given a fixed rest-frame wavelength $\lambda_\text{rest}$), compute the line-of-sight position $z$ of the ejecta as a function of time, following eq.~\eqref{eq:wavelength_relation}.
    \item \textbf{Compute optical depth and dilution factor:} Calculate the three-dimensional optical depth $\tau(x, y, \lambda)$ and the geometric dilution factor $W$ based on eqs.~\eqref{eq:tau_model} and~\eqref{eq:Wfunc}.
    \item \textbf{Generate 2D images:} For each spectral channel, create the 2D image by combining the photospheric emission, absorption, and recombination emission according to eqs.~\eqref{eq:Icont},~\eqref{eq:em_abs},~\eqref{eq:absorption}, and~\eqref{eq:em_model}.
    \item \textbf{Integrate over spatial coordinates:} Sum over the 2D image to obtain the spatially integrated spectrum, equivalent to evaluating the integral $\int \dd\Omega\, I_\lambda(\lambda, \vect{\theta})$.
    \item \textbf{Perform 2D Hankel transform:} Apply a two-dimensional Fourier transform in polar coordinates (a Hankel transform) to the images to obtain the intensity correlation function.
    \item \textbf{Conduct Fisher analysis:} Calculate the Hessian matrix of the negative log-likelihood function around the true model parameters to perform Fisher analysis and estimate parameter uncertainties~\cite{2024PhRvD.109l3029D}.

\end{enumerate}

\section{Effect of Model Parameters on Distance Measurement}
\label{app:spectral_index}

\begin{figure}
    \centering
    \includegraphics[width=\linewidth]{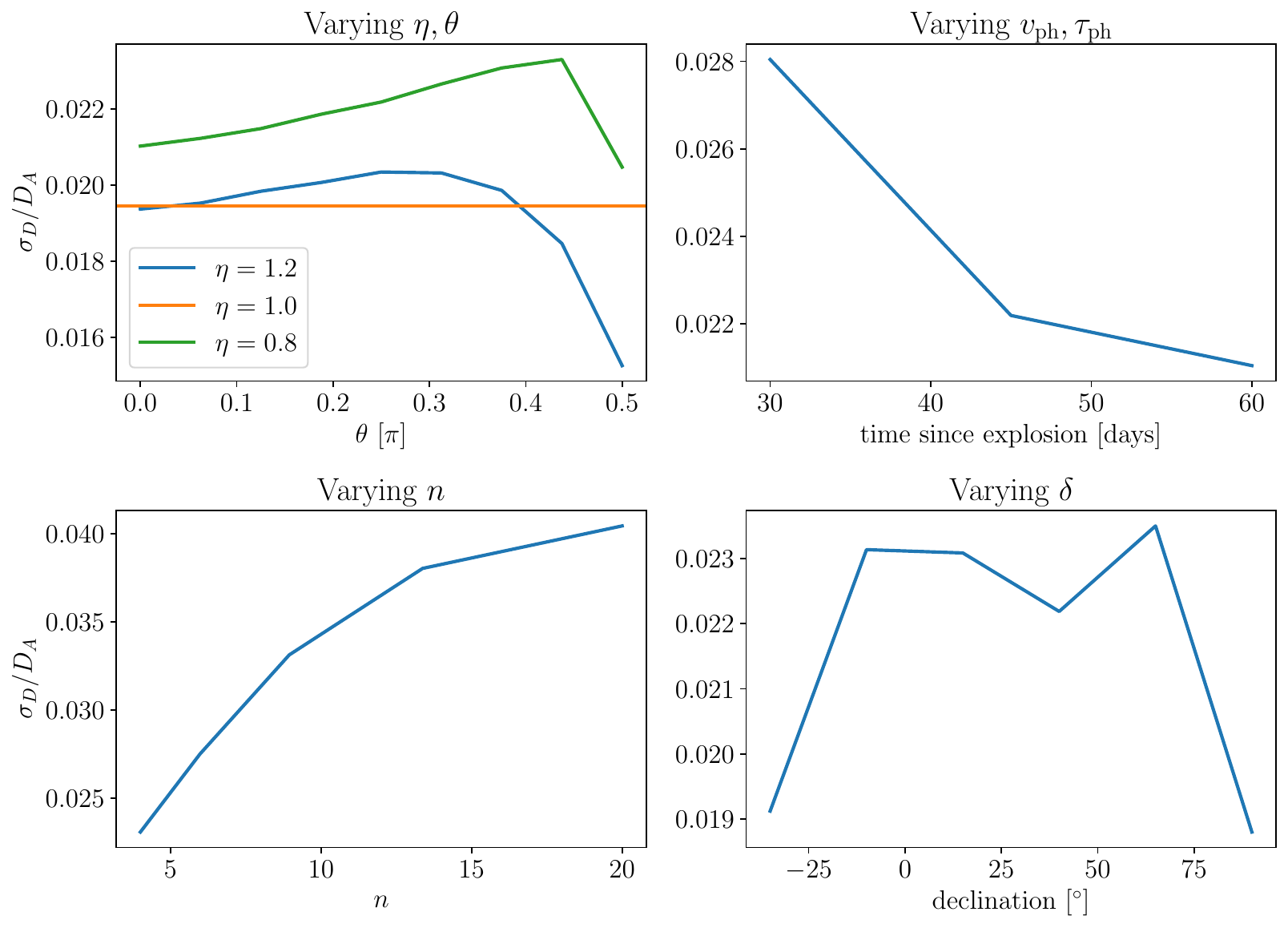}
    \caption{Impact of model parameters on angular diameter distance precision. Each panel shows the fractional uncertainty \( \sigma_D / D_A \) in the angular diameter distance measurement, computed using Fisher information, as individual model parameters are varied. \textbf{Top left:} Variation with asymmetry $(\eta)$ and inclination $(\theta)$. \textbf{Top right:} Evolution over time as the ejecta velocity ($v_{\text{ph}})$ and line optical depth $(\tau_{\text{ph}})$ change. \textbf{Bottom left:} Dependence on the spectral index $(n)$ of the optical depth profile. \textbf{Bottom right:} Variation with SN declination $(\delta)$, assuming a telescope array at 30$^\circ N$ latitude. \nblinkeem{plots/corner_plot_appendix.ipynb}}
    \label{fig:sensitivity}
\end{figure}

We explore how the variation of key parameters in the parametric model described in section~\ref{sec:parametric} affects the inference precision of the angular diameter distance  to SNe IIP using intensity interferometry. We compute the fractional uncertainty $\sigma_D/D_A$ using Fisher information and study the sensitivity to the following parameters: the degree of asymmetry $\eta$, inclination angle $\theta$, ejecta velocity at the photosphere $v_\text{ph}$, line optical depth at the photosphere $\tau_\text{ph}$, spectral index $n$ of the optical depth profile, and the declination of the SN $\delta$.
Figure~\ref{fig:sensitivity} summarizes the results:
\begin{itemize}
    \item \textbf{Top left panel:} Varying $\eta$ and $\theta$ (same parameters as in figure~\ref{fig:morphology}). The distance uncertainty is relatively insensitive to these parameters, though slightly improved precision is seen for $\eta = 1.2$ compared to $\eta = 0.8$. This is because a prolate deformation of the photosphere enlarges the emitting surface, improving the SNR in intensity correlation.
    
    \item \textbf{Top right panel:} Varying $v_\text{ph}$ and $\tau_\text{ph}$ while keeping the photosphere static (as in figure~\ref{fig:tauv}). The uncertainty improves at later times due to the increase in $\tau_\text{ph}$ which leads to deeper P Cygni lines and stronger contrast in both the spectrum and intensity correlation.

    \item \textbf{Bottom left panel:} Varying the spectral index $n$ between 4 and 20, consistent with previous P Cygni line modeling studies~\cite{castor1979atlas}. The distance uncertainty worsens as $n$ increases, since a steeper profile implies lower column density of line-producing atoms, reducing the depth and contrast of spectral features and intensity correlation. This again highlights that EEM’s sensitivity arises from expanding ejecta material at and beyond the photosphere, rather than from the  photosphere itself.

    \item \textbf{Bottom right panel:} Varying the SN's declination $\delta$ from $-35^\circ$ to $90^\circ$, assuming an interferometer array located at $30^\circ N$ latitude. The uncertainty remains fairly constant across declinations, though targets at lower $\delta$ may suffer from higher air mass and reduced nightly observing time.
\end{itemize}

Overall, this analysis shows that the proposed method is robust across a wide range of physical and observational configurations, with distance precision primarily limited by factors that reduce the depth in the P Cygni profile.

\section{Modeling Recombination Emission}
\label{app:recomb}

We model recombination emission as an additional scaling factor $\alpha_\text{em}$ applied to the emission component of the spectral radiance in eq.~\eqref{eq:em_abs}:

\begin{equation} I_\lambda(\lambda, \theta) = \alpha_{\text{em}} I^\text{em}(\lambda, \theta) + I^\text{abs}(\lambda, \theta), 
\end{equation}
where $\alpha_\text{em}$ controls the relative strength of emission compared to absorption in the P Cygni line profile. (No recombination emission corresponds to $\alpha_\mathrm{em} = 1$.)

\begin{figure}
    \centering
    \includegraphics[width = \linewidth]{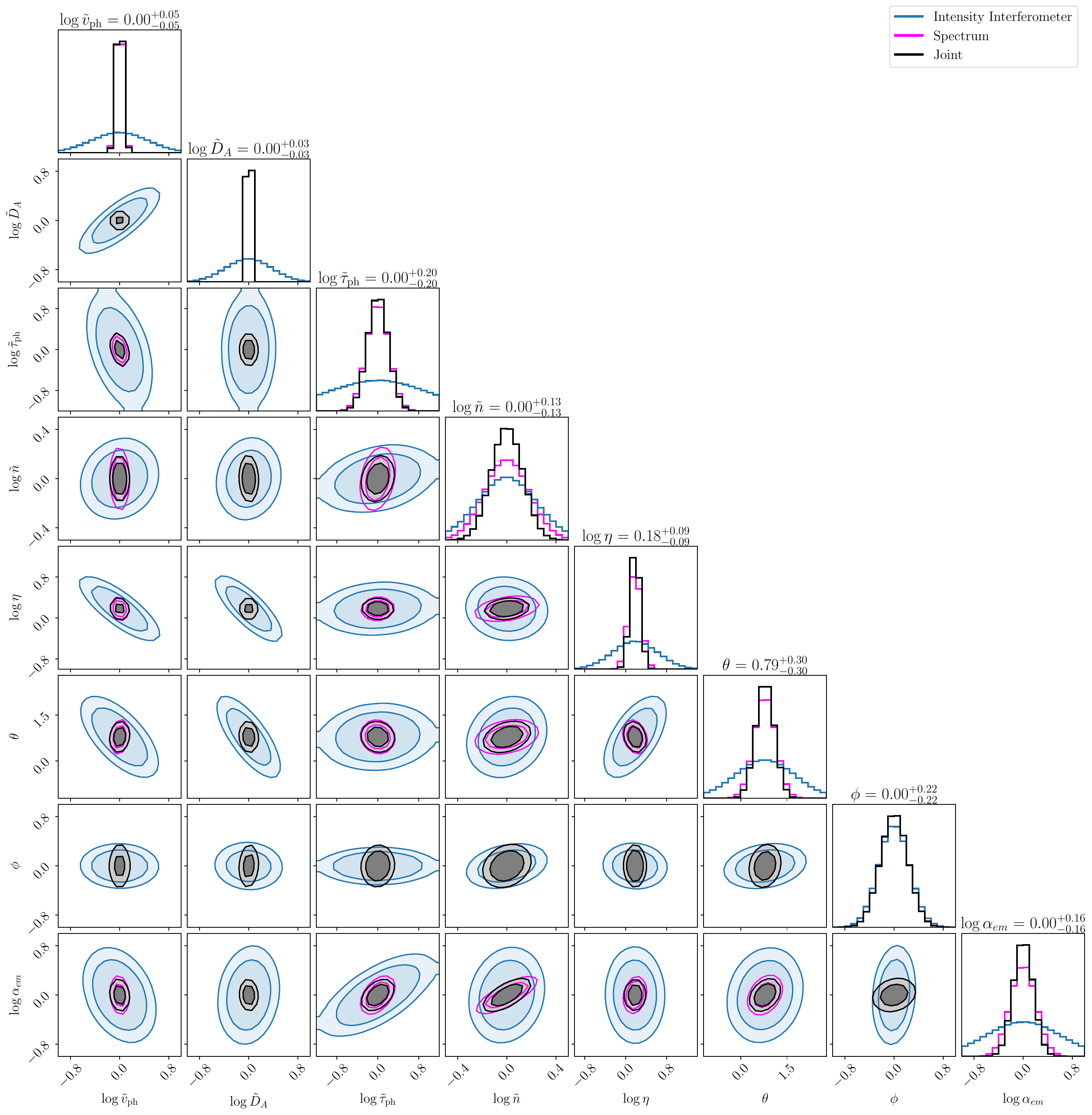}
    \caption{Corner plot showing parameter constraints for the extended parametric model that includes recombination emission as an additional scaling factor on the emission component of the P Cygni profile on the same benchmark SN as in figure~\ref{fig:corner}. The truth parameters are $v_\text{ph} = 6\times10^3~\text{kms/s},~D_A = 3~\text{Mpc},~\tau_\text{ph} = 2,~n = 4,~\eta = 1.2,~\theta = \pi/4,~\phi = 0,~\text{and}~\alpha_\text{em} = 1$. \nblinkeem{plots/corner_emission.ipynb}}
    \label{fig:recombination_corner}
\end{figure}

We perform a Fisher analysis on the benchmark SN described in section~\ref{sec:parameter-estimation}, extending the parametric model in section~\ref{sec:parametric} to include $\alpha_\text{em}$ as an additional free parameter. Figure~\ref{fig:recombination_corner} shows the resulting corner plot.
Notably, the inclusion of recombination emission degrades the precision of angular diameter distance $(D_A)$ and asymmetry $(\eta)$ measurements from intensity interferometry alone more significantly than from the spectral data alone. The joint likelihood analysis yields comparable uncertainties on $D_A$ and $\eta$ to the baseline model in figure~\ref{fig:corner}, but results in increased uncertainties for line profile parameters such as $\tau_\text{ph}, n$, and the Euler angles $\theta$ and $\phi$. This reflects the added degeneracy introduced by allowing for excess emission beyond that predicted by the base model.

\end{document}